\documentclass[12pt]{article}
\usepackage{amssymb,amsbsy,amsmath,amsfonts}
\usepackage{a4}
\usepackage{cite}
\usepackage{graphicx}
\usepackage[dvipsnames]{xcolor}
\usepackage{float}
\usepackage{rotating}

\newcommand{\epslash}{{\epsilon \!\!\!/}}
\newcommand{\pslash}{{p \!\!\!/}}
\newcommand{\kslash}{{k \!\!\!/}}
\newcommand{\qslash}{{q \!\!\!/}}
\newcommand{\Qslash}{{Q \!\!\!\!/\,}}
\newcommand{\npslash}{{n_+ \!\!\!\!\!\!\!/\,\,\,\,}}
\newcommand{\nmslash}{{n_- \!\!\!\!\!\!\!/\,\,\,\,}}

\newcommand{\eq}[1]{{(\ref{#1})}}
\newcommand{\bfp}{{\bf p}}
\newcommand{\bfe}{{\bf e}}

\begin{document}
\vspace{0.7cm}
\begin{center}
\Large\bf
\rm On $B \to V \ell \ell$
at small dilepton invariant mass, power corrections, and new physics
\unboldmath
\end{center}
\vspace{0.8cm}
\begin{center}
{\sc S.~J\"ager and J.~Martin Camalich}\\
\vspace{0.7cm}
{\sl Department of Physics, University of Sussex, Brighton BN1 9QH,
  UK}
\end{center}
\begin{abstract}
We investigate rare semileptonic $\bar B \to \bar K^* \ell^+ \ell^-$
decays, providing a comprehensive treatment of
theoretical uncertainties in the low-$q^2$ region as
needed for interpreting current and future LHCb and $B$-factory
data in terms of the new physics search.
We go beyond the usual focus on form-factor
uncertainties, paying proper attention to non-factorizable terms.

A central point is the systematic
exploitation of the $V-A$ structure of SM weak
interactions, which leads to the suppression of two helicity
amplitudes and some of the angular coefficients. We review how this
works at the level of (helicity) form factors, and show
that the hierarchies extend to non-factorizable terms. For virtual charm
effects, we give an argument for it in terms of light-cone QCD sum
rules that continues to hold at the level of ``long-distance''
$\Lambda_{\rm QCD}^2/{m_c^2}$ power corrections, reducing an
important source of theoretical uncertainty in any 
$\bar B, \bar B_s \to V \ell^+ \ell^-$ (or
$\bar B \to V \gamma$) decay.
The contributions of the remaining hadronic
weak Hamiltonian respect a similar hierarchy.
We employ a resonance model to preclude
(in the $\bar B \to \bar K^*$ case) large long-distance corrections to this.

A phenomenological part pays particular attention to the
region of lowest dilepton mass,  $4 m_\ell^2 \le q^2 \le 2$ GeV${}^2$.
Two observables remain theoretically clean, implying a
(theoretical) sensitivity to the real (imaginary) part of
the ``right-handed'' Wilson coefficient $C_{7}'$ to 10\% (1\%)
of $C_7^{\rm SM}$, both in the muonic and the
electronic mode. We also show that there are two near-exact
relations between angular coefficients, even in the presence of new
physics and when lepton masses are not neglected.
\end{abstract}

\newpage

\section{Introduction}
Rare $B$ decays are CKM and/or loop-suppressed in the Standard Model (SM), which
accounts for their small branching fractions and makes them sensitive
to small contributions from possible new degrees of freedom beyond the
SM (BSM). Unfortunately, they generally suffer from
non-perturbative strong-interaction effects,
the perennial bugbear of flavour (and collider) physics and
difficult to control theoretically.
An example where such uncertainties are small
is provided by the decay $B_s \to \mu^+ \mu^-$, for which LHCb has
just published first evidence \cite{:2012ct},
\begin{equation}
  BF(B_s \to \mu^+ \mu^-)_{\rm exp} =
     (3.2^{+1.4}_{-1.2}|_{\rm stat}{}^{+0.5}_{-0.3}|_{\rm syst} )
       \times 10^{-9} \quad \mbox{[$3.5 \sigma$ significance]} ,
\end{equation}
to be contrasted with the SM theory prediction,
\begin{equation}
  BF(B_s \to \mu^+ \mu^-)_{\rm SM} = (3.54 \pm 0.30) \times 10^{-9}, 
\end{equation}
which incorporates estimates of soft-photon emissions and
takes account of the finite lifetime difference of
the $B_s$ \cite{DeBruyn:2012wk,Buras:2012ru}.
This result strongly constrains new-physics contributions through
scalar $\times$ scalar semileptonic operators, such as may occur
in the MSSM at large $\tan\beta$. It also constrains physics entering
through a modified $Zbs$ vertex, although less strongly, and
in view of the statistics foreseen at LHCb, this
will remain so for the foreseeable future.
Beyond this single, exceptionally clean example, one generally has
to deal with amplitude ratios including strong rescattering phases, which are
currently not tractable with a first-principles method like lattice
QCD. One faces a trade-off between theoretical cleanness (as in
leptonic decays) and number of modes, which increases with the number
of hadrons in the final state, as does the theoretical complexity.

Rare semileptonic and radiative decays such as $\bar B \to \bar K^* \mu^+ \mu^-$
and $\bar B \to \bar K^* \gamma$ provide, in a sense, the best of both worlds.
On the one hand, they have a rich kinematic and helicity structure.
Experimental results on $\bar B \to \bar K^* \mu^+ \mu^-$
include measurements of the forward-backward
asymmetry and the longitudinal polarisation fractions
by the $B$-factories, CDF, and LHCb
\cite{Wei:2009zv,Aaltonen:2011cn,LHCbNote,Ritchie:2013mx}, of
the angular observables $A_T^{(2)}$ and $A_{\rm im}$ by
CDF \cite{Aaltonen:2011cn,Aaltonen:2011ja} and related
angular coefficients by LHCb \cite{LHCbNote},
and  measurements of non-angular observables (e.g.\
\cite{Aubert:2008ps,:2012vwa,Aaij:2012cq}).
Results on the full angular distribution are anticipated from LHCb. 
$\bar B_s \to \phi \mu^+ \mu^-$ has also been observed
\cite{Aaltonen:2011qs,Schaack:2012fsa}.
On the theory side, there have been a large number of conceptual
and phenomenological studies
(e.g.\ 
\cite{Ali:1991is,Burdman:1998mk,Melikhov:1998cd,Ali:1999mm,Kruger:1999xa, Kim:2000dq, Burdman:2000ku,Grossman:2000rk,Kim:2001xua, Beneke:2001at,Bosch:2001gv,Ali:2002qc,Faessler:2002ut,Kagan:2001zk,Feldmann:2002iw,Beneke:2004dp,Grinstein:2004vb,Grinstein:2005ud,Kruger:2005ep,Matias:2005rs, Ball:2006cva, Ball:2006eu,Lunghi:2006hc,Bobeth:2008ij,Egede:2008uy, Altmannshofer:2008dz,Egede:2009te,Egede:2009tp,Bobeth:2010wg,Bharucha:2010bb,Alok:2009tz, Egede:2010zc,Khodjamirian:2010vf,Alok:2010zd,Kou:2010kn,Chang:2010zy,ReeceThesis, Beylich:2011aq, Bobeth:2011gi,Becirevic:2011bp,Alok:2011gv, Lu:2011jm,Wang:2011aa,Matias:2012xw,Becirevic:2012dp,Korchin:2012kz,Das:2012kz, Matias:2012qz,Blake:2012mb,Altmannshofer:2011gn,DescotesGenon:2011yn,Beaujean:2012uj,DescotesGenon:2012zf,Altmannshofer:2012az,Behring:2012mv}).
On the other hand, involving only one hadron in the initial and final
states, these modes approximately factorize naively,
into products of nonperturbative
form factors and Wilson coefficients given in terms of short-distance
SM and BSM vertices. Naive factorization is, however, violated by
the hadronic weak hamiltonian. Schematically,
\begin{eqnarray}
 {\cal A}(\bar B \to V \ell^- \ell^+) &=&
    \sum_i C_i \langle \ell^- \ell^+ | \bar l \Gamma_i l | 0 \rangle
    \langle V | \bar s \Gamma'_i b | \bar B \rangle \nonumber \\
&&  + \frac{e^2}{q^2} \langle \ell^- \ell^+ | \bar l \gamma^\mu l | 0 \rangle
    F.T. \langle V | T\{ j_{\mu, \rm em}^{\rm had}(x) {\mathcal H}_{\rm
      eff}^{\rm had}(0) \} |
    \bar B \rangle ,    \label{eq:BVllschematic}
\end{eqnarray}
where $C_i$ are semileptonic Wilson coefficients, $j_{\rm em}^{\rm had}$
is the hadronic part of the electromagnetic current, and ${\mathcal
  H}_{\rm eff}^{\rm
  had}$ is the hadronic weak $\Delta F=1$ hamiltonian.
This equation is correct up to higher orders in the QED coupling
$\alpha_{\rm em}$. The form factors $\langle V | \bar s \Gamma_i' b | \bar B
\rangle$ are, in principle, accessible on the lattice.
For small dilepton invariant mass, they also lend themselves
to light-cone sum rules \cite{Ball:1998kk}, and satisfy certain constraints from
heavy-quark symmetry/large-energy relations \cite{Charles:1998dr}.
Making use of these relations, a number of observables can be shown
to have reduced sensitivity to hadronic (form factor) uncertainties
(see eg \cite{Burdman:1998mk,Ali:1999mm,Kruger:2005ep,Matias:2012xw}).
However, this cleanness is compromised by the non-factorizable terms
on the second line of \eq{eq:BVllschematic}, which pose the
most challenging theoretical problem.

A  breakthrough has been a comprehensive treatment of these effects
within QCD factorization (QCDF) based on the heavy-quark limit
\cite{Beneke:2001at,Bosch:2001gv},
valid for large hadronic recoil and in particular
in the ``low-$q^2$'' region of dilepton invariant mass
below the charmonium resonances, and for the radiative decay.
At leading power in an expansion in $\Lambda/m_b$, where $\Lambda
\equiv \Lambda_{\rm QCD}$ is the dynamical QCD scale, the
terms on the second line in \eq{eq:BVllschematic} are expressed in
terms of form factors, decay constants, and convolutions of light-cone
distribution amplitudes of the initial and final state mesons
with  perturbative hard-scattering kernels.
The same
framework also allows one to systematically compute perturbative
corrections to the form factor relations \cite{Beneke:2000wa}. QCD factorization
breaks at subleading powers, due to end-point divergences.
The power corrections that do not factorize have to be modeled or constrained
by some means, and those that do factorize are only partially known.
Phenomenological analyses tend to either ignore power corrections or employ ad hoc estimates.
On the other hand, a recent analysis  \cite{Khodjamirian:2010vf}
of power corrections within the framework of light-cone QCD sum
rules (LCSR) found the possibility of large effects,
of ${\cal O}(10 \%)$ at the amplitude level.
Importantly, some of the power corrections are of order
$\Lambda^2/(4 m_c^2)$, rather than $\Lambda/m_b$.
All this is sufficient to cast into doubt the cleanness of the optimized
observables.

In this paper, we revisit the issue of long-distance 
corrections, employing a systematic description
in terms of helicity (as opposed to transversity) amplitudes.
The point is that, when the treatment of \cite{Khodjamirian:2010vf} is
reformulated in terms of helicity amplitudes, one can show a
systematic suppression of the contributions of four-quark operators
involving charm quarks to one (out of three) long-distance-sensitive
helicity amplitudes, to be termed $H_V^+$. This suppression
in $\Lambda/m_b$ holds even at the level of $\Lambda^2/(4 m_c^2)$
power corrections. A suppression is likewise
observed for the contributions of the chromomagnetic penguin
operator $Q_{8g}$ to $H_V^+$. Moreover, if one adopts a resonance
dominance model of long-distance contributions at low $q^2$,
one finds again a suppression of the amplitude in question. This
eliminates the leading candidate of large ``duality-violating''
corrections to the QCDF results (and power-correction estimates)
for the latter contributions. 

The phenomenological relevance is as follows: Of the so-called ``clean''
observables, there are precisely two which vanish in the limit
$H_V^+ \to 0$. These emerge as truly theoretically clean null tests of
the Standard model, even with a conservative treatment of
contributions from the hadronic weak Hamiltonian.
Our results also show that this remains so at
the $q^2 \to 0$ end of the dilepton invariant mass spectrum, which
translates into good sensitivity to the ``right-handed'' magnetic
Wilson coefficient $C_7'$, as has been suggested
earlier \cite{Becirevic:2012dx}, even
when fully taking account of hadronic effects. For the other
observables, however, the new physics sensitivity becomes more
limited.

The remainder of this paper is organised as follows. In
Section \ref{sect:amps} we set up our notation and review
(and simplify) the decomposition of the decay amplitude and angular
distribution, and ``clean'' observables, in helicity amplitudes.
Section~\ref{sect:hadronic}
contains a detailed discussion of sources of hadronic uncertainties
both in the factorizable and non-factorizable contributions to
the helicity amplitudes, and establishes the suppression of $H_V^+$.
We also employ a parameterization of form factors at low $q^2$ which
transparently separates the constraints from kinematics and the
heavy-quark limit from the issue of modelling power corrections.
Section~\ref{sec:Pheno} comprises a detailed phenomenology
of the ``clean'' observables, with particular attention to
the low end of the low-$q^2$ region, which has traditionally been
cut off ad hoc at $q^2 = 1$ GeV${}^2$. We find that the observables
$P_1$ and $P_3^{\rm CP}$ (in the notation of \cite{Matias:2012xw})
in $\bar B \to \bar K^* \mu^+ \mu^-$
stand out as theoretically cleanest, translating to very good sensivity
to right-handed currents, via the Wilson coefficients $C_7'$, $C_9'$,
and $C_{10}'$. Specifically, we assess the (theoretical)
sensitivity to the real and imaginary parts of $C_7'$ to be on the
order of $10 \%$ and $1
\%$, respectively, with the sensitivity coming entirely from
the region $q^2 < 3\, {\rm GeV}^2$, and dominated by the $q^2$-interval
$ [0.1, 2]\, {\rm GeV}^2$.
We also comment on the electronic mode, which shows a
theoretical sensitivity to $C_7'$ very similar to the muonic mode.
Throughout we take into account both the
small but nonzero values of right-handed Wilson coefficients
in the SM and the effect of a nonzero muon mass, and show that two
known algebraic relations in the massless case can be modified such
that they hold, to excellent accuracy, in the presence of a finite
muon mass all the way down to the kinematic end point.
Section \ref{sect:conclusions} contains our conclusions.

\section{Amplitudes and kinematic distribution}
\label{sect:amps}
\subsection{Weak Hamiltonian}
The process $\bar B(p) \to M(k) \ell^+ \ell^-$, where $M$ is a charmless
final state (not necessarily a single meson),
is mediated by the $\Delta B=1$ weak effective Hamiltonian,
which is a sum of hadronic and semileptonic parts (where
``semileptonic'' is understood to include the magnetic penguin terms),
\begin{equation}
  {\cal H}_{\rm eff} = {\cal H}_{\rm eff}^{\rm had} + {\cal H}_{\rm
      eff}^{\rm sl} ,
\end{equation}
with
\begin{equation}
  {\cal H}_{\rm eff}^{\rm had} =
     \frac{4 G_F}{\sqrt{2}} \sum_{p=u,c} \lambda_p \left[
       C_1 Q_1^p + C_2 Q_2^p + \sum_{i=3\dots 6} C_i P_i + C_{8g}
       Q_{8g} \right] ,
\end{equation}
\begin{equation}
 \begin{aligned}
  {\cal H}_{\rm eff}^{\rm sl} &=& - \frac{4 G_F}{\sqrt{2}} \lambda_t \Big[
     C_{7} Q_{7\gamma} + C'_{7} Q'_{7\gamma}
    + C_{9} Q_{9V} + C'_{9} Q'_{9V}
    + C_{10} Q_{10A} + C'_{10} Q'_{10A} \\
    &&+ C_S Q_S + C'_S Q'_S
    + C_P Q_P + C'_P Q'_P
    + C_T Q_T + C'_T Q'_T  \Big] .
 \end{aligned}
\end{equation}
The operators $P_i$ are given in~\cite{Chetyrkin:1996vx}, the
$Q_i$ are defined as
\begin{equation}
\begin{aligned}
  Q_{7\gamma} &= \frac{e}{16\pi^2}\,\hat{m}_b\,
    \bar s\sigma_{\mu\nu} P_R  F^{\mu\nu} b \,, \\
  Q_{9V} &= \frac{\alpha_{\rm em}}{4\pi} (\bar s \gamma_\mu P_L b) (\bar l
  \gamma^\mu l) \,, \\
  Q_{S} &= \frac{\alpha_{\rm em}}{4\pi} \frac{\hat m_b}{m_W} (\bar s P_R b) (\bar l l) \,, \\
  Q_{T} &= \frac{\alpha_{\rm em}}{4\pi} \frac{\hat m_b}{m_W} (\bar s
  \sigma_{\mu\nu} P_R b) (\bar l \sigma^{\mu\nu} P_R l) \,,
\end{aligned}
\qquad\quad
\begin{aligned}
\\
  Q_{8g} &= \frac{g_s}{16\pi^2}\,\hat{m}_b\,
    \bar s\sigma_{\mu\nu} P_R G^{\mu\nu} b \,, \\
  Q_{10A} &= \frac{\alpha_{\rm em}}{4\pi} (\bar s \gamma_\mu
  P_L b) (\bar l \gamma^\mu \gamma^5l)_{A} \,, \\
  Q_{P} &= \frac{\alpha_{\rm em}}{4\pi} \frac{\hat m_b}{m_W} (\bar s P_R b) (\bar l
  \gamma^5 l) \,, \\
\qquad
\end{aligned}
\label{operators}
\end{equation}
and the primed operators $Q'_i$ are obtained from these by
$P_R \to P_L, P_L \to P_R$ in the quark bilinears.
$g_s$ ($e$) denotes the strong (electromagnetic) coupling constant
coming from the covariant derivative
$D_\mu=\partial_\mu+ieQ_fA_\mu+ig_sT^AA_\mu^A$ ($Q_f=-1$ for the leptons),
$\alpha_{\rm em}= e^2/(4 \pi)$  and
$\hat{m}_b$ the $b$-quark mass defined in the $\overline{\rm MS}$
scheme.

The contribution of the semileptonic Hamiltonian $\mathcal H_{\rm eff}^{\rm sl}$
to the decay amplitude factorizes (in the ``naive'' sense)
into a sum of products of hadronic and leptonic currents,
\begin{equation}
\mathcal A^{\rm sl} = \langle M \ell^+ \ell^- | {\cal H}_{\rm eff}^{\rm sl} | \bar B \rangle
=  L_V^\mu \, a_{V\mu} + L_A^{\mu} \, a_{A \mu}
  + L_{S} \, a_{S} + L_{P} \, a_{P}
  + L_{TL}^{\mu} \, a_{TL,\mu} + L_{TR}^\mu \, a_{TR,\mu} ,
\label{eq:slamplitude}
\end{equation}
where
\begin{equation}
 \begin{aligned}
  L_V^\mu &= \langle \ell^+ \ell^- | \bar l \gamma^\mu l | 0 \rangle , \\
  L_S &= \langle \ell^+ \ell^- | \bar l l | 0 \rangle , \\
  L_{TL}^\mu &= \frac{i}{\sqrt{q^2}} \langle \ell^+\ell^- | q_\nu\bar l \sigma^{\mu\nu} P_L
   l | 0 \rangle ,
 \end{aligned}
 \qquad
 \begin{aligned}
  L_A^{\mu} &= \langle \ell^+ \ell^- | \bar l \gamma^\mu \gamma^5 l | 0
    \rangle , \\
  L_P &= \langle \ell^+ \ell^- | \bar l \gamma^5 l | 0 \rangle , \\
  L_{TR}^\mu &= \frac{i}{\sqrt{q^2}} \langle \ell^+\ell^- | q_\nu\bar l \sigma^{\mu\nu} P_R
  l | 0 \rangle ,
 \end{aligned}
\end{equation}
and we have made use of the relation
\begin{equation}
(\bar s \sigma_{\mu\nu} P_{R(L)} b) (\bar l \sigma^{\mu\nu} P_{R(L)} s) =
\frac{4}{q^2} (\bar s q_\nu \sigma^{\mu\nu} P_{R(L)} b)
  (\bar l q_\rho \sigma^{\mu\rho} P_{R(L)} l) ,
\label{eq:tensorrelation}
\end{equation}
where $q = p-k$ is the dilepton
four-momentum.\footnote{Equation \eq{eq:tensorrelation} holds
for arbitrary time-like four-vector $q_\mu$.}
The hadronic currents $a_{V\mu}, \dots$ are expressed in terms of
form factors and Wilson coefficients, and enter the helicity
amplitudes given below.

The hadronic Hamiltonian $\mathcal H_{\rm eff}$ requires in addition two
insertions of the electromagnetic current (one hadronic and one
leptonic) to mediate the semileptonic decay,
\begin{equation}
\begin{aligned}
\mathcal A^{\rm (had)} &= - i \frac{e^2}{q^2} \!\! \int \!\! d^4x e^{- i q \cdot x}
   \langle \ell^+ \ell^- | j_\mu^{\rm em, lept}(x) | 0 \rangle
 \!\! \int \!\! d^4 y\, e^{i q \cdot y}   \langle M | T \{ j^{\rm em, had, \mu}(y)
 {\mathcal H}^{\rm had}_{\rm eff}(0) \} | \bar B \rangle
   \\
&\equiv  \frac{e^2}{q^2} L^\mu_V a_\mu^{\rm had} \; ,     \label{eq:amuhad}
\end{aligned}
\end{equation}
where $j^{{\rm em, had},\mu} = \sum_q e_q \bar q \gamma^\mu q$.
Hence, while this
contribution does not naively factorize, it can be absorbed into
$a_{V\mu}$ in \eq{eq:slamplitude}.
Before discussing the amplitudes in more detail,
we comment on the approximations implicit in and some consequences of
\eq{eq:slamplitude}, \eq{eq:amuhad}
 
\begin{itemize}
\item The semileptonic weak Hamiltonian is the most general one up
to dimension six
and can accomodate arbitrary new physics with a heavy mass scale. This
includes all the standard scenarios, such as supersymmetry,
extra dimensions and little Higgs. In the Standard Model,
$C_{7}$, $C_{9}$ and $C_{10}$ are sizable,
$C_{7}'$ is suppressed by $m_s/m_b$, and the remaining
Wilson coefficients are negligible.
\item The hadronic weak Hamiltonian is the Standard Model one,
  neglecting the small electroweak penguin terms. Beyond
the Standard Model, there is a large number of extra operators;
however unless new physics effects are dramatic their impact
(through $a_\mu^{\rm had}$) will be very small and we will ignore them
below. Such scenarios are also constrained by hadronic $B$ decay data.
\item We work to leading order in the electromagnetic
coupling, but all formulae so far are exact in
the strong coupling, with non-factorizable effects confined to
$a_\mu^{\rm had}$.
\item The leptonic currents can be decomposed into spin-0 and spin-1
  terms ($L^\mu_V$, $L^\mu_A$) or are pure spin-1 objects ($L^\mu_{TL}$,
  $L^\mu_{TR}$). It follows that the dilepton can only be created in a
  spin-0 or spin-1 state.
  Angular momentum conservation then implies that $\lambda$ is also the
  helicity of $M$, which is thus constrained to the values $\pm1$ or
  $0$ even if $M$ has spin greater than one.\footnote{
  This statement is exact, rather than a
  consequence of naive factorization, following from the well-known
  fact that a particle's orbital angular momentum does not contribute
  to its helicity.
  If $M$ is a multiparticle state, eg $K\pi$, we mean by ``spin''
  the total angular momentum of $M$ in its cm frame and by ``helicity'' the
  projection of the $M$ angular momentum onto the total $M$ momentum
  in the $\bar B$ rest frame.}
\end{itemize}

\subsection{Helicity amplitudes and helicity form factors}
We now carry out the decomposition of the leptonic currents in spins
and helicities. The resulting coefficients give the well-known
``helicity amplitudes'' \cite{Melikhov:1998cd}.
This is easily achieved \cite{Altmannshofer:2008dz}
through the completeness relation (see Appendix~\ref{app:polavectors}
for our conventions)
\begin{equation}
  \eta_{\mu\nu} = \epsilon_{t,\mu} \epsilon_{t,\nu}^*
     - \sum_{\lambda =\pm1,0} \epsilon_\mu(1, \lambda) \epsilon^*_\nu(1,
     \lambda) .
\end{equation}
Here  $\epsilon(1, \lambda)$,
$\lambda = \pm1, 0$, denotes a (spin-1) helicity triplet of
polarisation 4-vectors for a vector particle of four-momentum $q^\mu$
and mass $\sqrt{q^2}$, and $\epsilon_t^\mu =
q^\mu/\sqrt{q^2}$.
We may picture the latter as the ``time-like'' polarization four-vector of
an auxiliary virtual gauge boson of mass $\sqrt{q^2}$, but the
decomposition works independently of the origin of the weak
Hamiltonian, and also for the tensorial currents. The result is
\begin{equation}
 \begin{aligned}
 \mathcal A &= - \sum_{\lambda=\pm1,0} \mathcal L_V(\lambda) H_V(\lambda)
  - \sum_{\lambda=\pm1,0} \mathcal L_A(\lambda) H_A(\lambda)
  \; + L_{S} H_{S} \; + L_{P} H_{P}
\\
  & \quad\;
  - \sum_{\lambda=\pm1,0} \mathcal L_{TL}(\lambda) H_{TL}(\lambda)
  - \sum_{\lambda=\pm1,0} \mathcal L_{TR}(\lambda) H_{TR}(\lambda)  ,
 \end{aligned}
\end{equation}
where
\begin{equation}
 \begin{aligned}
  \mathcal L_V(\lambda) &= \epsilon_\mu(\lambda) L_V^\mu \,, \\
  \mathcal L_A(\lambda) &= \epsilon_\mu(\lambda) L_A^\mu \,, \\[0.5mm]
  \mathcal L_{TL}(\lambda) &= \epsilon_\mu(\lambda) L_{TL}^\mu \,, \\
  \mathcal L_{TR}(\lambda) &= \epsilon_\mu(\lambda) L_{TR}^\mu \,, \\[0.5mm]
  \mathcal L_S &= L_S \,, \\[2.5mm]
  \mathcal L_P &= L_P \,, \\[2.5mm]
 \end{aligned}
\qquad 
 \begin{aligned}
  H_V(\lambda) &= \epsilon^*_\mu(\lambda) a_V^\mu \,, \\
  H_A(\lambda) &= \epsilon^*_\mu(\lambda) a_A^\mu \,, \\
  H_{TL}(\lambda) &= \epsilon^*_\mu(\lambda) a_{TL}^\mu \,, \\
  H_{TR}(\lambda) &= \epsilon^*_\mu(\lambda) a_{TR}^\mu \,, \\
  H_S &= a_S \,\\
  H_P &= a_P + \frac{2 m_\ell}{q^2} q_\mu a_A^\mu \,.
 \end{aligned}
\end{equation}
We have made use of the fact that all leptonic
currents except for $L^\mu_A$ are conserved, so $\epsilon_{t,\mu}$
contracts to zero with them. Moreover, the axial current obeys
$q_\mu L^\mu_A = 2 m_\ell L_P$, which allowed us to absorb
the spin-zero axial vector amplitude into $H_P$
\cite{Altmannshofer:2008dz}.\footnote{We do not distinguish between the
lepton mass $m_\ell$ and the lepton field mass parameter, as we will
work to leading order in the electromagnetic coupling.}

The helicity amplitudes $H_V, H_A, H_P, H_S$ are related to
the ``standard'' helicity
amplitudes~\cite{Kruger:1999xa,Altmannshofer:2008dz} as follows,
\begin{equation}
H_{\lambda L/R} =  i\,\sqrt{f}\, \frac{1}{2} (H_V(\lambda) \mp H_A(\lambda)) , \qquad 
A_{t} =  i\, \frac{\sqrt{q^2}}{2m_\ell} \sqrt{f}\, H_P , \qquad 
A_{S} = -i\,\, \sqrt{f} \, H_S ,
\end{equation}
where $f$ is a normalization factor, which for $M = K^*$
and the conventions of \cite{Altmannshofer:2008dz} is equal to $F$
defined in Section \ref{sect:kindist} below.
The helicity amplitudes $H_{\pm 1, L(R)}$ are often
expressed in terms of transversity amplitudes,
\begin{equation}
A_{\parallel L(R)} = \frac{1}{\sqrt{2}} (H_{+1,L(R)} + H_{-1,L(R)}), \qquad
A_{\perp L(R)} = \frac{1}{\sqrt{2}} (H_{+1,L(R)} - H_{-1,L(R)}) .
\end{equation}
However, we will work with helicity amplitudes throughout this paper,
for reasons to become clear below. 
Explicitly, we have
\begin{eqnarray}     \label{eq:helampfirst}
  H_V(\lambda) &=&-i\, N \Big\{
      C_{9} \tilde{V}_{L\lambda} +C_{9}'  \tilde{V}_{R\lambda}
      + \frac{m_B^2}{q^2} \Big[\frac{2\,\hat m_b}{m_B} (C_{7} \tilde{T}_{L\lambda} +
            C_{7}'  \tilde{T}_{R\lambda})
            - 16 \pi^2 h_\lambda \Big] \Big\} , \nonumber \\ && \\
  H_A(\lambda) &=& -i\, N (C_{10}  \tilde{V}_{L\lambda} + C_{10}'\tilde{V}_{R\lambda}) , \\
  H_{TR}(\lambda) &=&-i\, N \frac{4\, \hat m_b\, m_B}{m_W \sqrt{q^2}}\,C_{T} \tilde{T}_{L\lambda} , \\
  H_{TL}(\lambda) &=&-i\, N \frac{4\, \hat m_b\, m_B}{m_W \sqrt{q^2}}\,C_{T}' \tilde{T}_{R\lambda} , \\
  H_S &=& i\, N \frac{\hat m_b}{m_W} (C_S \tilde{S}_L + C_S' \tilde{S}_R), \\
  H_P &=& i\, N \Big\{ \frac{\hat m_b}{m_W} (C_P \tilde{S}_L + C_P' \tilde{S}_R)
\nonumber \\ && \qquad                + \frac{2\,m_\ell \hat m_b}{q^2} \left[
                     C_{10} \Big(\tilde{S}_L - \frac{m_s}{m_b} \tilde{S}_R \Big)
                     + C_{10}' \Big(\tilde{S}_R - \frac{m_s}{m_b} \tilde{S}_L\Big)
                \right] \Big\} ,
 \label{eq:helamplast}
\end{eqnarray}
where
$$
N = -\frac{4 G_F m_B}{\sqrt{2}} \frac{e^2}{16\pi^2} \lambda_t 
$$
is a normalisation factor,
\begin{equation}  \label{eq:hlambda}
h_\lambda \equiv \frac{i}{m_B^2} \epsilon^{\mu*}(\lambda) a^{\rm
  had}_\mu
\end{equation}
contains the contribution from the hadronic hamiltonian, i.e.\ all
non-factorizable effects, and we have defined helicity form factors
\begin{eqnarray}
  -i m_B \tilde{V}_{L(R)\lambda}(q^2) &=& \langle M (\lambda)| \bar s \epslash^{*}(\lambda)
                     P_{L(R)} b | \bar B \rangle , \\
  m_B^2 \tilde{T}_{L(R)\lambda}(q^2) &=& \epsilon^{*\mu}(\lambda) q^\nu
     \langle M (\lambda)| \bar s \sigma_{\mu\nu} P_{R(L)} b | \bar B \rangle , \\
  i m_B \tilde{S}_{L(R)}(q^2) &=& \langle M(\lambda=0) | \bar s P_{R(L)}  b | \bar B \rangle .
\end{eqnarray}
These expressions are still general enough to describe an arbitrary
charmless final state $M$. Concretely, for a
two-spinless-meson final state, not necessarily originating
from a resonance, the form factors will
carry dependence on the dimeson invariant mass $k^2$
and its angular momentum $L$, in addition to the dilepton invariant
mass $q^2$.

Note that parity invariance of strong interactions implies the relations
\begin{eqnarray}
  \tilde{V}_{L\lambda} &=& - \eta (-1)^{L} \tilde{V}_{R,-\lambda} \equiv \tilde{V}_{\lambda}, \\
  \quad \tilde{T}_{L\lambda} &=& - \eta (-1)^L \tilde{T}_{R,-\lambda} \equiv \tilde{T}_{\lambda}, \\
  \quad \tilde{S}_L &=& - \eta (-1)^L \tilde{S}_R \equiv \tilde{S} ,
\end{eqnarray}
where $s$ and $\eta$ are (respectively) the angular momentum
and intrinsic parity of $M$.
For a resonance, its spin $s$ replaces $L$.
Hence there are seven independent helicity form factors for spin $\ge 1$ and
three for spin 0 (when $\lambda = 0$).
Helicity form factors have previously been used in the literature
as a technical vehicle in constraining form factors from unitarity
\cite{Bharucha:2010im}.
As we will explain in detail below, helicity form factors
are also preferable over the standard basis for form factors in weak
decays: Not only do they simplify the expressions, but some of them
are systematically suppressed, which can and should be exploited to
reduce important sources of uncertainty.

We also find it convenient to define rescaled helicity-$0$
form factors as
\begin{eqnarray}
V_0(q^2)&=&\frac{2m_B\sqrt{q^2}}{\lambda^{1/2}}\tilde{V}_0(q^2),\nonumber\\
T_0(q^2)&=&\frac{2m_B^3}{\sqrt{q^2}\lambda^{1/2}}\tilde{T}_0(q^2),\nonumber\\
S(q^2)&=&- \frac{2m_B(m_b+m_s)}{\lambda^{1/2}}\tilde{S}(q^2), \label{eq:FFRedef}
\end{eqnarray}
where $\lambda = 4 m_B^2 |\vec k|^2$ ($\vec k$ is the 3-momentum of
the recoiling meson in the $\bar B$ rest frame), and also
define
$V_{\pm1}(q^2) \equiv \tilde V_{\pm 1}(q^2)$,
$T_{\pm1}(q^2) \equiv \tilde T_{\pm 1}(q^2)$.
The helicity form factors can be expressed in terms of the traditional
form factors. For a vector, we then have
(conventions for polarisation vectors and form factors in Appendix
\ref{app:conventions})
\begin{eqnarray}
  V_\pm(q^2) &=& \frac{1}{2} \left[ \Big(1 + \frac{m_V}{m_B} \Big)
    A_1(q^2) \mp \frac{\lambda^{1/2}}{m_B (m_B + m_V)} V(q^2) \right], \nonumber\\
  V_0(q^2) &=& \frac{1}{2 m_V \lambda^{1/2} (m_B + m_V)} \left[
    (m_B + m_V)^2 (m_B^2-q^2-m_V^2) A_1(q^2) - \lambda A_2(q^2)
  \right], \nonumber\\
  T_\pm(q^2) &=& \frac{m_B^2-m_V^2}{2 m_B^2} T_2(q^2) \mp
    \frac{\lambda^{1/2}}{2 m_B^2} T_1(q^2) , \nonumber\\
  T_0(q^2) &=& \frac{m_B}{2\,m_V\lambda^{1/2}}
       \left[ (m_B^2+3 m_V^2-q^2) T_2(q^2) -
       \frac{\lambda}{(m_B^2-m_V^2)} T_3(q^2) \right], \nonumber\\
  S(q^2) &=&  A_0(q^2),\label{eqs:transvTohelic}
\end{eqnarray}
 We also have
$V_{R\lambda} = - V_{-\lambda}$, $T_{R\lambda} = - T_{-\lambda}$,
$S_{R}=-S_L$.

For a pseudoscalar, we have
\begin{eqnarray}
  V_0(q^2) &=& i f_+(q^2), \\
  T_0(q^2) &=& i \frac{2m_B}{(m_B+m_P)} f_T(q^2), \\
  S(q^2) &=& \frac{1+\frac{m_s}{m_b}}{1-\frac{m_s}{m_b}}\frac{m_B^2-m_M^2}{\lambda^{1/2}} f_0(q^2) .
\end{eqnarray}
In this case, $V_{R0} = V_0$, $T_{R0}=T_{0}$, $S_R = S$.

\subsection{Kinematic distribution}
\label{sect:kindist}
We now consider the process
\begin{equation}
  \bar B(p) \to V(k) [\to \bar K(k_1) \pi(k_2)]\; \ell^-(q_1) \ell^+(q_2) ,
\end{equation}
i.e.\ decays to a vector decaying further into two pseudoscalars
(for definiteness, a kaon and a pion, with
$\bar K = \bar K^0$ or $K^-$, $\pi = \pi^+$ or $\pi^0$), where all four
final-state particles carry definite four-momenta. These are the
states in which detection is made.
Following \cite{Kruger:2005ep}, we define angles $\theta_K, \theta_l, \phi$
as follows. We first define, in the $\bar B$ rest frame,
\begin{equation}
 \bfe_l = \frac{\bfp_{l^-} \times \bfp_{l^+}}
                 {|\bfp_{l^-} \times \bfp_{l^+}|}, \qquad
 \bfe_K = \frac{\bfp_{\bar K} \times \bfp_{\pi}}
                 {|\bfp_{\bar K} \times \bfp_{\pi}|} , \qquad
 \hat z = \frac{\bfp_{\bar K} + \bfp_{\pi}}
                 {|\bfp_{\bar K} + \bfp_{\pi}|} \,.
\end{equation}
Then define $\phi$, in the interval $[0, 2 \pi]$, through
$
 \sin\phi = (\bfe_l \times \bfe_K) \cdot \hat z$ and
$ \cos\phi = \bfe_l \cdot \bfe_K .
$
Moreover, $\theta_l$ is defined as the angle between the direction of
flight of the $\bar B$ and the $\ell^-$ in the dilepton rest frame
and $\theta_K$ as the angle between the direction of motion of the
$\bar B$ and the $\bar K$ in the dimeson ($\bar K^*$) rest frame,
both in the interval $[0, \pi)$. For a $B$ decay, we define the angles
in the same way, in particular
$\theta_l$ is the angle between the $\ell^-$ (rather than the
$\ell^+$) and the $B$. This convention agrees with
\cite{Altmannshofer:2008dz} and
leads to simple expressions for untagged observables.

Next, we assume a resonant decay through an on-shell vector meson.
(This means we are making a narrow-width approximation.\footnote{
Off-resonance effects vanish in the limit of an infinitely narrow
$\bar K^*$. They can be included in the framework by introducing
dependence on the hadronic final state invariant mass $k^2$ and total angular
momentum $L$ in the helicity amplitudes \cite{Lu:2011jm}.
In particular, off-resonant
$L=0$ ($S$-wave) contributions have been recently studied in
\cite{Becirevic:2012dp,Matias:2012qz,Blake:2012mb}.
They modify some of the angular coefficients, but do not
impact on those that involve only $\lambda=\pm1$ amplitudes.
}) We
should then make the replacement
\begin{equation}  \label{eq:vectordecay}
   |\bar K^* ; \lambda \rangle \longrightarrow
     \sqrt{b} \int d\Omega_K Y_1^\lambda(\theta, \phi_K)
       |\theta_K; \phi_K \rangle ,
\end{equation}
where $\theta_K$ is the angle between the $+z$ direction and the $\bar K$
direction in the $\bar K^*$ cm frame and $\phi_K$ is the angle between
the $x$ axis and the projection of the former onto the $xy$ plane,
and $b \equiv BF(K^* \to K \pi) \approx 1$.
(Except for the zero point of the angle $\phi_K$, this is entirely
fixed by conservation of probability and of
angular momentum and is independent of the details of
the $\bar K^*$ decay vertex. See also \cite{Lu:2011jm}.)
Squaring the amplitude  and summing over lepton spins
the fully differential decay rate is obtained as
\begin{eqnarray}
\lefteqn{\frac{d^{(4)} \Gamma}
  {dq^2\,d(\cos\theta_l)d(\cos\theta_k)d\phi}= \frac{9}{32\,\pi} } \nonumber \\[2mm]
& \times& \Big(\! I^s_1\sin^2\theta_k+I^c_1\cos^2\theta_k     \label{Eq:DR4body1}
+(I^s_2\sin^2\theta_k+I^c_2\cos^2\theta_k)\cos2\theta_l \nonumber \\
&&+ I_3\sin^2\theta_k\sin^2\theta_l\cos2\phi
+I_4\sin2\theta_k\sin2\theta_l\cos\phi  \\ 
&&+ I_5\sin2\theta_k\sin\theta_l\cos\phi
+(I_6^s\sin^2\theta_k + I_6^c \cos^2\theta_K) \cos\theta_l\nonumber\\
&&+ I_7\sin2\theta_k\sin\theta_l\sin\phi+I_8\sin2\theta_k\sin2\theta_l\sin\phi
+I_9\sin^2\theta_k\sin^2\theta_l\sin2\phi \Big) . \nonumber
\end{eqnarray}
The angular coefficients $I_i$ are functions exclusively of $q^2$.
They can be expressed in terms of the helicity or transversity
amplitudes Eqs.~\eq{eq:helampfirst}--\eq{eq:helamplast} as
\begin{eqnarray}
I_1^c&=&F \left\{
  \frac{1}{2}\left(|H_V^0|^2+|H_A^0|^2\right)+ 
  |H_P|^2+\frac{2m_\ell^2}{q^2}\left(|H_V^0|^2-|H_A^0|^2\right)
  + \beta^2 |H_S|^2
  \right\},
\label{Eq:I1c} \nonumber\\
I_1^s&=&F
\left\{\frac{\beta^2\!+\!2}{8}\left(|H_V^+|^2+|H_V^-|^2+(V\rightarrow
    A)\right)+\frac{m_\ell^2}{q^2}\left(|H_V^+|^2+|H_V^-|^2-(V\rightarrow
    A)\right)\right\},\,\
\label{Eq:I1s}\nonumber \\
I_2^c&=&-F\, \frac{\beta^2}{2}\left(|H_V^0|^2+|H_A^0|^2\right)
, \label{Eq:I2c} \nonumber\\
I_2^s&=&F\, \frac{\beta^2}{8}\left(|H_V^+|^2+|H_V^-|^2\right)+(V\rightarrow A)
,\label{Eq:I2s} \nonumber\\
I_3&=&-\frac{F}{2}{\rm Re} \left[H_V^+(H_V^-)^*\right]+(V\rightarrow A)
,\label{Eq:I3}\nonumber \\
I_4&=&F\, \frac{\beta^2}{4}{\rm
  Re}\left[(H_V^-+H_V^+)\left(H_V^0\right)^*\right]+(V\rightarrow A)
,\label{Eq:I4} \nonumber\\
I_5&=&F\left\{ \frac{\beta}{2}{\rm
  Re}\left[(H_V^--H_V^+)\left(H_A^0\right)^*\right]+(V\leftrightarrow
A) - \frac{\beta\,m_\ell}{\sqrt{q^2}} {\rm Re} \left[H_S^* (H_V^+ + H_V^-)\right]
\right\}\,,\label{Eq:I5} \nonumber\\
I_6^s&=&F \beta\,{\rm Re}\left[H_V^-(H_A^-)^*-H_V^+(H_A^+)^*\right]
,\label{Eq:I6s} \nonumber\\
I_6^c&=&2 F \frac{\beta\, m_\ell}{\sqrt{q^2}} {\rm Re} \left[ H_S^* H_V^0 \right]
,\label{Eq:I6c} \nonumber \\
I_7&=&F \left\{ \frac{\beta}{2}\,{\rm
  Im}\left[\left(H_A^++H_A^-\right)(H_V^0)^* \, +(V\leftrightarrow A) \right]
- \frac{\beta\, m_\ell}{\sqrt{q^2} }\, {\rm Im} \left[ H_S^*(H_V^{-} -
  H_V^{+}) \right] \right\}
,\label{Eq:I7} \nonumber\\
I_8&=&F\, \frac{\beta^2}{4}{\rm Im}\left[(H_V^--H_V^+)(H_V^0)^*\right]+(V\rightarrow A)
,\label{Eq:I8} \nonumber\\
I_9&=&F\, \frac{\beta^2}{2}{\rm Im}\left[H_V^+(H_V^-)^*\right]+(V\rightarrow A)
,\label{Eq:I9}
\end{eqnarray}
where
$$
F=\frac{ \lambda^{1/2}\beta\, q^2}{3 \times 2^{5} \,\pi^3\, m_B^3} BF
(K^* \to K \pi),
\qquad
\beta = \sqrt{1 - \frac{4 m_\ell^2}{q^2} } .
$$
We have omitted the terms involving the tensor amplitudes $H_{TL}$ and
$H_{TR}$, which will be considered elsewhere.
The phenomenology in Section \ref{sec:Pheno}
will be incomplete in this sense, but as explained there,
we expect the effect on the considered observables to be very small.

The analysis of the CP-partner decay $B\rightarrow K^*\mu^+\mu^-$ gives rise to an equivalent distribution, 
$d^{(4)} \bar{\Gamma}/(dq^2\,d(\cos\theta_l)d(\cos\theta_k)d\phi)$,
which is obtained from~(\ref{Eq:DR4body1}) by replacing 
\begin{equation}
I_{1s(c),2s(c),3,4,7}\rightarrow \bar{I}_{1s(c),2s(c),3,4,7},\hspace{1cm}\bar{I}_{5,6,8,9}\rightarrow -\bar{I}_{5,6,8,9},\label{Eq:DR4bodyCP2}
\end{equation}
when one uses the angles defined as in the $\bar{B}$ decays with $K^-\rightarrow K^+$ ~\cite{Kruger:1999xa}. In the later equation the $\bar{I_i}$'s are equal to the $I_i$'s but with all the weak phases conjugated. 

\subsection{``Clean'' observables, helicity hierarchies, and new physics }
The 12 angular coefficients, together with their CP-conjugates, provide
complete information about the decay distribution. In practice,
however, certain combinations of these observables are
more useful. Two important examples are the
forward-backward asymmetry \cite{Ali:1991is,Burdman:1998mk,Ali:1999mm}
and the transverse asymmetry
$A_T^{(2)}$ \cite{Kruger:2005ep},
\begin{eqnarray}
A_{FB}(q^2) &\equiv&    \label{eq:AFB}
  \frac{\left[\int_0^1 - \int_{-1}^0\right]
      d(\cos \theta_l) d^2 \Gamma'/(dq^2 d\cos\theta_l) }
       {\int_{-1}^1 d(\cos \theta_l) d^2 \Gamma'/(dq^2 d\cos\theta_l)}
   = - \frac{3 \Sigma_6^s}{4 \Gamma'} \, , \\
A_T^{(2)} &\equiv& \frac{\Sigma_3}{2 \Sigma_{2s}} \equiv P_1 \, ,
\end{eqnarray}
where $\Sigma_i \equiv (I_i + \bar I_i)/2$ and $d\Gamma' = (d\Gamma +
d\bar\Gamma)/2$ denote CP-averages. These have two potential advantages.
First, taking ratios leads to cancellations of form factor uncertainties
between the numerator and denominator. In particular, $P_1$ only
involves the amplitudes $H_V^\pm$ and $H_A^\pm$. In the factorizable
approximation and neglecting the small Wilson coefficients $C_7'$,
$C_9'$, and $C_{10}'$, this leaves dependence only on $T_-$, $V_-$, $T_+$, and
$V_+$. For energetic $E_V \sim m_b$ (small $q^2$), one has the
heavy-quark/large-energy relations $T_- = V_-, T_+ = V_+ = 0$ (see below),
which are broken only by (calculable) $\alpha_s$-corrections and
(incalculable) $\Lambda/m_b$ power corrections. Neglecting these, too,
the single remaining form factor
then cancels out between the numerator and denominator. In the
case of $A_{FB}$, the cancellation is incomplete, as the denominator
involves further amplitudes. However, the location of the zero-crossing
$q^2_0$ is determined entirely by the numerator, and (under the same
approximations) is free of form factor uncertainties. For these reasons,
$P_1$, $q^2_0$, and similar observables are often termed ``clean''. An
optimized set has been recently defined in \cite{Matias:2012xw} and will be
studied in the phenomenological part below.

A second point is that $P_1$ actually vanishes under the stated approximations,
as a consequence of all terms being proportional to either $V_+$ or
$T_+$. Hence, it is an approximate null-test of the Standard Model,
and a probe of any new physics that generates the Wilson coefficients
$C_7'$, $C_9'$, or $C_{10}'$. The same is true of $I_9$ and certain
combinations constructed from it.

Clearly, the actual theoretical cleanness of the observables
will depend on the size
of the radiative and  power corrections and non-factorizable effects.
The following
section is devoted to a thorough study of these effects, and their
impact on the ``wrong-helicity'' amplitudes $H_V^+$ and $H_A^+$ in
particular. We will show that, under very conservative assumptions,
$H_V^+$ and $H_A^+$ remain suppressed, such that
the clean character of $I_3$ and $I_9$ as null tests, but not of
other observables, is preserved by non-factorizable and power
corrections.

Finally, let us recall that the radiative decay $\bar B \to V \gamma$
is described in terms of a subset of the amplitudes for $\bar B \to V
\ell^+ \ell^-$. The precise relation is ($\lambda = \pm 1$)
\begin{eqnarray}
  {\cal A}(\bar B \to V(\lambda) \gamma(\lambda))
    &=& \lim_{q^2 \to 0} \frac{q^2}{e} H_V(q^2=0; \lambda) \nonumber \\
    &=& \frac{i N m_B^2}{e} \left[\frac{2 \hat m_b}{m_B}
          (C_7 \tilde T_\lambda(0) - C_7' \tilde T_{-\lambda})(0)
          - 16 \pi^2 h_\lambda(q^2=0) \right] . \nonumber \\
\end{eqnarray}

\section{Helicity amplitudes: anatomy, hierarchies, and hadronic uncertainties}

\label{sect:hadronic}
The helicity amplitudes governing the observables
involve form factors and the
nonlocal objects $h_\lambda$, all of which carry hadronic uncertainties,
limiting the sensitivity of rare $B$ decays to new physics.
However, hadronic uncertainties can be constrained by means of the
equations of motion, the $V-A$ structure of the weak hamiltonian,
and an expansion in $\Lambda/m_b$ (QCD factorization). Our main point is that
this results in the suppression of entire helicity amplitudes,
including non-factorizable effects, such that the discussion is  indeed
best framed in terms of helicity (rather than transversity) amplitudes
and helicity form factors.
We first translate what is known about the form factors to the
helicity basis, including the fact that the heavy-quark limit
implies the suppression of two of them \cite{Burdman:2000ku}.
We next survey how this bears out in various theoretical
approaches to form factor determinations, concluding with a brief
argument for the suppression of the positive-helicity form factors
in the framework of light-cone sum rules, at the level
of the correlation function.
We then show that the $V-A$ structure
also implies suppression of the ``charm-loop'' contribution to the nonlocal
positive-helicity amplitude $h_{+1}$, building on a method
introduced in \cite{Khodjamirian:2010vf}. In addition, we show
that the same conclusion applies to hadronic resonance models
for the ``light-quark'' contributions to $h_\lambda$, once known
experimental facts about the helicity structure of $\bar B \to V V$ are
incorporated (which can be theoretically understood on the same
basis).

\subsection{Form factors}
\label{sect:FFs}

The $\bar B \to M$ form factors are nonperturbative objects.
In the following, we restrict ourselves to the $\bar B \rightarrow
V$ case.
 First-principles lattice-QCD computations are becoming
available~\cite{Becirevic:2006nm,Liu:2011raa}, although they will
be restricted for the foreseeable future to the region
of slow-moving $V$ (high $q^2$). A state-of-the-art method of
obtaining form factors at low $q^2$ is given by  QCD sum rules
on the light cone (see~\cite{Ball:1998kk,Ball:2004rg}). This involves,
unfortunately, certain
irreducible systematic uncertainties which are difficult to quantify.
Sum rules are also useful in guiding extrapolations of high-$q^2$
lattice-QCD results~\cite{Bharucha:2010im}.

\subsubsection{Theoretical constraints on form factors at low $q^2$}

The form factors fulfil two exact relations that in the helicity basis
take the form
\begin{eqnarray}
   T_+(q^2=0) &=& 0   \label{eq:Tplusq20} , \\
   S(q^2=0) &=&  V_0(0).  \label{eq:Sq20}
\end{eqnarray}
At large recoil, i.e.\ small $q^2$, one has further relations
which hold up to corrections of ${\cal O}(\Lambda/m_b)$ but to
all orders in $\alpha_s$.
As a result, the seven form factors are given,
at leading power in $\Lambda/m_b$ and $\Lambda/E$
(where $E\equiv E_{V}$ is itself of order $m_b$ for low $q^2$),
in terms of only two independent {\it soft}
form factors~\cite{Charles:1998dr}, $\xi_\perp$ and $\xi_\parallel$,
with radiative corrections systematically calculable  in 
QCDF~\cite{Beneke:2000wa}
as a perturbative expansion in $\alpha_s$. These corrections also involve nonperturbative objects such as decay
constants and light-cone distribution amplitudes (LCDAs) of the initial
and final mesons.
The factorization properties and calculation of radiative corrections
become particularly transparent when formulated
as a matching of QCD to soft-collinear effective theory
(SCET)~\cite{Bauer:2000yr,Bauer:2001yt,Beneke:2002ph,Beneke:2002ni}.
Corrections at ${\cal O}(\Lambda/m_b)$
violate factorization and need to be modeled, or estimated or
constrained by another method.

The soft form factors can be chosen to coincide
with two ``physical'', i.e.\ full-QCD form factors, which makes them
well-defined to all orders in $\Lambda/m_b$.
Appropriate choices are $V$ and $A_0$, for $\xi_\perp$ and
$\xi_\parallel$ respectively~\cite{Beneke:2000wa}, given that they
are matrix elements of conserved currents and, as such, free from
renormalization scale ambiguities. 
For $\xi_\perp$, we find it convenient to use the transversity form factor
$T_1$ instead. In this case, however, the transversal soft
form factor depends on the factorization scale $\mu$ as (to LL accuracy)
\begin{equation}
T_1(q^2,\mu)\equiv\xi_\perp(q^2,\mu)=
\xi_\perp(q^2,m_b)\left(\frac{\alpha_s(\mu)}{\alpha_s(m_b)}\right)^{4/23} .
\label{Eq:xiperpRenSc}
\end{equation}
Setting
\begin{eqnarray}
T_1(q^2, \mu) &\equiv&\xi_\perp(q^2, \mu),\nonumber\\
S(q^2) &\equiv&\frac{E}{m_{V}}\xi_\parallel(q^2),\nonumber
\end{eqnarray}
with $E\simeq(m_B^2-q^2)/(2 m_B)$,
the symmetry relations in Ref.~\cite{Beneke:2000wa} can be expressed
in the helicity basis as
\begin{eqnarray}
T_-&=& \frac{2E}{m_B} \xi_\perp,\nonumber\\ 
T_+&=&0,\nonumber\\
T_0&=&\frac{E}{m_{V}}\xi_\parallel\left(1+ \frac{\alpha_s \, C_F}{4\pi}\left[\ln\frac{m_b^2}{\mu^2}-2+4L\right]\right)+\frac{\alpha_s \, C_F}{4\pi}\Delta T_0,\nonumber\\
V_-&=&\frac{2E}{m_B}\;\xi_\perp \, \left(1 +
    \frac{\alpha_s \, C_F}{4\pi} \,
    \left[  \ln \frac{\mu^2}{m_b^2} + L \right]
  \right)
+ \frac{\alpha_s \, C_F}{4\pi} \, \Delta V , \nonumber\\
V_+&=&0\nonumber,\\
V_0&=&\frac{E}{m_{V}}\xi_\parallel\left(1+ \frac{\alpha_s \, C_F}{4\pi}\left[-2+2L\right]\right)+\frac{\alpha_s \, C_F}{4\pi}\Delta V_0,\label{Eq:HQLRrelsII}
\end{eqnarray}
with 
 $C_F=4/3$ and $L=-2E/(m_B-2E)\ln(2E/m_B)$. These expressions hold up
to higher-order corrections in $\alpha_s$, which augment the
$\alpha_s$ terms shown, and power corrections.

The  $\alpha_s$-contributions multiplying the soft form
factors come from the hard-vertex corrections, and
the remaining ones  originate from 
hard scattering with the spectator quark \cite{Beneke:2000wa},
\begin{eqnarray}
\Delta T_0&=&-\frac{m_B^2}{4E^2}\Delta\,F_\parallel, \\
\Delta V_0&=&-\frac{q^2}{4E^2}\Delta\,F_\parallel, \label{Eq:T0V0HS} \\
\Delta V&=&-\frac{1}{2} \Delta F_\perp,
\end{eqnarray}
where $\Delta F_\perp$ and $\Delta
F_\parallel$ involve (finite) convolutions of hard-scattering kernels with
light-cone distribution amplitudes and can be found
in Ref.~\cite{Beneke:2000wa}.
Higher-order corrections in $\alpha_s$ do not change
the structure or infrared safety of \eq{Eq:HQLRrelsII},
i.e. factorization can be proven to all orders \cite{Beneke:2003pa}.
In particular, the vanishing of $T_+$ and $V_+$ at leading power
is an all-orders result \cite{Burdman:2000ku,Beneke:2005gs}, which
looks this simple only in the helicity basis. The
${\cal O}(\alpha_s^2)$ contributions have been calculated in
\cite{Beneke:2004rc,Becher:2004kk,Beneke:2005gs,Kirilin:2005xz}, and their numerical impact
was found to be small, mainly reducing the residual (unphysical)
scale dependence.

Thus, in the heavy-quark/large-recoil limit,
the form factors $T_+$ and $V_+$ exactly vanish.
Combining \eq{eq:Tplusq20} and \eq{Eq:HQLRrelsII}, we have at low $q^2$
\begin{eqnarray}
  T_+(q^2) &=& {\cal O}(q^2/m_B^2) \times {\cal O}(\Lambda/m_b) , \label{eq:TplusTaylor}\\
  V_+(q^2) &=& {\cal O}(\Lambda/m_b) .  \label{eq:VplusTaylor}
\end{eqnarray}
On the other hand, $T_0$ and $V_0$ are not suppressed, and are
independent of any hadronic 
information related to the transversal polarizations of the vector meson. 
Notice also that, with our choice of soft form factors, the vector
form factor $V$,
and hence $V_-$, has a purely residual (higher-order) scale dependence $\mu$
at any given order of perturbation theory, from the factorization
into the scale-dependent $\xi_\perp$ and a scale-dependent
perturbative factor.
One can explicitly check that this produces a
relative change in the form factor of no more than a 1.5\% in the
range $m_b/2\leq\mu\leq2m_b$.

\subsubsection{Numerical values of the $B\rightarrow K^*$ soft form factors}

Although the symmetry relations reduce (at leading power)
the number of independent non-perturbative functions,
for a quantitative treatment one still has to compute the soft
form factors by a nonperturbative method and
estimate (or calculate) the power corrections.
 Sum rules formulated on the light cone
(LCSR) are customarily used in exclusive $B$ decays to obtain
numerical values of the form factors in the large-recoil
domain~\cite{Ball:1998kk,Ball:2004rg,Khodjamirian:2010vf}. Other
approaches that have been used to calculate
the form factors in this regime include local QCD sum rules
(QCDSR)~\cite{Colangelo:1995jv} and (truncated) Dyson-Schwinger
equations (DSE)~\cite{Ivanov:2007cw}. We list in Table~\ref{Table:FFsValues} the results on the
$B\rightarrow K^*$ form factors at $q^2=0$ and in the transversity
basis for the different calculations considered in this
paper. The central values of most of the form factors are
quite similar, except some (prominently $A_0$) for which different methods
disagree.
\begin{table}[h]
\centering
\caption{Values of the $B\rightarrow K^*$ QCD form factors at $q^2=0$ in the conventional basis and in the different approaches considered in our study. The uncertainties are those quoted or suggested in the corresponding references (a 15\% relative error for the DSE results). 
\label{Table:FFsValues}}
\begin{tabular}{|c|cccc|}
 \hline
& LCSR~\cite{Ball:2004rg}&LCSR~\cite{Khodjamirian:2010vf}& QCDSR~\cite{Colangelo:1995jv}& DSE~\cite{Ivanov:2007cw}\\
\hline
$V(0)$&0.411(33)&0.36$^{+23}_{-12}$&0.47(3)&0.37(6)\\
$A_0(0)$&0.374(34)&0.29$^{+0.10}_{-0.07}$&0.30(3)&0.25(4)\\
$A_1(0)$&0.292(28)&0.25$^{+0.16}_{-0.10}$&0.37(3)&0.28(4)\\
$A_2(0)$&0.259(27)&0.23$^{+0.19}_{-0.10}$&0.40(3)&0.30(5)\\
$T_1(0)$&0.333(28)&0.31$^{+0.18}_{-0.10}$&0.38(6)&0.30(5)\\
$T_3(0)$&0.202(18)&0.22$^{+0.17}_{-0.10}$&0.3&0.27(4)\\
\hline
\end{tabular}
\end{table}

A well known issue between the generic results of LCSR calculations
(e.g. those of Ref.~\cite{Ball:2004rg}), QCD factorization, and the SM
value of the Wilson coefficient $C_7$ is that they lead to a branching
fraction of the decay $B\rightarrow K^*\gamma$ that is larger than the
experimental value. Given that $C_7$ is constrained to be
close to its SM value by the inclusive $B\rightarrow X_s\gamma$ decay
rate (at least when assuming $C_7^\prime=0$ and $C_7$ real), it is often
assumed that this discrepancy is due to a systematic error in the LCSR
model which produces an overestimation in the value of the relevant
form factor $T_1$ at $q^2=0$~\cite{Beneke:2001at}. A possible solution
to this problem is to re-scale the form factors such that $T_1(0)$, in
combination with the SM value for $C_7$, lead to the experimental
branching fraction of the radiative
decay~\cite{Altmannshofer:2008dz}. Although this procedure discards in
part the
sensitivity  of the $B\rightarrow K^*\ell^+\ell^-$ decay rate to new
physics,
it does not affect any physical information extracted from asymmetries or ratios, which are, indeed, the most interesting observables in the semileptonic decay~\cite{Altmannshofer:2008dz}.   

Using the SM value for $C_7$ and
$\mathcal{B}(B\rightarrow K^*\gamma)=(4.33\pm0.15)\times10^{-5}$, one obtains   
\begin{eqnarray}
\xi_\perp(0)=T_1(0)=0.277(13),\label{Eq:xiperp0} 
\end{eqnarray}
where the error is obtained adding linearly the experimental and the theoretical uncertainties, which come from the SM parameters and the hadronic parameters entering the non-factorizable contributions to the amplitude (shown in Appendix~\ref{app:SMparam}). Re-scaling the values of the form factors in Table~\ref{Table:FFsValues} and taking the average of the results given by the different models, we obtain
\begin{eqnarray}
\xi_\parallel(0)=\frac{2m_{K^*}}{m_B}S(0)=0.09(2),\label{Eq:xiparallel0}
\end{eqnarray}
where we have estimated the uncertainty to be such that it includes
the two extreme values coming from the LCSR~\cite{Ball:2004rg} and
DSE~\cite{Ivanov:2007cw}. For the $q^2$ dependence we choose the pole
forms introduced in~\cite{Beneke:2001at} based on pure heavy-quark-limit arguments,
\begin{eqnarray}
 \xi_\perp(q^2)=\xi_\perp(0)\left(\frac{1}{1-q^2/m_B^2}\right)^2,\hspace{2cm}\xi_\parallel(q^2)=\xi_\parallel(0)\left(\frac{1}{1-q^2/m_B^2}\right)^3. \label{Eq:softFFq2}
\end{eqnarray}

\begin{figure}[h]
\centering
\includegraphics[scale=0.4]{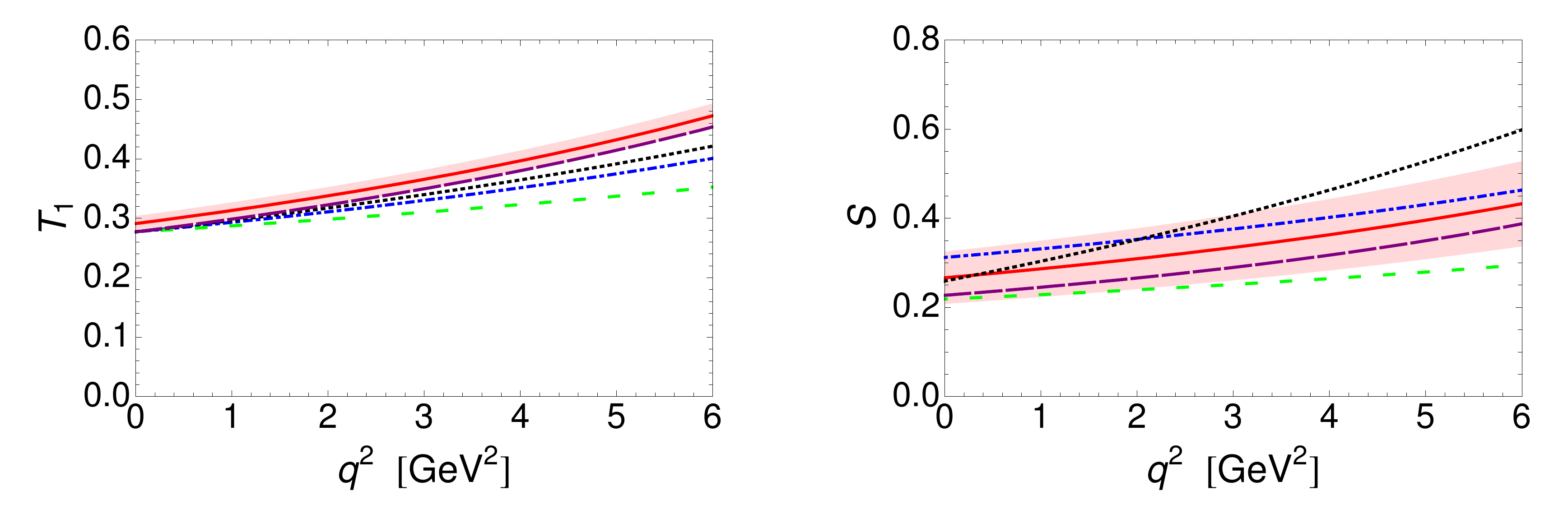}
\caption{Form factors $T_1$ and $S$ in the large-recoil region used in this work as a reference for the symmetry relations (thick red line) with the error bands produced by the uncertainties on $\xi_\perp(0)$ and $\xi_\parallel(0)$. These are compared with the re-scaled (see in the main text) central values of the results obtained in the LCSRs (blue dot-dashed~\cite{Ball:2004rg} and black dotted~\cite{Khodjamirian:2010vf}), in QCDSRs~\cite{Colangelo:1995jv} (green short-dashed) and DSEs~\cite{Ivanov:2007cw} (purple long-dashed).\label{Fig:plotst1a0}}
\end{figure}

In Fig.~\ref{Fig:plotst1a0} we plot the form factors $T_1$ and $S$ in the large-recoil region used in this work as a reference for the symmetry relations in Eqs.~(\ref{Eq:HQLRrelsII}) with the error bands produced by the uncertainties on $\xi_\perp(0)$ and $\xi_\parallel(0)$. These are compared with re-scaled central values of the results reported using LCSR~\cite{Ball:2004rg,Khodjamirian:2010vf}, QCDSR~\cite{Colangelo:1995jv} and DSE~\cite{Ivanov:2007cw}. 

\subsubsection{Power corrections to the large-recoil form  factor relations}
The main factor limiting the utility of QCD factorization for the
form factors (and elsewhere) are unknown power corrections
${\cal O}(\Lambda/m_b)$. Although one might naively expect such
power corrections to be $\sim(5-10)\%$,
the possibility of having larger corrections cannot be dismissed.
Let us refer to effects breaking the (QCD-corrected) symmetry relations
in Eqs.~(\ref{Eq:HQLRrelsII}) as power corrections, even though
these also include the numerically unimportant perturbative
$\alpha_s^n$, ${n\geq2}$ contributions. They
 govern the so-called factorizable power corrections
in $B\rightarrow K^*\ell^+\ell^-$ decay. As they do not
cancel out in the ``clean'' observables defined in~\cite{Matias:2012xw},
estimating them is important to assess hadronic uncertainties.
The conventional procedure in
phenomenological analyses is to use the results of some
technique that automatically includes power corrections (most commonly
LCSR). However, in doing so, the systematic errors coming from the
assumptions and approximations implied by the particular
approach are not transparent. A related issue is that the $q^2$-dependence
of the form factors is often ``hard-coded'' and not treated as an uncertainty.

Instead, in this paper we parameterize power corrections to the
form factors in a model-independent fashion. The uncertain
parameters can then be estimated by various methods.
For a given (helicity) form factor $F$, we parameterize the corrections
to~(\ref{Eq:HQLRrelsII}) as
\begin{eqnarray}
F^{\rm p.c.}= a_F + b_F\frac{q^2}{m_B^2}
  + {\cal O}\left( \left(\frac{q^2}{m_B^2}\right)^2;
  \Lambda^2/m_b^2\right),\label{Eq:pcsFFs} 
\end{eqnarray}
where $a_F$ and $b_F$ are dimensionless numbers
of order ${\cal O}(\Lambda/m_b)$.
Importantly, Eqs.~\eq{eq:Tplusq20} and \eq{eq:Sq20}
imply that $a_{T_+}=0$ and $a_{V_0}=0$.

Ideally, the coefficients $a_F$ and $b_F$ should be estimated by
a direct calculation of helicity form factors, for example through
LCSR (see discussion below). For lack of present availability of such
results, we compare available results (in the conventional basis) of
the LCSR, QCDSR and DSE approaches to give an estimate of
the uncertainty given by the break-down of the symmetry relations. 
We estimate ranges for the nonvanishing $a_F$ and $b_F$ by taking the
average deviations between the results given by the symmetry relations
and those obtained in the different models considered in this paper.
The resulting ranges are given in Table~\ref{Table:VFactPCs}.
Note that the $a_F$ are all at the few-percent level,
well in line with ``naive'' expectations about the size of power corrections.
The (less important) $b_F$ coefficients are also around the 10-percent
level or below, with the exception of the helicity-zero form factors
$V_0$ and, to smaller extent, $T_0$.
\begin{figure}
\centering
\includegraphics[scale=0.4]{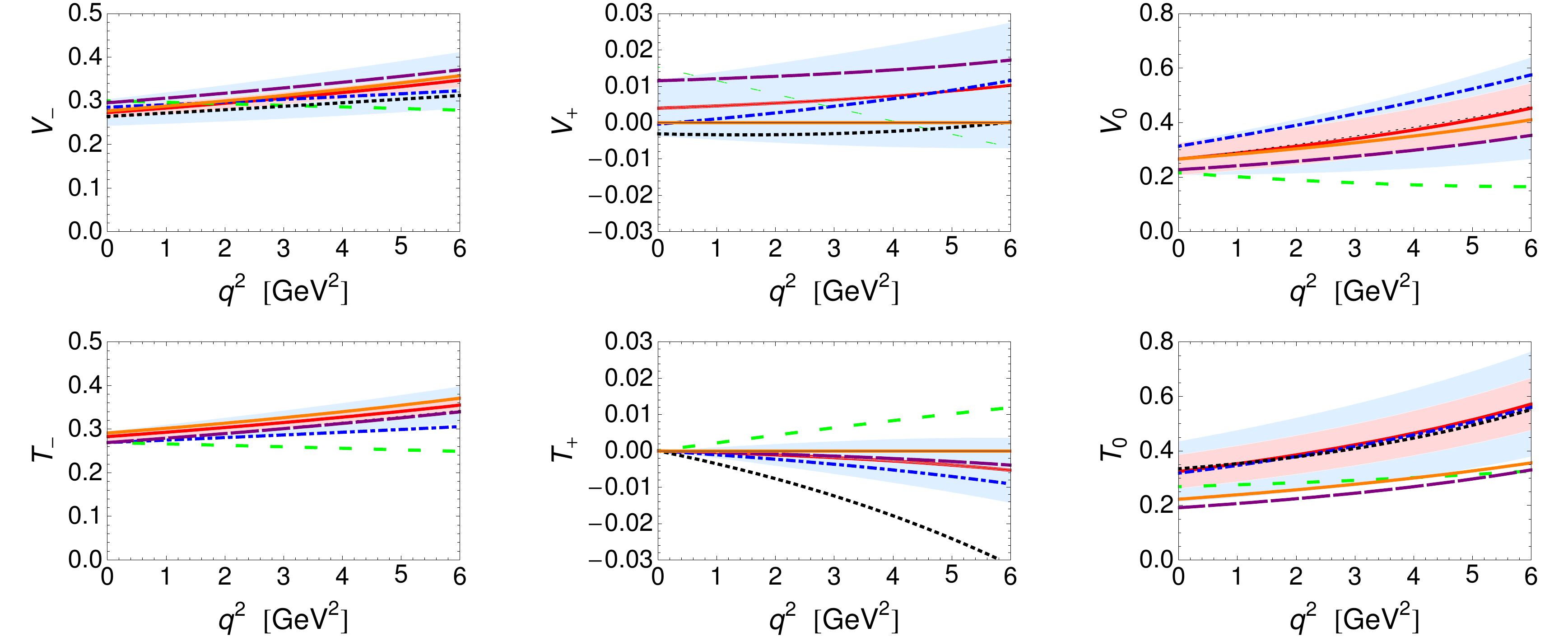}
\caption{Form factors in the helicity basis and in the large-recoil region used in this work (red thick and solid line). The inner (red) error band is produced by the uncertainties on $\xi_\perp(0)$ and $\xi_\parallel(0)$, while the outer (blue) includes also the estimated uncertainties on the factorizable power corrections. These curves are compared with the re-scaled central values of the results obtained in the LCSR (blue dot-dashed~\cite{Ball:2004rg} and black dotted~\cite{Khodjamirian:2010vf} lines), in QCDSR~\cite{Colangelo:1995jv} (green short-dashed line) and DSE~\cite{Ivanov:2007cw} (purple long-dashed line). For the sake of completeness we also include the curves (orange solid and thin line) obtained in the strict heavy-quark/large recoil limit, Eqs.~(\ref{Eq:HQLRrelsII}). \label{Fig:plotsFFs}}
\end{figure}
In Fig.~\ref{Fig:plotsFFs} we plot our ranges for the form factors,
including  power corrections described by
 \eq{Eq:pcsFFs}, with the ranges for $a_F$ and $b_F$ given
in Table~\ref{Table:VFactPCs}. We also show the
central values of the results obtained from
LCSR~\cite{Ball:2004rg,Khodjamirian:2010vf},
QCDSR~\cite{Colangelo:1995jv} and in DSE~\cite{Ivanov:2007cw},

\begin{table}
\centering
\caption{Bounds on the dimensionless constants $a_F$ and $b_F$
parameterizing the power corrections to the form factor
heavy-quark relations. 
\label{Table:VFactPCs}}
\begin{tabular}{c|cccccc|}
 \cline{2-7}
 &\multicolumn{1}{|c}{$V_-$}&$V_+$&$T_-$&$T_+$&$V_0$&$T_0$\\
 \hline
 \multicolumn{1}{|c|}{$|a|_{\rm max}$}&0.027&0.008&0&0&0&0.050\\
 \hline
 \multicolumn{1}{|c|}{$|b|_{\rm max}$}&0.136&0.042&0.125&0.043&0.434&0.206\\
 \hline
 \end{tabular}
 \end{table}
Our method may over- or underestimate the power corrections.
For instance, one would obtain larger uncertainties by combining the LCSR
uncertainties on the standard form factors entering $V_+$ and $T_+$
in quadrature (or linearly). Hence, our numbers
should be considered preliminary until dedicated calculations appear.
Nevertheless, it is reassuring that the nonperturbative approaches appear to
be consistent with
expectations from the heavy-quark limit. In fact, for the
LCSR for the form factors the heavy-quark limit has been analyzed
a long time ago in \cite{Bagan:1997bp}, and the result is fully consistent
with the structure later obtained in \cite{Beneke:2000wa} and
expressed in the helicity basis through \eq{Eq:HQLRrelsII}.

To reduce the LCSR sum-rule uncertainty manifestly, one should
derive sum rules directly for the helicity form factors.\footnote{Such direct
  calculations have been recently advocated, independently, by T.\
  Feldmann \cite{TFtalk}.}
To demonstrate this, consider the correlation function
\begin{equation}
  G_{F\lambda}(q^2; p^2) = i \int d^4y e^{-i p\cdot y}
       \langle K^*(k) | T \{ \epsilon^*_\mu(q; \lambda)  (\bar s
       \Gamma_F^\mu b)[0] \, j_B^\dagger(y) \} | 0 \rangle ,
\end{equation}
where $j_B= i m_b \bar q \gamma_5 b$, $q$ is a light quark field,
and $\Gamma^{\mu}_F$ a Dirac structure appearing in the definition of
a given helicity form factor.
Its hadronic representation,
\begin{equation}
  G_{F\lambda} = \frac{ \langle K^*(k, \lambda) | \epsilon^*_\mu(q; \lambda)  \bar s
       \Gamma_F^\mu b | B \rangle}{p^2 - m_B^2} \frac{f_B m_B^2}{m_b} +
     \dots ,
  \label{eq:helsr}
\end{equation}
obtained by inserting a complete set of hadronic states between
the two currents,
contains the helicity form factor $F_\lambda(q^2) \propto \langle
K^*(k) | \epsilon^*_\mu(q; \lambda)  \bar s  \Gamma_F^\mu b | B
\rangle$, with the ellipsis denoting contributions from higher
resonances and continuum states with the quantum numbers
of $j_B$.
A sum rule is obtained from \eq{eq:helsr} by taking $p^2 < m_B^2$
off-shell by an amount ${\cal O}(1\, {\rm GeV}^2)$, evaluating $G_{F
  \lambda}$ perturbatively in a light-cone expansion, and
Borel-transforming both sides to suppress the hadronic states above the
$B$. This is completely  analogous to
the standard sum rules \cite{Ball:1998kk}, and in fact simpler as
the correlation function for a given $\Gamma_F$ and $\lambda$
directly gives the respective helicity form factor, with no need for
Lorentz decomposition. The light-cone expansion of
$G_{F\lambda}$ results in convolutions of perturbative kernels with
light-cone distribution amplitudes, which are organised in
terms of increasing twist. Let us demonstrate the suppression of
the positive-helicity form factors in this framework.  At tree level,
we have for $T_+$
$$
   i \!\! \int \!\! d^4y e^{-i p\cdot y} T \{ \epsilon^*_\mu(q; +)  (\bar s
       \Gamma_T^\mu b)[0] \, j_B^\dagger(y) \} =
   \bar s(k_1) \epsilon^*_\mu(q; +) q_\nu \sigma^{\mu\nu} P_R
      \frac{\pslash-\kslash_2 + m_b}{(p-k_2)^2 - m_b^2} q(k_2)
$$
Now, to leading-twist accuracy we can replace $\bar s(k_1)$ and
$q(k_2)$ by collinear fields satisfying $\nmslash q = 0$ and
$\bar s \npslash = 0$.\footnote{We are assuming the $K^*$ to move in
the $+ \hat z$ direction here, as throughout the present paper.
Note that in the relevant
literature on LCSR for form factors, the final-state hadron is often taken
to be moving in $-\hat z$ direction instead.}
As a consequence, we can insert a projector
$\bar s \to \bar s \frac{\nmslash\npslash}{4}$. We then encounter
in the Dirac structure a factor
$$
   \nmslash\npslash \epslash^*(q; +) \qslash P_R
$$
It is easy to verify that this vanishes as a consequence of
$$
   \npslash \epslash^*(q; +) P_L = 0 
$$
This shows that the correlation function, and thus the form factor $T_+$,
is suppressed as ${\cal O}(\Lambda/E_{K*}) = {\cal O}(\Lambda/m_b)$ as
a higher-twist effect.
Independently,
$$
   \qslash = \frac{n_+ \cdot q}{2} \nmslash + \frac{n_- \cdot q}{2}
   \npslash ,
$$
where at low $q^2 \ll m_b^2 \sim E_{K^*}^2$ we have $n_+ \cdot q \sim
m_b$, while $n_- \cdot q \sim q^2/m_b^2$. Hence up to corrections of
${\cal O}(q^2)$, $T_+$ vanishes as a consequence of
$$
  \epslash^*(q; +) P_L \nmslash = 0 .
$$
In combination, we obtain a behaviour that respects
\eq{eq:TplusTaylor}. In the analogous case of $V_+$, one obtains the
behaviour \eq{eq:VplusTaylor} in an analogous way, but no
$q^2$-suppression (for there is no $\qslash$ factor in the definition
of $V_+$).
For a more quantitative treatment, one needs to
go to higher-twist accuracy; this requires going beyond the collinear
approximation for the external momenta and quark fields.
In principle, the necessary ingredients are known up to twist-4.
We leave such an analysis for future work. Note that the argument
does not imply a suppression of the helicity-$0$ form factors.

\subsection{Non-Factorizable effects: helicity hierarchies for $h_\lambda$}

We next turn to the contribution of the hadronic weak hamiltonian in
rare semileptonic $B$-decays. As explained above, these
contributions enter only in $H_{V}(\lambda)$,
via  $h_\lambda$ defined through \eq{eq:amuhad} and \eq{eq:hlambda}.
They do not naively factorize into form factors
and leptonic currents, but involve the $B$ to $V$ matrix elements
of a $T$-product of the weak hamiltonian and an electromagnetic
current $ \langle V | T(j^{\rm em, had, \mu}(y) {\mathcal H}_{\rm eff}^{\rm had}(0)) | \bar B \rangle$.
A systematic treatment based on QCD factorization exists
\cite{Beneke:2001at,Beneke:2004dp},
to leading power in an expansion in $\Lambda/m_b$.
Contributions can be divided into ``form factor correction'',
``annihilation'', and
``spectator-scattering'' terms and are given in terms of (soft) form
factors, hard-scattering kernels, and light-cone distribution
amplitudes, as in the form factor relations discussed above.
Power corrections do not, in general, factorize. Only a subset
of $\Lambda/m_b$ power corrections to $B \to V \ell^+ \ell^-$
have been studied in the literature, in the context of isospin asymmetries
\cite{Kagan:2001zk,Feldmann:2002iw}. Some of these contributions
violate factorization and
were modelled in a standard way \cite{Beneke:2001ev,Kagan:2001zk}.

Importantly, QCD factorization  predicts that the suppression of the
positive-helicity amplitudes continues to hold
in the presence of non-factorizable terms
\cite{Beneke:2001at},
and it is also true for the known power
corrections \cite{Kagan:2001zk,Feldmann:2002iw}. That is, $h_+$ still
vanishes at this order, and this is again a direct consequence of
the $V-A$ structure of SM weak interactions. For $h_-$, the
known power corrections do not vanish, and result in an isospin
asymmetry of about 10\% at $q^2 = 0$, rapidly falling with increasing
$q^2$. For the isospin-averaged branching fraction, this reduces
to about a 3\% effect [from a replacement $e_u - e_d \to e_u + e_d$].
However, the modelling of (incalculable) power corrections is not necessarily
accurate, and it is in particular important to investigate to what
extent $h_+$ may be generated by them.

In the following, we separate the hadronic weak hamiltonian
into a charm part (with large CKM and Wilson coefficients),
a gluonic part ($Q_{8g}$, with a large CKM and Wilson coefficient), and
a light-quark part (involving only operators which are suppressed
by small CKM elements or by small Wilson coefficients), and
investigate in each case the possible sizes of power corrections, in
particular to $h_+$, by methods complementary to QCDF.

\subsubsection{Charm loop}
Within the context of LCSR, a study of charm loop effects
at low $q^2$ has been given recently by Khodjamirian et al
\cite{Khodjamirian:2010vf}, and the analogous contributions
to $B \to K^* \gamma$ have been considered earlier in \cite{Ball:2006cva}.
In \cite{Khodjamirian:2010vf}, long-distance
charm-loop effects are estimated to be sizable (and
with large uncertainties); these effects correspond in part to power
corrections in QCDF. Unfortunately, the results in
\cite{Khodjamirian:2010vf} are only presented in
a numerical form and only for transversity, not helicity, amplitudes,
expressed through effective (amplitude-dependent)
shifts of the Wilson coefficients $C_9$.
Nevertheless, central values and uncertainties given there
are suggestive of a strong correlation of the charm loop contributions
to $A_{\parallel}$ and  $A_{\perp}$
and a suppression $h_+ \ll h_-$. The
computation in \cite{Ball:2006cva} provides directly
results for $h_+$ and $h_-$ at $q^2=0$, both of them numerically
smaller, implying also $h_-$ much smaller than
suggested by \cite{Khodjamirian:2010vf}.

The aim of this section is to argue that a hierarchy
$h_+ \ll h_-, h_0$ results, as far as the charm loop goes,
from the light-cone dominance of
the amplitude at $q^2 \ll m_B^2$. To this end, let us recast the
strategy of \cite{Khodjamirian:2010vf} in terms of helicity amplitudes,
picking out the charm loop in $h_\lambda$,
\begin{eqnarray}  \label{eq:hlambdacc}
h_\lambda|_{c\bar c}
 &=& \frac{1}{m_B^2} \frac{2}{3} \, \epsilon_{\mu}^*(\lambda) \!
 \int d^4 y \, e^{i q \cdot y}   \langle M | T \big\{ (\bar c \gamma^\mu c)(y)
 (C_1^c Q_1^c + C_2^c Q_2^c)(0)  \big\} | \bar B \rangle .
\end{eqnarray}
Next, \cite{Khodjamirian:2010vf} shows that the Fourier integral
is dominated by the light-cone $y^2 \approx 0$. A light-cone OPE
is then performed. To leading order, this results in a local operator
whose matrix elements can be identified with the charm-loop
contribution to the form factor term in QCDF (ie those charm-loop
effects that do not involve the spectator quark).
At the one-gluon level, one has the expression
\begin{equation} \label{eq:hccLD}
  h_\lambda|_{c \bar c, \rm LD} =
     \epsilon^{\mu*}(\lambda) \langle M(k, \lambda) | \tilde {\cal
       O}_{\mu}| \bar B \rangle ,
\end{equation}
where
\begin{equation}
 \tilde {\cal O}_{\mu} =    \label{eq:hlamnonloc}
  \int d\omega I_{\mu\rho\alpha\beta}(q,
 \omega) 
   \bar s_L \gamma^\rho \delta \Big(\omega - \frac{i n_+ \cdot D}{2} \Big)
   \tilde G^{\alpha\beta} b_L  ,
\end{equation}
with $D$ the covariant derivative and  $I_{\mu\rho\alpha\beta}$
given in \cite{Khodjamirian:2010vf}.
The nonlocal operator \eq{eq:hlamnonloc}
is the first subleading term in an expansion in
$\Lambda^2/(4 m_c^2 - q^2)$, with terms involving
two and more gluon fields contributing only at higher
orders \cite{Khodjamirian:2010vf}. Eq.~\eq{eq:hccLD} hence provides
an approximation to the long-distance charm-loop contributions.
It \textit{can} be further expanded in local operators,
\begin{equation}
  \tilde {\cal O}_{\mu}^{(n)} =
 \frac{1}{n!} \frac{d^n}{d\omega^n} I_{\mu\rho\alpha\beta}(q, \omega)
 \Big|_{\omega=0} \bar s_L \gamma^\rho
   \Big(\frac{i n_+ \cdot D}{2} \Big)^n \tilde G^{\alpha\beta} b_L .
\end{equation}
The result of  \cite{Ball:2006cva} corresponds to
keeping only the $n=0$ term, and evaluating its matrix
element by means of a LCSR for a correlation function 
\begin{equation}
   i \int \! d^4y \, e^{-i p\cdot y}    \label{eq:BJZcorrfun}
     \langle K^*| [\tilde {\cal O}_{\mu}^{(0)}(q)](0)\, j_B^\dagger(y)| 0 \rangle.
\end{equation}
Ref.\ \cite{Ball:2006cva} argued the suppression of higher terms
in the local OPE by a \textit{larger}
expansion parameter of order $m_B \Lambda / (4 m_c^2)$, which has
been taken as $(20 - 40)\%$ and used to justify truncating the
OPE after the leading term.
This numerical value corresponds to taking
$\Lambda \sim 300-650 $ GeV (for $\overline{\rm MS}$ quark masses),
and should hold up to an ${\cal O}(1)$ factor, which if large could
in principle spoil the convergence of the OPE.
More seriously, the power counting itself was obtained
by appealing to inclusive $B \to X_s \gamma$ decay, where similar
matrix elements
$\langle B | \bar b (q \cdot D)^n G_{\alpha\beta} \Gamma b | B \rangle $
occur as part of power corrections to the charm
loop \cite{Voloshin:1996gw,Ligeti:1997tc}. ($\Gamma$ denotes a Dirac
structure which is irrelevant to the present discussion.)
There, the softness of the
$B$ meson constituents provides one power of $\Lambda$ in the
numerator, which can be seen via $q \cdot D \supset - i q\cdot k_G \sim
m_b \Lambda$, where $k_G$ is the gluon momentum \cite{Voloshin:1996gw}.
(The resulting `suppression' factor is estimated as 0.6 in
\cite{Ligeti:1997tc}.) 
However, with an energetic $K^*$ in the final state as in
\eq{eq:BJZcorrfun} the constituents
have energies ${\cal O}(m_b)$, so $n_+ \cdot D \supset n_+ \cdot k_G
\sim m_b$ and a scaling $m_b^2/(4 m_c^2)$ of the
putative expansion parameter seems appropriate; at least,
establishing a suppression requires a new argument. We therefore
will not rely on the estimate of \cite{Ball:2006cva} in this paper.
Ref. \cite{Khodjamirian:2010vf} estimates instead the full nonlocal
operator matrix element from a LCSR for a different correlation function
\begin{equation}
 \langle 0 | T \{ j_\nu^{K^*}(y) \tilde O_{\mu}(0) \} | B \rangle ,
\end{equation}
where $j_\nu^{K^*} = \bar d \gamma_\nu s$,
which yields the matrix element in terms of $B$-meson LCDAs.

\begin{figure}[t]
\centering
\includegraphics[scale=0.8]{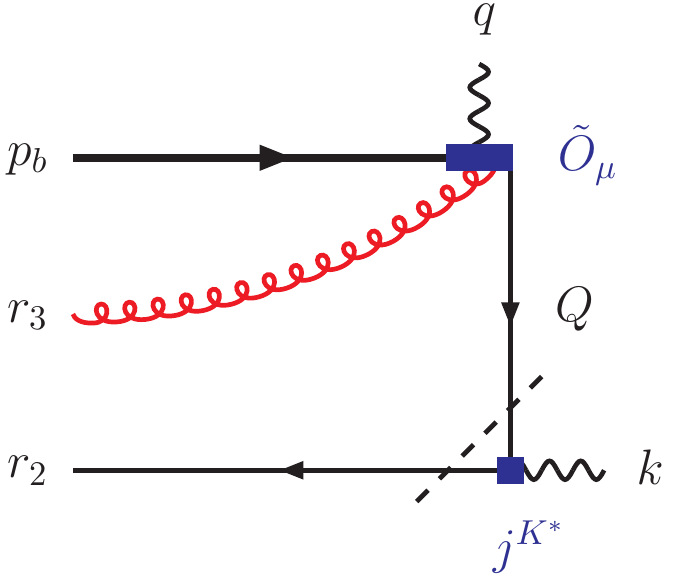}
\caption{Tree-level diagram for the light-cone OPE for the
correlation function \eq{eq:nonloccorrfun}.
The dashed line indicates a two-particle cut contributing to the
(perturbative) spectral density.
 \label{fig:nlsumrule}}
\end{figure}

To show the suppression of $h_+$, note that $h_{\pm}$ can be obtained
directly from
\begin{equation}    \label{eq:nonloccorrfun}
 G_{h\lambda}(q^2; k^2) = - i \int d^4 y e^{i k y}
    \langle 0 | T \{ \epsilon^{\nu*}(\hat z; \lambda) j_\nu^{K^*}(y)
       \epsilon^{\mu*}(-\hat z; \lambda) {\tilde O}_{\mu}(0) \} | B
       \rangle .
\end{equation}
To be precise, we take $k = (k^0, 0, 0, |{\bf k}|)$, as well as $q$,
in the $(tz)$ plane.
Note that for $\lambda=\pm$ the polarisation 4-vectors are
(with these conventions) independent of ${\bf k}$, hence the
rhs indeed defines a Lorentz-invariant function of $k^2$ and $q^2$.
(The formalism could, with appropriate care, be extended to $\lambda =0$.)
The hadronic representation contains the desired matrix element,
\begin{equation}
 G_{h\lambda}(q^2; k^2) = \frac{f_{K^* \parallel} \, m_{K^*}}{m^2_{K^*} - k^2}
   \langle K^*(\tilde k; \lambda) | \epsilon^{\mu*}(-\hat z; \lambda)
   {\tilde O}_{\mu}(0) | B \rangle
 \; + \mbox{continuum contributions} .
\end{equation}
Here  $\tilde k = (\sqrt{m^2_{K^*} + {\bf k}^2}, 0, 0, |{\bf k}|)$
is the physical (on-shell)  4-momentum of the $K^*$ corresponding to
the given $q^2$.
To obtain a LCSR, following \cite{Khodjamirian:2010vf} we take $k^2
\sim - 1 {\rm GeV}^2 \sim - m_b \Lambda$
(corresponding to a Borel parameter $\sim
\sqrt{m_b \Lambda}$) and consider the light-cone OPE of $G_{h\lambda}$.
The leading (tree) diagram is shown in Fig.\ \ref{fig:nlsumrule}.
Defining $n_+ \cdot q \equiv m_B + l$, and taking the $b$-quark
momentum to be $p_b = m_b/2 (n_+ + n_-) + r_1$,
we have $l \sim \Lambda > 0$ and
\begin{eqnarray}
   q &=& m_b \frac{n_-}{2} + \frac{q^2}{m_b + l} \frac{n_+}{2} , \\
   k &=& m_b \frac{n_+}{2} + {\cal O}(\Lambda), \\
   Q &=&  m_b \frac{n_+}{2} + {\cal O}(\Lambda) ,
\end{eqnarray}
such that $Q^2 \sim m_b \Lambda$, making the internal propagator,
in QCDF/SCET terminology, a
hard-collinear line. The operator product, to tree level, becomes
\begin{equation}  \label{eq:OmutildeDirac}
  \epsilon_\nu^{*}(\hat z, \lambda) \gamma^\nu(k) \frac{\Qslash +
    m_s}{Q^2-m_s^2} P_R \epsilon^{\mu*}(-\hat z; \lambda)
   {\tilde O}_{\mu}(q) ,
\end{equation}
where $P_R$ has been inserted for convenience [note that
${\tilde O}_{\mu}(q)$ contains a chiral projector]. Neglecting
terms ${\cal O}(\Lambda, m_s)$ in the propagator, this vanishes
for $\lambda = +$. Hence the nonlocal charm loop contribution to $h_+$ is
${\cal O}(\Lambda/m_b)$-suppressed relative to $h_-$, and altogether
\begin{equation}
  h_+|_{c \bar c, \rm LD} = {\cal O}\left( \frac{\Lambda^3}{4 m_c^2 m_b} \right) 
\end{equation}
relative to the leading-power amplitude $H_{V}^-$. 
Nevertheless, we will see below that it constitutes the dominant
remaining source of theoretical uncertainty on
several ``clean'' observables, thanks to the strong suppression of the
helicity form factor $T_+$. This shows at the same time the importance
of taking it into account, and motivates further, more quantitative
work on the nonlocal power-corrections to (mainly) the
positive-helicity amplitudes.

Unlike in the form factor case, the more traditional LCSR of 
Ref.\ \cite{Ball:2006cva}, involving the correlation function
\eq{eq:BJZcorrfun} with an on-shell $K^*$ rather than
a $B$, does not lend itself to an analogous argument. Essentially,
the reason is that that sum rule is given in terms of (most importantly)
chiral-even twist-3 3-particle LCDAs, where, loosely speaking, the
helicity of the meson is determined by that of the gluon
field-strength tensor, whereas the helicities of the quark and
antiquark fields cancel out. Hence the presence of a chiral projector
has no direct impact on the helicity of the $K^*$. Of course the
suppression should still be seen in the numerical result for the
full (rather than lowest-order local OPE) result. However,
higher-order terms in the local OPE are not known, nor do they
seem to be suppressed, as argued above.

\subsubsection{Chromomagnetic penguin operator}
The operator $Q_{8g}$ comes with a large CKM factor and its Wilson
coefficient is not small. In QCDF, it enters exclusively
through the hard spectator scattering contributions. These factorize
at leading power, but at  order $\Lambda/m_b$, attempting to factorize
them results in endpoint divergences
\cite{Kagan:2001zk}, implying a breakdown of factorization. Cutting
off the end point divergence and modelling the end-point contribution
as in \cite{Kagan:2001zk,Feldmann:2002iw},  $h_+$ still vanishes
at subleading power. However, on general grounds one would expect
chirality-violating QCD (and strange mass) effects to generate a
contribution to $h_+$ at order $\Lambda/m_b$ and $m_s/m_b$.
It is possible to estimate the $Q_{8g}$ contribution to $h_\lambda$
with a LCSR, similarly to the sum rule for the
nonlocal operator $\tilde O_\mu$. For $B \to K \ell^+ \ell^-$ (where the
dilepton helicity $\lambda = 0$), such a calculation has been very
recently performed in \cite{Khodjamirian:2012rm}, wherein a ``soft''
contribution is identified and expressed in terms of three-particle
$B$-meson light-cone distribution amplitudes. The operator $\tilde O_\mu$ is
replaced by the simpler (local) $Q_{8g}$, but the
required correlation function involves an extra insertion of the
electromagnetic current anywhere on the quark line in
Fig.~\ref{fig:nlsumrule}.

Here, we note that our above argument showing a helicity hierarchy
of the long-distance charm loop contributions likewise can
be applied to this sum rule for the matrix element of $Q_{8g}$.
The operator $Q_{8g}$ still provides a chiral projector, and the
fermion line entering $j^{K*}$ in Fig.~\ref{fig:nlsumrule} is
still ``hard-collinear'', such that the first four factors in
\eq{eq:OmutildeDirac} are unchanged, even if the electromagnetic
current insertion occurs on that line (if the strange quark
mass is neglected). The result is again that
the long-distance (soft) contribution $h_+|_{Q_8g, \rm LD}$ 
is suppressed by $\Lambda/m_b$ or $m_s/m_b$ relative to $h_-|_{Q_8g, \rm LD}$.
As the latter is already suppressed by a power of $\Lambda/m_b$ relative
to  the leading-power amplitude $H_V^-$,
the impact on $Q_{8g}$ on $h_+$ should
be negligible. (Note also that the effect of the ``soft''
$Q_{8g}$ contributions in $B \to K \ell^+ \ell^-$ was found
to be well below 1 \% of the total hadronic contribution
in \cite{Khodjamirian:2012rm}. It is difficult to see how
a much larger contribution could occur in the present case, even for
the non-helicity-suppressed amplitude $h_-$.)

\subsubsection{Light quarks and resonance structure}
\label{sec:lightquarks}
The remaining contributions of the hadronic weak Hamiltonian
to the decay amplitude coming from the
QCD penguin operators and the double Cabibbo-suppressed
current-current operators involving up quarks,
\begin{eqnarray}
a_\mu^{\rm had,\,lq}=\int d^4x\,e^{-iq\cdot x}\langle
\bar{K}^*|T\{j^{\rm em}_{\mu}(x),\,{\mathcal H}^{\rm had,\,lq}_{\rm eff}(0)\}|\bar{B}\rangle,\label{Eq:amuhadlightqs}
\end{eqnarray}
and are either doubly Cabibbo-suppressed 
or weighted by the small Wilson coefficients $C_{3-6}$. Again,
a systematic description exists within QCDF \cite{Beneke:2001at},
with a vanishing contribution to $h_+$ at leading power and a
breakdown of factorization at subleading powers. Because of the
multiple suppression factors, the contributions to $H_V^+$ arising
in this fashion are negligible. 

However, long-distance non-perturbative effects may
manifest themselves partly as resonances or poles in the complex-$q^2$
plane, implying a resonance structure which we do not expect to be
accounted for at any order in $\Lambda/m_b$. 
Therefore we employ a hadronic description to estimate both
power corrections and the possibility of large ``duality-violating''
effects in $B \to K^* \ell^+ \ell^-$ observables.
In order to do this, let us consider instead the object
\begin{eqnarray}
\tilde a_\mu^{\rm had,\,lq}=\int d^4x\,e^{-iq\cdot x}\langle
\bar{K}^*|T\{j^{\rm em,lq}_{\mu}(x),\,{\mathcal H}^{\rm had}_{\rm eff}(0)\}|\bar{B}\rangle,\label{Eq:amuhadtilde}
\end{eqnarray}
where we only keep the light-quark part of the electromagnetic
current, relevant for resonance structure in the low-$q^2$ region
(but revert to the full weak Hamiltonian).
Ideally, we would like to compute $\tilde a_\mu^{\rm had,\,lq}$
taking into account the
fact that pions and other light hadrons are the relevant
degrees of freedom of QCD in this domain, in a systematic fashion as, for example, using chiral perturbation theory ($\chi$PT)~\cite{Weinberg:1978kz,Gasser:1983yg}, together with any of the methods that extend its range of applicability up to the region of the light resonances~\cite{Ecker:1988te,Ecker:1989yg,Oller:1998hw,Oller:1998zr}. In fact, this program is attainable for kaon decays in which the energies and masses are all small compared with the chiral symmetry breaking scale $\Lambda_{\chi S B}\sim$ 1 GeV~\cite{Cirigliano:2011ny}. In case of the $B$-meson decays, one encounters heavier scales and, at present, it is not clear how to integrate them out in order to extract the long-distance effects in a model-independent way~\cite{Bijnens:2010ws,Colangelo:2012ew}.

\begin{figure}[t]
\centering
\includegraphics[scale=0.3]{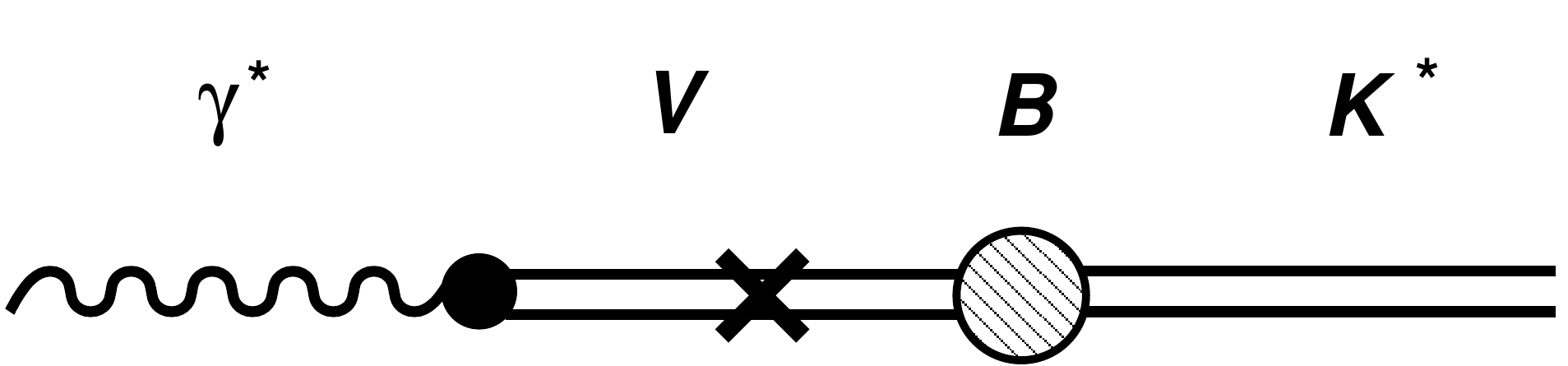}
\caption{Graphical representation of the VMD model. The filled bulb
  represents the $\bar{B}\rightarrow V\bar{K}^* $ decay vertex, as
  obtained in QCD factorization, the solid bulb $f_V$, as obtained
  from experiment, and the double lines resonance propagators, with
  the cross indicating the multi-particle dressing of the respective pole.  \label{Fig:BVVll}}
\end{figure}

In this paper we use a model to estimate the contribution of the light
hadronic degrees of freedom in the low-$q^2$ region. We start by
making a factorization approximation of the correlation
function Eq.~(\ref{Eq:amuhadtilde}),
using a basis of hadronic states $|P(0)\rangle$ and $|P^\prime(x)\rangle$,
\begin{eqnarray}
\tilde a_\mu^{\rm had,\,lq} = \int d^4x\,e^{-iq\cdot x}\sum_{P,P^\prime}
\langle 0|j^{\rm em,lq}_{\mu}(x)|P^\prime\rangle \langle
P^\prime(x)|P(0)\rangle \langle \bar{K}^*P|\,{\mathcal H}^{\rm
  had}_{\rm eff}(0)|\bar{B}\rangle, \label{Eq:amuhadlightHad}
\end{eqnarray} 
where the sums include further integrations for multi-particle
states. We next assume that these sums are saturated by the lightest
neutral vector resonances $V=\rho(770)$, $\omega(782)$ and
$\phi(1020)$, i.e.\ vector meson dominance (VMD). This hypothesis has
proven very fruitful in modelling the electromagnetic structure of
light hadrons at low energies. It finds microscopic justification in
the large $N_c$ limit of QCD~\cite{'tHooft:1973jz} and it has been
successfully implemented to connect the short-range part of the
low-energy interactions of pions with
QCD~\cite{Ecker:1988te,Ecker:1989yg}. (For a compilation of
phenomenological applications of the model in the weak decays of
mesons see Ref.~\cite{Lichard:1997ya}.) In the VMD, the first factor
in the RHS of Eq.~(\ref{Eq:amuhadlightHad}) is a semileptonic decay
constant, $f_V$, the second the vector-meson propagator and the third
a $\bar{B}\rightarrow V\bar{K}^*$ decay amplitude.
Finally, we (partially) take
into account the effect of the continuum of multi-particle hadronic states by dressing the poles of the resonance by their (off-shell) width. 
All in all, the estimate for the hadronic contribution at low $q^2$
 can be pictured as in Fig.~\ref{Fig:BVVll}.

In order to carry out the computation, it is convenient to use an
effective Lagrangian containing fields which serve as interpolators
for the vector resonances. We choose the anti-symmetric representation
advocated in Refs.~\cite{Ecker:1988te,Ecker:1989yg} for applications
in $\chi$PT. Other Lagrangian formulations consistent with chiral
symmetry and electromagnetic gauge invariance\footnote{Notice that in
  a previous VMD analysis \cite{Korchin:2011ze} of the vector-meson
  contribution to the $B\rightarrow K^*\ell^+\ell^-$ decay,
  electromagnetic gauge and non-gauge invariant Lagrangians were
  considered in the same footing and large differences between the two
  approaches have been reported at low $q^2$. In this paper we work
  exclusively with approaches consistent with electromagnetic gauge
  symmetry (and QCD, as stated in the main text).} are equivalent to
this one, once consistency with QCD asymptotic behavior of 2-point spectral functions is demanded~\cite{Ecker:1989yg}. 
  We address the reader to Appendix~\ref{app:resonancesdetails} for
 the details and conventions used in the model.

As for the $\bar{B}\rightarrow V\bar{K}^*$ decay amplitude,
it is natural, in the present context, to use the QCD factorization
calculation reported in Ref.~\cite{Beneke:2006hg}. 
In fact, as already discussed in \cite{Beneke:2001at}, there is
a one-to-one correspondence between a subclass of
diagrams in the QCDF calculation of $\bar B \to \bar K^* \ell^+ \ell^-$ and
of the diagrams appearing in the QCDF calculation for $\bar B \to V
\bar K^*$, where $V$ is one of the light-quark resonances. For
our model calculation, we employ the QCDF results for the $B \to V
\bar K^*$ amplitudes, while at the same time omitting the
corresponding terms from the $\bar B \to \bar K^* \ell^+ \ell^-$ amplitude.
Next, the heavy-quark limit predicts a hierarchy of helicity amplitudes,
$H_0 : H_- : H_+ = 1 : \Lambda/m_b : \Lambda^2/m_b^2$, i.e.\ in
relative terms $H_+ / H_- \sim \Lambda/m_b$. Although
$H_-$ and $H_+$ do not factorize in QCDF,
as the hard-scattering contributions develop
end-point singularities, and the hierarchy $H_0 \gg H_-$ is
numerically removed by large penguin corrections (obviating the
so-called polarization puzzle~\cite{Kagan:2004uw}), the hierarchy
$H_+ \ll H_-, H_0$ remains and is in good agreement
\begin{figure}
\centering
\includegraphics[scale=0.4]{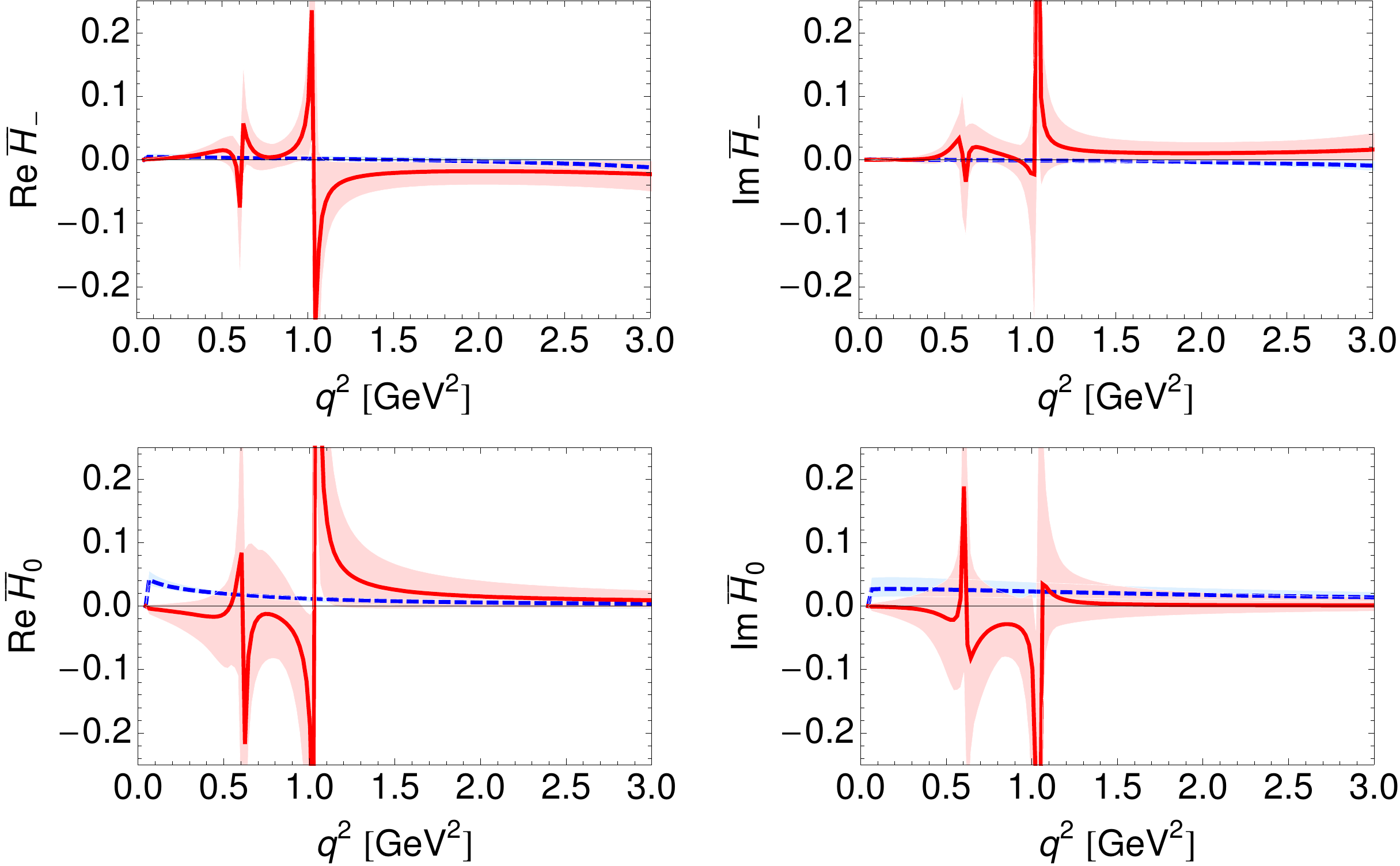}
\caption{Relative contribution (see main text) of the light-quark component of the electromagnetic current to the helicity amplitudes $H_-^V$ and $H_0^V$. The dashed (blue) line is the result in QCD factorization while the solid (red) line is the result in the hadronic model used in this work.    \label{Fig:PlotsCompLQ}}
\end{figure}
with experiment in $\bar{B}^0\rightarrow\bar{K}^{0*}\phi$ decays, for
which complete angular analyses
exist~\cite{Aubert:2004xc,Chen:2005zv}. An important consequence of
this is that both in QCDF and in our model,
the contribution to the $\bar{B}\rightarrow
\bar{K}^*\ell^+\ell⁻$ ``wrong-helicity'' amplitude is suppressed as  
\begin{equation}
h_+^{\rm had,\,lq}/h_0^{\rm had,\,lq} = \mathcal{O}(\Lambda/m_b)^2,
\qquad
h_+^{\rm had,\,lq}/h_-^{\rm had,\,lq} = \mathcal{O}(\Lambda/m_b),
\end{equation}
on top of the  smallness already implied by the CKM
factor and Wilson coefficients. As a result, we expect the
light-quark hadronic pollution of observables sensitive to the
chirality-flipped operator $Q_7^{\prime}$ to be negligible, and
preclude in particular large ``duality-violating'' corrections to
the leading-power
QCDF results for $\bar B \to \bar K^* \ell^+ \ell^-$.
Thus, in the phenonomenological section we
set $h_+^{\rm had,\,lq} = 0$.

In Fig.~\ref{Fig:PlotsCompLQ} we compare the model calculation to
that obtained from the corresponding subclass of QCDF diagrams.
More precisely, in order to show their relative impact on
the helicity amplitudes, we plot
$\bar{H}_-=H_-^{\rm lq}/H_-^{V}$ and $\bar{H}_0=H_0^{\rm
  lq}/H_0^{V}$. The two approaches agree on the relative smallness of
these terms, of a few percent, across most part of the low $q^2$
region. Remarkably, this continues to be true in the upper part,
despite the fact we are neglecting the excited vector mesons $\rho^*$,
$\omega^*$ and $\phi^*$ that populate the region $q^2\gtrsim2$
GeV$^2$. In this regard, we point out that these states have a
sub-dominant effect in the hadronic structure of the electromagnetic
current as it is, indeed, probed in $e^+e^-$ annihilation
experiments. Therefore, one would expect their contribution to produce
some oscillations in the hadronic determination converging,
approximately, to the QCDF result at larger $q^2$. In
the numeric computations we ignore these higher-mass resonance
contributions and we enforce a continuous matching onto the QCDF
result at the threshold $q^2\sim2$ GeV$^2$ by means of a
suitable $q^2$-dependent smearing function
(see Appendix \ref{app:resonancesdetails}).

On the other hand, in the regions around the resonance poles, the
discrepancy between the two approaches is maximal and the hadronic
model gives contributions which are comparable in magnitude to the
leading ones. In order to discuss this in more detail, consider the
schematic form of the semileptonic decay amplitude (see Appendix~\ref{app:resonancesdetails}),
\begin{eqnarray}
\mathcal{M}_T=\mathcal{M}+\epsilon \frac{8\pi\,Q_V\,f_{K^*}\, f_V}{\left(q^2-m_V^2+im_V\Gamma_V\right)}\mathcal{M}^\prime,\nonumber
\end{eqnarray}
given by the interference of the leading piece $\mathcal{M}$ including the effects of the electromagnetic and semileptonic penguin coefficients $C_{7,9,10}$ and $\mathcal{M}^\prime$, the $\bar{B}\rightarrow V \bar{K}^*$ amplitude normalized such that $|\mathcal{M}|\sim|\mathcal{M}|^\prime$ and $\epsilon\sim\lambda_{\rm CKM}^2\sim 0.05$.  Then,
\begin{eqnarray}
|\mathcal{M}_T|^2&=&|\mathcal{M}|^2+\epsilon^2\frac{(8\pi\,Q_V\,f_{K^*}\, f_V)^2}{(q^2-m_V^2)^2+m_V^2\Gamma_V^2}|\mathcal{M}^\prime|^2\nonumber\\
&+&2\epsilon\frac{8\pi\,Q_V\,f_{K^*}\, f_V}{(q^2-m_V^2)^2+m_V^2\Gamma_V^2}\left\{(q^2-m_V^2){\rm Re}[\mathcal{M}^*\mathcal{M}^\prime]+m_V\,\Gamma_V\,{\rm Im}[\mathcal{M}^*\mathcal{M}^\prime]\right\}.\nonumber
\end{eqnarray}
The contribution of the resonances to the decay rate is sizable and
potentially of the same order as the leading one only within a $q^2$-region
of width $\sim m_V\Gamma_V$ around the position of their poles, in which
$\epsilon\times (f_V f_{K^*}/m_V\Gamma_V)\sim1$. This means, at the
same time, that integrating over a large enough region of $q^2$,
$\Delta q^2\gg m_V\,\Gamma_V$, suppresses their relative contributions
by a factor $m_V\,\Gamma_V/\Delta q^2$. Therefore, we conclude that
although the effects of the light resonances could alter the line
shape of observables, they are largely washed out by binning in
$q^2$. We will see this explicitly in the results presented in the
next section. 

\subsection{Summary and phenomenological implementation}
In the preceding two subsections we have studied hadronic effects in
the contributions of the various parts of the weak $\Delta B=1$
effective hamiltonian to the helicity amplitudes. Our main outcome is
a strong suppression of the helicity amplitudes $H_V^+$ and $H_A^+$,
which in turn suppresses the coefficients $I_3$ and $I_9$ in the
angular distribution.
\begin{itemize}
\item
At the factorizable level, this comes about through a double
suppression of $T_+$ by  $q^2/m_B^2$ and $\Lambda/m_b$, and a
single suppression of $V_+$ by $\Lambda/m_b$. In the absence
of right-handed currents (primed Wilson coefficients) this translates
to suppressed $H_V^+$ and $H_A^+$.
\item The contributions of the hadronic weak Hamiltonian enter only
into $H_V^+$, via $h_+$ defined in \eq{eq:hlambda}. They are
calculable at the leading power in an expansion in $\Lambda/m_b$ in
QCDF. At this order, $h_+ = 0$.
\item Long-distance ``charm loop'' effects can be studied in an expansion
$\Lambda^2/(4 m_c^2 - q^2)$. At zeroth order, one recovers
the non-factorizable form factor QCDF
expressions, which give vanishing contributions to $h_+$.
The first nonzero, long-distance, contribution to
$h_+$ arises at order $\Lambda^3/(4 m_b m_c^2)$ relative to the amplitudes
$H_V^-$ and $H_V^0$.
Long-distance contributions of the chromomagnetic operator $Q_{8g}$ are
expected to be small, with the contribution to $h_+$ suppressed
(at least) as $\Lambda^2/m_b^2$.
\item The remaining effects are suppressed by
small CKM and Wilson coefficients. We estimate long-distance
corrections to QCDF by means of a resonance (VMD) model, which
embodies a suppression of the contribution to $h_+$, as a consequence of
the helicity structure in $\bar B \to V \bar K^*$ decay. After binning
the model predicts very small corrections in the other amplitudes
$h_-$ and $h_0$.
\end{itemize}
In our phenomenological analysis, we employ our model-independent
parameterisation of form factors through \eq{Eq:softFFq2} and 
\eq{Eq:pcsFFs} (keeping terms up to and including $\mathcal
O(q^2/m_B^2)$),
together with the numerical values and ranges given in Sect.\ \ref{sect:FFs}.
For the contributions of the hadronic weak Hamiltonian, we employ
the leading-power QCD factorization expressions
\cite{Beneke:2001at,Beneke:2004dp}. To estimate the long-distance
corrections, we employ the following model for long-distance charm loop
effects,
\begin{eqnarray}
  h_-|_{c \bar c, \rm LD} &=& 0.1\, e^{i \phi_-} C_7^{\rm SM} , \\
  h_+|_{c \bar c, \rm LD} &=& 0.02\, e^{i \phi_+} C_7^{\rm SM} ,\label{eq:modelPCcharm}
\end{eqnarray}
where $\phi_\pm$ are arbitrary soft rescattering phases, comprising
a conservative interpretation of the
numerical findings of \cite{Khodjamirian:2010vf} and the hierarchy
$h_+ \ll h_-$. We will also allow for an extra long-distance
contribution
$h_0|_{c \bar c, \rm LD} = 0.2 \, e^{i \phi_0} C_7^{\rm SM} $, as
there is no power suppression of $h_0$; this should be considered
an ad hoc model but  does not impact on the observables emerging as
clean in the phenomenological analysis below, which only involve
helicities $\pm 1$. We have increased the magnitude of these effects
beyond the error estimates of \cite{Khodjamirian:2010vf}, such
as to accomodate within it the small long-distance contributions from the
chromomagnetic penguin operators. For the light-quark Hamiltonian,
we estimate possible long-distance corrections by means of the
model described in Sect.\ \ref{sec:lightquarks}. Input parameters
are summarized in Appendix \ref{app:SMparam}.

\section{Phenomenology at low $q^2$}
\label{sec:Pheno}
\subsection{Observables for $\bar{B}\rightarrow\bar{K}^*\ell^+\ell^-$}

In general, there are twelve $q^2$-dependent observables
(shown in \cite{Altmannshofer:2008dz} neglecting tensor operators,
but this remains true in the their presence) that
are accessible through a full-angular analysis of the
$\bar{B}\rightarrow\bar{K}^*\ell^+\ell^-$ decay rate and which
correspond to the angular coefficients $I_i(q^2)$ in
Eq.~(\ref{Eq:DR4body1}).
In the absence of scalar and tensor operators, which includes
the SM, $I_6^c=0$, and there is one relationship among the remaining
coefficients, reducing the number of
independent observables to ten~\cite{Egede:2010zc}.
If one furthermore assumes $m_\ell=0$, two more relations
can be established,
\begin{equation}
3I_{1s}=I_{2s},\hspace{3cm}I_{1c}=-I_{2c},\label{Eq:RelsI121}    
\end{equation}
leading to eight independent observables.

The analysis of the CP-partner decay $B\rightarrow K^*\ell^+\ell^-$ gives a same amount of independent observables as in the $\bar{B}$ decay, the $\bar{I}_i$'s.  In this sense, it is useful to define the following combinations of $I_i$'s and $\bar{I}_i$'s,
\begin{equation}
\Sigma_i=\frac{I_i+\bar{I}_i}{2},\hspace{2cm}
\Delta_i= \frac{I_i-\bar{I}_i}{2},\label{Eq:CPobservables}
\end{equation}
which can be used to construct a variety of
CP-averages and asymmetries~\cite{Bobeth:2008ij,Altmannshofer:2008dz}.         

\subsection{The $\bar{B}^0\rightarrow\bar{K}^{*0}\mu^+\mu^-$ decay}

\label{sect:observables}
\begin{figure}
\begin{center}
\includegraphics[scale=0.5]{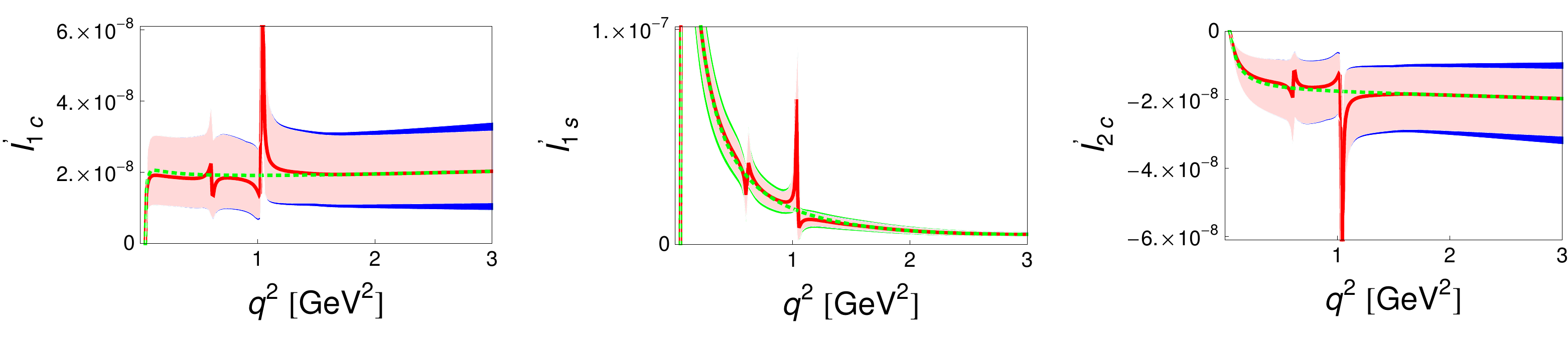}
\includegraphics[scale=0.5]{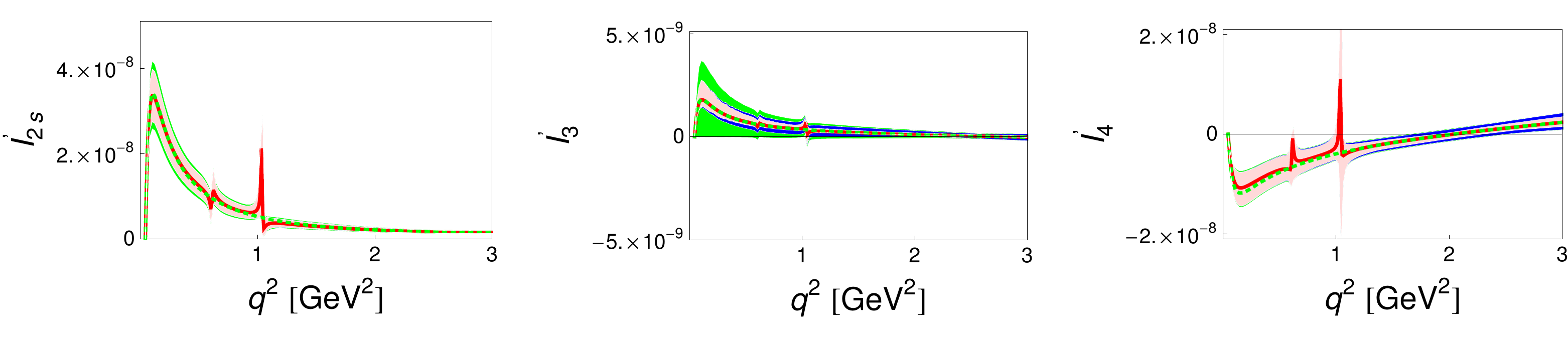}
\includegraphics[scale=0.5]{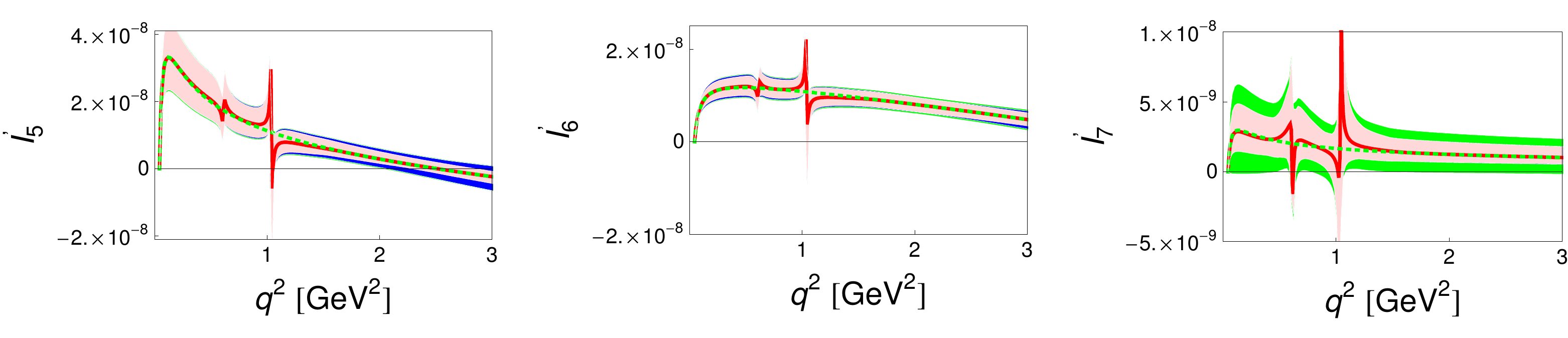}
\includegraphics[scale=0.47]{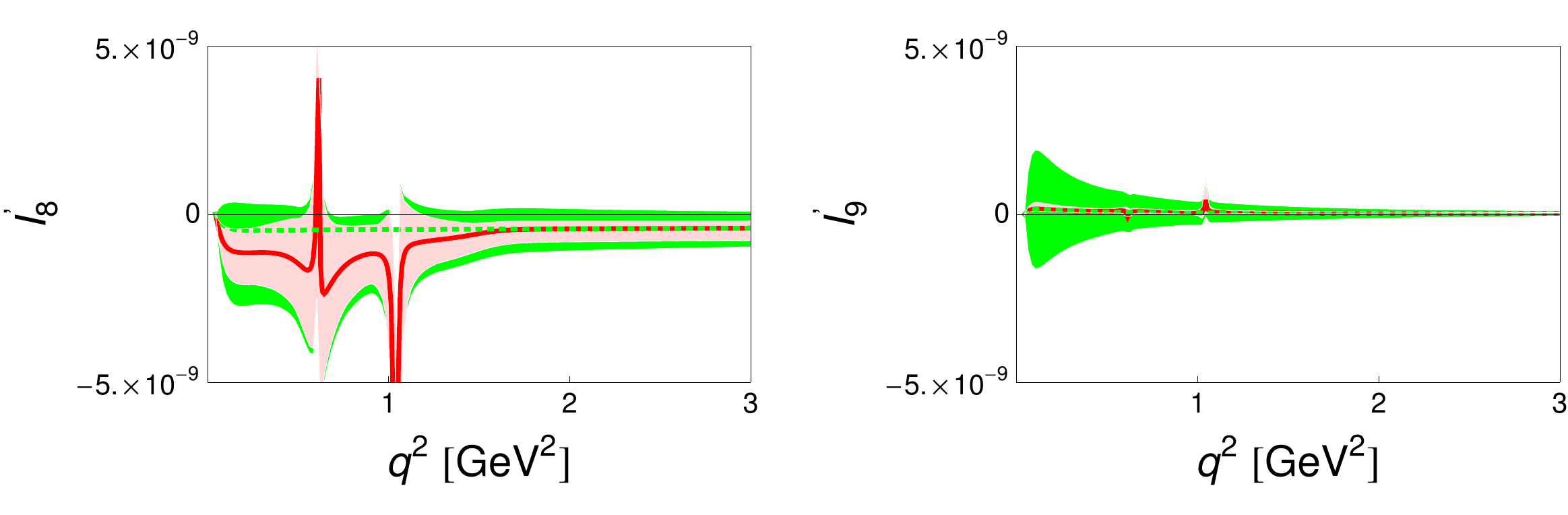}
\caption{Angular coefficients normalized by the $\bar{B}^0$ decay rate
  ($I^{\prime}_i$) around the low-$q^2$ end-point. Solid (red) and dashed (green) lines correspond to the SM prediction including the light-quark contributions in the hadronic model or QCD factorization, respectively. The error bands stem from the hadronic and CKM uncertainties and renormalization scale dependence (inner-red), plus the unknown factorizable power-corrections (internal-blue) and plus the non-factorizable charm-loop uncertainty (outer-green). The errors are added, subsequently, in quadratures.\label{Fig:PlotIs}}
\end{center}
\end{figure}

In Fig.~\ref{Fig:PlotIs}, we show the SM predictions for the eleven angular coefficients available in this case and normalized by the $\bar{B}^0$ decay rate at low $q^2$. The solid (red) and the dashed (green) lines correspond to the prediction including the light-quark contributions in the hadronic model or in QCD factorization, respectively (see Sec.~\ref{sect:hadronic}). The inner (red) error band is the uncertainty derived from the hadronic parameters (soft form factors, decay constants,$\ldots$), the CKM parameters and the renormalization scale. The intermediate (blue) and outer (green) bands result from the addition in quadratures of the unknown factorizable and charm-loop power corrections, subsequently. Factorizable corrections are estimated using Eq.~(\ref{Eq:pcsFFs}), and the charm-loop uncertainty are modelled according to Eqs.~(\ref{eq:modelPCcharm}).

The main source of uncertainties in the $I_i$'s stem from the soft form factors and, in some cases, from the charm-loop. In particular, for the coefficients proportional to $H_{V,A}^+$, $I_3$ and $I_9$, the latter source is, by far, the most important. On the other hand, it is remarkable that the uncertainties in the coefficients arising from the unknown factorizable power corrections are negligible at low $q^2$. This effect is due to the constraints imposed by the exact relations~(\ref{eq:Tplusq20}). Finally, notice that the vector-meson resonances alter significantly the line shape of most of the $I_i$'s, except for those $\propto H_{V,A}^+$  due to the suppression of the corresponding helicity amplitude in the $\bar{B}\rightarrow \bar{K}^*V$ decays (see Sec.~\ref{sect:hadronic}).   

\subsubsection{CP-averages: The branching fraction, $F_L$ and the $P$-basis}
\label{sec:CPaverages}

\begin{figure}
\begin{center}
\includegraphics[scale=0.47]{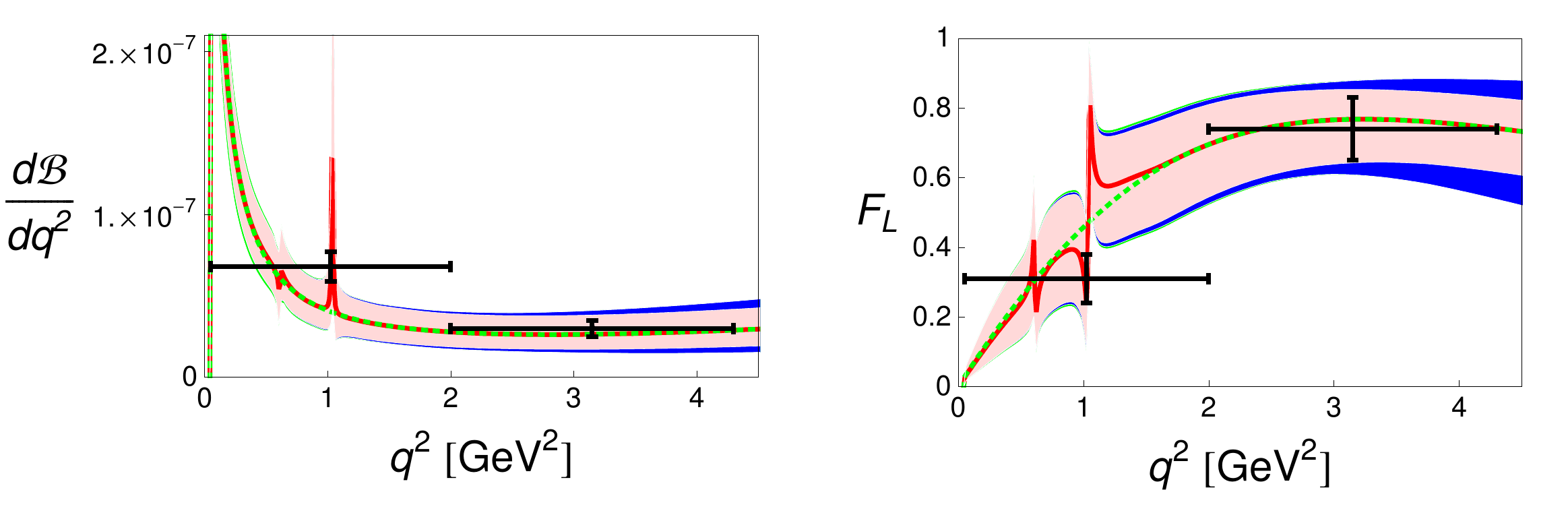}
\includegraphics[scale=0.5]{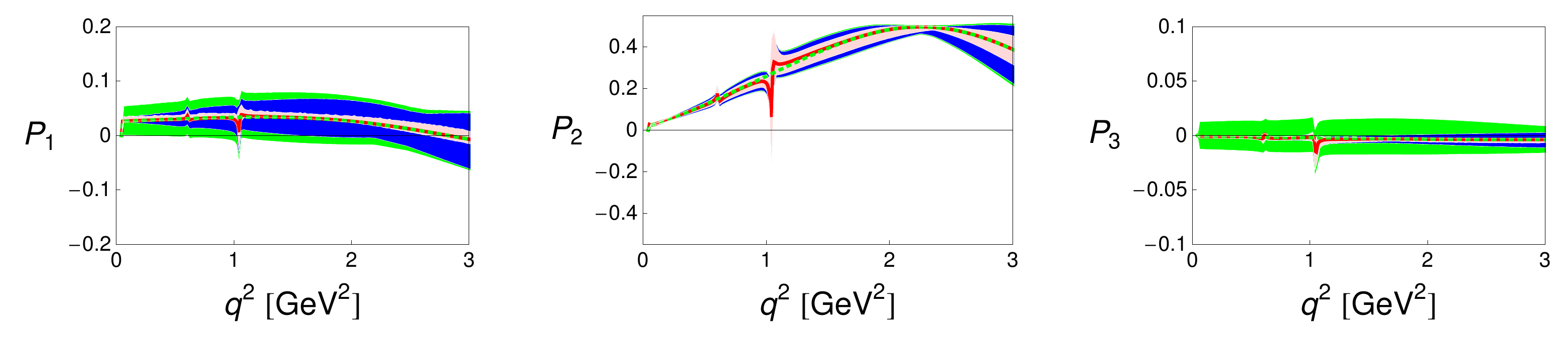}
\includegraphics[scale=0.5]{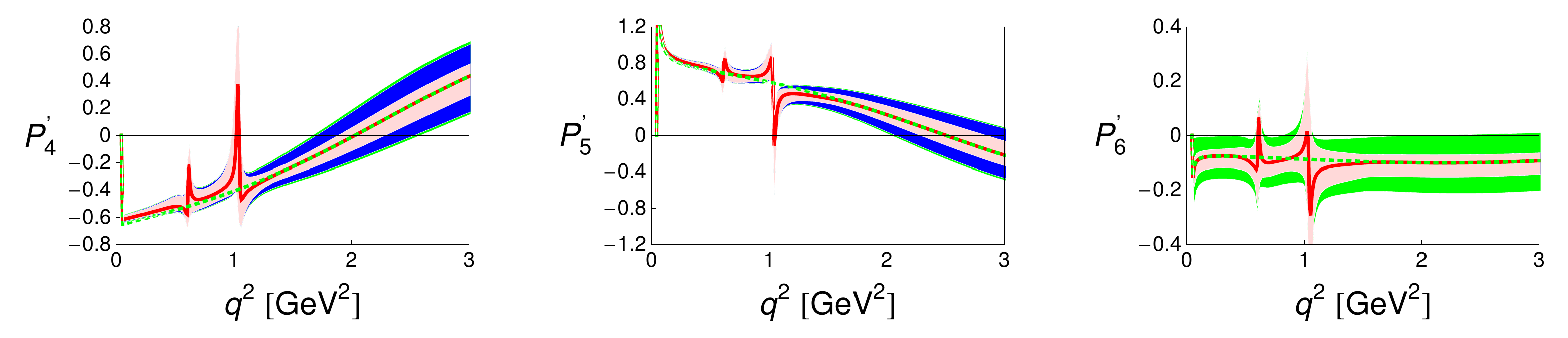}
\caption{Differential branching fraction, $F_L$ and the ``clean'' observables $P_i^{(\prime)}$ around the low-$q^2$ end-point. We show in black the experimental results for the two first observables in the bins [0.05, 2] GeV$^2$ and [2, 4.3] GeV$^2$~\cite{LHCbNote}. The color code is as in Fig.~\ref{Fig:PlotIs}.\label{fig:PlotsClean}}
\end{center}
\end{figure}

Each of the observables that can be constructed out of the CP
combinations in Eqs.~(\ref{Eq:CPobservables}) has a different
sensitivity to the various standard and non-standard Wilson
coefficients. In order to maximize these sensitivities, it is important to find a set observables with reduced dependence to the uncertain hadronic parameters underpinning the theoretical predictions, in particular the $B\rightarrow K^*$ form factors~\cite{Kruger:2005ep,Egede:2008uy}. Indeed, a proper set of  ``clean'' observables can be obtained using suitable ratios of angular coefficients~\cite{Matias:2012xw,DescotesGenon:2012zf,Becirevic:2011bp}. Following this strategy, we use the following set of CP-averaged observables~\cite{DescotesGenon:2012zf},
\begin{equation*}
P_1=\frac{\Sigma_3}{2\Sigma_{2s}},\hspace{1cm}P_2=\frac{\Sigma_{6}}{8\Sigma_{2s}},\hspace{1cm}P_3=-\frac{\Sigma_9}{4\Sigma_{2s}},
\end{equation*}
\begin{equation*}
P_4^\prime=\frac{\Sigma_4}{\sqrt{-\Sigma_{2s}\Sigma_{2c}}},\hspace{1cm}P_5^\prime=\frac{\Sigma_5}{2\sqrt{-\Sigma_{2s}\Sigma_{2c}}},\hspace{1cm}P_6^\prime=-\frac{\Sigma_7}{2\sqrt{-\Sigma_{2s}\Sigma_{2c}}}.
\end{equation*}
These observables remain independent of each other when the leptons are assumed to be massless. As it was explained above, two more independent observables can be defined in case one lifts this assumption and incorporates lepton-mass dependent contributions to the decay rate. These other observables have some special features and we discuss them separately in Sec~\ref{sec:M1andM2}. Nevertheless, we keep all the lepton-masses at their physical values in the differential decay rate and, in particular, in the expressions of the angular coefficients.   
  
The set of observables in Eqs.~(\ref{Eq:Pis}) is considered clean in the sense that the soft form factors cancel at LO in the ratios so that the uncertainty stemming from the form factors is suppressed in $\alpha_s$ (as computed at NLO in QCD factorization) or $\Lambda/m_b$ (unknown factorizable power corrections). Notice that some of these observables can be related with others previously defined in the literature~\cite{Kruger:2005ep,Egede:2008uy,Becirevic:2011bp}. In order to form a complete set of independent observables, in the (pseudo)scalar-less case, we add the decay rate $\Gamma^\prime$ and $F_L$,
\begin{eqnarray}
\Gamma^\prime=\frac{1}{2}\frac{d\Gamma+d\bar{\Gamma}}{dq^2}&=&\frac{1}{4}\left(\left(3\Sigma_{1c}-\Sigma_{2c}\right)+2\left(3\Sigma_{1s}-\Sigma_{2s}\right)\right)\label{Eq:DRdef}
\end{eqnarray}
\begin{equation}
F_T=\frac{3\Sigma_{1s}-\Sigma_{2s}}{2\Gamma^\prime},\hspace{1cm}F_L=\frac{3\Sigma_{1c}-\Sigma_{2c}}{4\Gamma^\prime},\label{Eq:DefFTL} 
\end{equation}
satisfying $F_T=1-F_L$. Another option would involve, e.g., the forward-backward asymmetry~(\ref{eq:AFB}) although notice that it can be obtained straightforwardly combining $P_2$, $F_L$ and the decay rate.

In Fig.~\ref{fig:PlotsClean} we plot the differential branching
fraction, $F_L$ and the clean observables $P_{1-6}^{(\prime)}$, and we
follow the same color and line code as the one used in
Fig.~\ref{Fig:PlotIs}. Also, we show in black boxes the experimental
results for the two first observables in the bins [0.05, 2] GeV$^2$
and [2, 4.3] GeV$^2$ that have been measured by the LHCb
collaboration~\cite{LHCbNote}. These measurements agree very well with
the SM predictions. By comparing the first row with the second and
third rows of the panel in Fig.~\ref{fig:PlotsClean}, we ratify the
advantage of using a set of observables with reduced theoretical
uncertainties~\cite{Kruger:2005ep,Matias:2012xw,DescotesGenon:2012zf}. While
for the differential branching fraction and $F_L$ the limited
knowledge of the hadronic parameters (specially the soft form factors)
is the dominant source of uncertainty, for the $P_i^{(\prime)}$ the
power corrections become much more important. In the latter case, the
enforcement of the form factor relations~(\ref{eq:Tplusq20}) is
essential to constrain the size of the factorizable power corrections
at very low $q^2$, cf. $P_1$ and $P_3$. As a consequence, the
uncertainty in this region is dominated by the power corrections to
the charm loop. This has to be interpreted as an outcome of the
not-very-precise knowledge we have of these contributions, as compared
with the information we can gather for the form factors. As for the
effect of long-distance effects in the light-quark contributions we
see that, indeed, they can modify abruptly the line shape of most of
the observables in the neighborhood of the vector-resonance
poles. Nevertheless, these effects are again tiny in $P_1$ and $P_3$
(which are $\propto H_{V,A}^+$) and, as argued in
Sec.~\ref{sect:hadronic}, they dilute after binning in
$q^2$. Moreover, $P_1$ and $P_3$ as pure helicity-1 objects are
free from $S$-wave contamination. They emerge as especially clean
null tests of the Standard Model.

\begin{table}
\centering
\caption{Results and error budget on the binned CP-averaged observables of the muonic mode. \label{Table:binnedRes}}
\begin{tabular}{|cc|c|cccc|}
\hline
Obs.& $[q^2_{min},\,q^2_{max}]$&Result&Hadronic& Fact.& $c$-quark&Light-quark\\
\hline
&\raisebox{0.3ex}[15.pt]{[0.1, 1]}&0.81$^{+0.23}_{-0.20}$&$^{+0.20}_{-0.17}$&$^{+0.03}_{-0.03}$&$^{+0.10}_{-0.10}$&$\pm0.00$\\
&[0.1, 2]&1.13$^{+0.39}_{-0.38}$&$^{+0.36}_{-0.24}$&$^{+0.08}_{-0.07}$&$^{+0.13}_{-0.12}$&$\pm0.02$\\
\multicolumn{1}{|c}{\raisebox{1ex}[0.pt]{$10^7\times\langle \frac{d\mathcal{B}}{dq^2}\rangle$}}&[2, 4.3]&0.62$^{+0.33}_{-0.26}$&$^{+0.27}_{-0.21}$&$^{+0.19}_{-0.15}$&$^{+0.02}_{-0.01}$&$\pm0.00$\\
&[1, 6]&1.5$^{+0.8}_{-0.6}$&$^{+0.6}_{-0.5}$&$^{+0.46}_{-0.37}$&$^{+0.05}_{-0.05}$&$\pm0.02$\\
\hline
&\raisebox{0.3ex}[15.pt]{[0.1, 1]}&0.20$^{+0.11}_{-0.10}$&$^{+0.10}_{-0.09}$&$^{+0.02}_{-0.02}$&$^{+0.03}_{-0.02}$&$\pm0.01$\\
&[0.1, 2]&0.31$^{+0.16}_{-0.12}$&$^{+0.15}_{-0.11}$&$^{+0.04}_{-0.04}$&$^{+0.04}_{-0.03}$&$\pm0.01$\\
\multicolumn{1}{|c}{\raisebox{1ex}[0.pt]{$\langle F_L\rangle$}}&[2, 4.3]&0.75$^{+0.11}_{-0.16}$&$^{+0.09}_{-0.13}$&$^{+0.07}_{-0.9}$&$^{+0.02}_{-0.02}$&$\pm0.00$\\
&[1, 6]&0.70$^{+0.14}_{-0.17}$&$^{+0.11}_{-0.13}$&$^{+0.09}_{-0.11}$&$^{+0.02}_{-0.02}$&$\pm0.00$\\
\hline
&\raisebox{0.3ex}[15.pt]{[0.1, 1]}&2.9$^{+3.2}_{-3.1}$&$^{+0.8}_{-0.1}$&$^{+1.2}_{-1.3}$&$^{+2.9}_{-2.8}$&$\pm0.0$\\
&[0.1, 2]&3.0$^{+3.5}_{-3.4}$&$^{+0.8}_{-0.2}$&$^{+1.7}_{-1.7}$&$^{+2.9}_{-2.9}$&$\pm0.1$\\
\multicolumn{1}{|c}{\raisebox{1ex}[0.pt]{$10^2\times\langle P_1\rangle$}}&[2, 4.3]&$-1^{+7}_{-5}$&$^{+1.6}_{-0.8}$&$^{+7}_{-5}$&$^{+1.8}_{-1.6}$&$\pm0.0$\\
&[1, 6]&$-2^{+8}_{-6}$&$^{+1.3}_{-0.8}$&$^{+8}_{-6}$&$^{+1.6}_{-1.4}$&$\pm0.0$\\
\hline
&\raisebox{0.3ex}[15.pt]{[0.1, 1]}&1.02$^{+0.15}_{-0.17}$&$^{+0.08}_{-0.13}$&$^{+0.10}_{-0.09}$&$^{+0.08}_{-0.07}$&$\pm0.00$\\
&[0.1, 2]&1.57$^{+0.19}_{-0.26}$&$^{+0.08}_{-0.20}$&$^{+0.13}_{-0.13}$&$^{+0.11}_{-0.10}$&$\pm0.04$\\
\multicolumn{1}{|c}{\raisebox{1ex}[0.pt]{$10\times\langle P_2\rangle$}}&[2, 4.3]&$3.1^{+1.4}_{-1.6}$&$^{+0.8}_{-0.8}$&$^{+1.0}_{-1.2}$&$^{+0.5}_{-0.7}$&$\pm0.0$\\
&[1, 6]&$1.4^{+1.5}_{-1.5}$&$^{+0.8}_{-0.7}$&$^{+1.2}_{-1.1}$&$^{+0.5}_{-0.6}$&$\pm0.0$\\
\hline
&\raisebox{0.3ex}[15.pt]{[0.1, 1]}&$-0.1^{+1.5}_{-1.2}$&$^{+0.0}_{-0.2}$&$^{+0.1}_{-0.1}$&$^{+1.5}_{-1.2}$&$\pm0.0$\\
&[0.1, 2]&$-0.2^{+1.6}_{-1.3}$&$^{+0.0}_{-0.2}$&$^{+0.1}_{-0.1}$&$^{+1.6}_{-1.2}$&$\pm0.0$\\
\multicolumn{1}{|c}{\raisebox{1ex}[0.pt]{$10^2\times\langle P_3\rangle$}}&[2, 4.3]&$-0.3^{+1.2}_{-1.2}$&$^{+0.1}_{-0.3}$&$^{+0.7}_{-0.8}$&$^{+1.0}_{-0.9}$&$\pm0.0$\\
&[1, 6]&$-0.3^{+1.0}_{-1.0}$&$^{+0.1}_{-0.3}$&$^{+0.6}_{-0.6}$&$^{+0.8}_{-0.7}$&$\pm0.0$\\
\hline
&\raisebox{0.3ex}[15.pt]{[0.1, 1]}&$-5.1^{+0.9}_{-0.4}$&$^{+0.8}_{-0.0}$&$^{+0.2}_{-0.2}$&$^{+0.3}_{-0.2}$&$\pm0.3$\\
&[0.1, 2]&$-3.8^{+1.4}_{-0.6}$&$^{+1.2}_{-0.0}$&$^{+0.3}_{-0.3}$&$^{+0.4}_{-0.4}$&$\pm0.4$\\
\multicolumn{1}{|c}{\raisebox{1ex}[0.pt]{$10\times\langle P_4^\prime\rangle$}}&[2, 4.3]&$4.6^{+1.8}_{-2.2}$&$^{+0.9}_{-1.4}$&$^{+1.3}_{-1.4}$&$^{+1.}_{-0.9}$&$\pm0.0$\\
&[1, 6]&$4.6^{+1.6}_{-1.9}$&$^{+0.9}_{-1.2}$&$^{+1.1}_{-1.2}$&$^{+0.8}_{-0.8}$&$\pm0.0$\\
\hline
&\raisebox{0.3ex}[15.pt]{[0.1, 1]}&$6.9^{+0.8}_{-0.5}$&$^{+0.6}_{-0.1}$&$^{+0.3}_{-0.3}$&$^{+0.2}_{-0.2}$&$\pm0.3$\\
&[0.1, 2]&$5.5^{+0.7}_{-1.1}$&$^{+0.1}_{-0.8}$&$^{+0.6}_{-0.6}$&$^{+0.3}_{-0.3}$&$\pm0.2$\\
\multicolumn{1}{|c}{\raisebox{1ex}[0.pt]{$10\times\langle P_5^\prime\rangle$}}&[2, 4.3]&$-2.5^{+3.1}_{-2.7}$&$^{+1.5}_{-1.1}$&$^{+2.5}_{-2.3}$&$^{+0.9}_{-1.0}$&$\pm0.0$\\
&[1, 6]&$-2.8^{+3.0}_{-2.6}$&$^{+1.3}_{-1.1}$&$^{+2.5}_{-2.2}$&$^{+0.8}_{-0.9}$&$\pm0.0$\\
\hline
&\raisebox{0.3ex}[15.pt]{[0.1, 1]}&$-0.8^{+0.7}_{-0.8}$&$^{+0.3}_{-0.4}$&$^{+0.0}_{-0.0}$&$^{+0.7}_{-0.7}$&$\pm0.0$\\
&[0.1, 2]&$-0.8^{+0.7}_{-0.8}$&$^{+0.2}_{-0.5}$&$^{+0.0}_{-0.0}$&$^{+0.7}_{-0.7}$&$\pm0.0$\\
\multicolumn{1}{|c}{\raisebox{1ex}[0.pt]{$10\times\langle P_6^\prime\rangle$}}&[2, 4.3]&$-0.9^{+1.0}_{-1.0}$&$^{+0.3}_{-0.5}$&$^{+0.1}_{-0.1}$&$^{+0.9}_{-0.9}$&$\pm0.0$\\
&[1, 6]&$-0.7^{+0.8}_{-1.0}$&$^{+0.2}_{-0.5}$&$^{+0.1}_{-0.1}$&$^{+0.8}_{-0.8}$&$\pm0.0$\\
\hline
\end{tabular}
\end{table}
On the experimental side~\cite{LHCbNote}, what is measured are ratios
of binned observables (rather than binned ratios), hence it becomes
necessary to define the observables not as $q^2$-integrals of
functions of the $I_i$'s but rather as the same functions of the
corresponding binned angular coefficients; only in case the coefficients are slowly varying functions of $q^2$ in the bin considered, the two methods give approximately the same result. As one can deduce from Figs.~\ref{fig:PlotsClean} and~\ref{Fig:PlotIs}, this {\it is not} the case around the low-$q^2$ end-point. With these considerations, we define the following set of CP-averaged binned observables~\cite{DescotesGenon:2012zf},
\begin{equation*}
\langle\Gamma^\prime\rangle=\frac{1}{2}
\left\langle\frac{d\Gamma+d\bar{\Gamma}}{dq^2}\right\rangle,\hspace{1cm}\langle F_L\rangle=\frac{\langle3\Sigma_{1c}-\Sigma_{2c}\rangle}{4\langle\Gamma^\prime\rangle}, 
\end{equation*}     
\begin{equation*}
\langle P_1\rangle=\frac{\langle \Sigma_3\rangle}{2\langle \Sigma_{2s}\rangle},\hspace{0.3cm}\langle P_2\rangle=\frac{\langle \Sigma_{6}\rangle}{8\langle \Sigma_{2s}\rangle},\hspace{0.3cm}\langle P_3\rangle=-\frac{\langle \Sigma_9\rangle}{4\langle \Sigma_{2s}\rangle},
\end{equation*}
\begin{eqnarray}
\langle P_4^\prime\rangle=\frac{\langle \Sigma_4\rangle}{\sqrt{-\langle \Sigma_{2s}\rangle\langle \Sigma_{2c}\rangle}},\hspace{0.2cm}\langle P_5^\prime\rangle=\frac{\langle \Sigma_5\rangle}{2\sqrt{-\langle \Sigma_{2s}\rangle\langle \Sigma_{2c}\rangle}},\hspace{0.2cm}\langle P_6^\prime\rangle=-\frac{\langle \Sigma_7\rangle}{2\sqrt{-\langle \Sigma_{2s}\rangle\langle \Sigma_{2c}\rangle}},\nonumber\\
\label{Eq:BinnedObs}
\end{eqnarray}
where $\langle \Sigma_i\rangle=\int^{q^2_{max}}_{q^2_{min}}\Sigma_i(q^2)dq^2$. In Table~\ref{Table:binnedRes}, we present the predictions for the binned observables, as defined in Eq.~(\ref{Eq:BinnedObs}), in some bins of interest~\cite{SerraPrivate}. In particular, we highlight the results on the low-$q^2$ bin [0.1, 1] GeV$^2$, which presents an overall small theoretical (relative) error, as compared with the higher $q^2$ bins, due to the small form factor uncertainties in this region. It is also remarkable that the estimate on the contributions from the light hadronic resonances comes to be, for the CP averages, negligible. In this regard, the largest pollution appears in the bins including the $\phi(1020)$ resonance and in the observable $P_4^\prime$. As we will see in Sec.~\ref{sec:sensitivity}, low-$q^2$ bins are also important for their sensitivity to the Wilson coefficients $C_7$ and $C_7^\prime$.

\subsubsection{CP asymmetries: $A_{\rm CP}$ and $P_3^{{\rm CP}}$}
\label{sec:CPasymmetries}

\begin{figure}
\begin{center}
\includegraphics[scale=0.6]{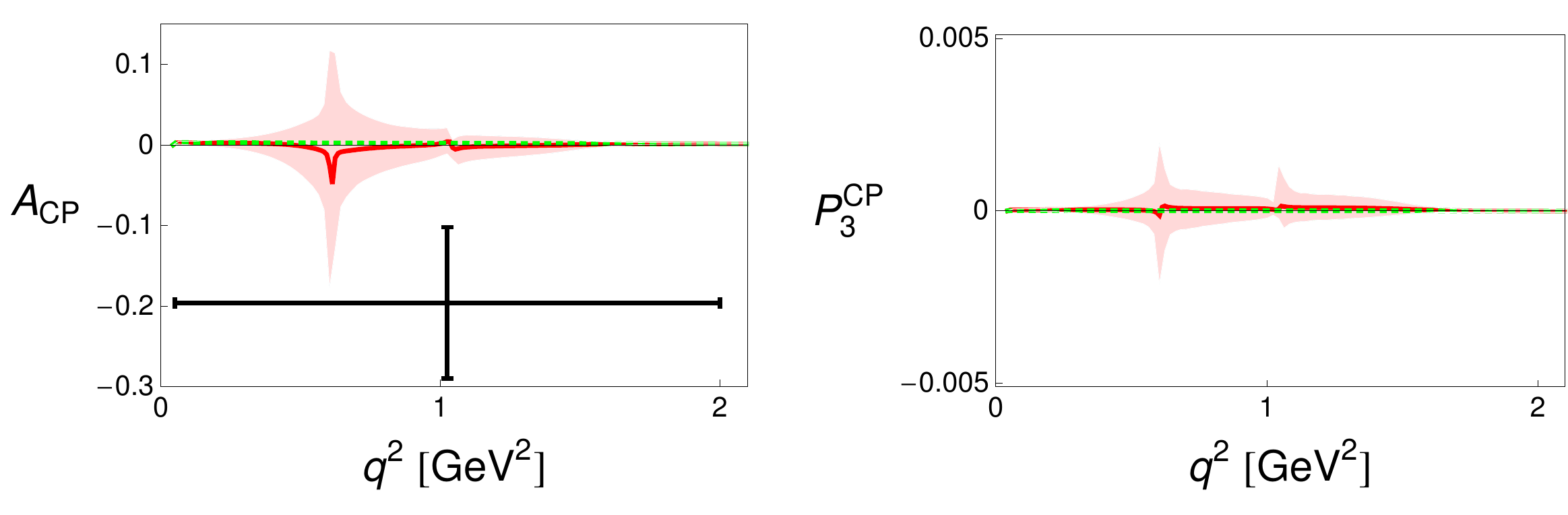}
\caption{Differential $A_{\rm CP}$ and $P_3^{{\rm CP}}$ at low $q^2$. For $A_{\rm CP}$ we show the experimental data point at the low-$q^2$ bin [0.05, 2] GeV$^2$. The charm-loop uncertainty and factorizable power corrections are negligible in this case.   \label{fig:PlotACP}}
\end{center}
\end{figure}

The CP asymmetries are observables suitable for searching for sources
of CP-violation beyond the CKM mechanism. In $\bar{B}\rightarrow
\bar{K}^*\ell^+\ell^-$, a weak phase arises from the interference
between the contributions weighed by $V_{tb}V_{ts}^*$ and
$V_{ub}V_{us}^*$, so all the CP-asymmetries for this decay in the SM
are suppressed by a factor
$\lambda_{ut}=V_{ub}V_{us}^*/V_{tb}V_{ts}^*$, which imaginary part is
of order
$\sim\bar{\eta}\lambda^2\sim10^{-2}$~\cite{Kruger:1999xa}. Another
important observation is that the sensitivity to a CP-violating phase
is modulated by a factor $\sin\delta_s\sin\delta_w$ or
$\cos\delta_s\sin\delta_w$ depending on whether the asymmetry is
T-even or T-odd (i.e\ odd under the transformation $\phi\rightarrow-\phi$). In this sense, QCD factorization predicts the strong phases, $\delta_s$, to be very small~\cite{Beneke:2001at,Beneke:2004dp} and, consequently, the latter asymmetries have been specifically singled out as interesting observables for the detection of new CP-violating phases beyond the SM~\cite{Bobeth:2008ij}.

Nonetheless, a hadronic, rather than a partonic treatment, is suited
for the description of the contribution of the light-quarks at
low-$q^2$ (see Sec.~\ref{sect:hadronic}). In the context of VMD
that is developed in this work, we have concluded that the overall
contribution to the CP-averaged observables is negligible and
consistent with the intrinsic suppression induced by small Wilson
coefficients or by the CKM factor $\lambda_{ut}$. For the
CP-asymmetries, though, large differences between the two approaches
can be expected since the leading contribution is given by the latter
pieces. To be more precise, large strong phases can be obtained in the
hadronic picture as they naturally arise from rescattering and
``dressing'' of the resonance poles. Therefore, at low $q^2$ a
suppression $\propto\sin\delta_s\approx0$ of the T-even asymmetries
cannot be expected and sensitivity to the weak phase(s) in these
observables can be generated by long-distance effects. Conversely, a
factor $\cos\delta_s$ close to 1 is not guaranteed on general grounds
at low $q^2$, which can hinder the efficient use of the T-odd observables for probing CP-violation in this regime.\footnote{In any case, we want to stress that a full treatment of the hadronic uncertainties on the CP-asymmetries is beyond the scope of the present work. For instance, neglecting the effect of higher-mass $\rho$, $\omega$ and $\phi$ resonances present across all the lowest-$q^2$ region might not be as safe approximation for the CP-asymmetries as it is for the averages.}

As an illustration, we show in the left-hand side of Fig.~\ref{fig:PlotACP} the CP-asymmetry
\begin{equation}
A_{\rm CP}=\frac{\Gamma-\bar{\Gamma}}{\Gamma+\bar{\Gamma}}=\frac{3\Delta_{1c}-\Delta_{2c}+2\left(3\Delta_{1s}-\Delta_{2s}\right)}{4\Gamma^\prime}, 
\end{equation}
which is T-even and, on the right hand-side, the T-odd asymmetry $P_3^{\rm CP}$~\cite{MatiasVirtoCP},
\begin{equation}
P_3^{\rm CP}=-\frac{\Delta_9}{4\Sigma_{2s}},
\end{equation}
for the decays of the neutral mesons in the muonic mode. The latter
observable is a redefinition of
$A_9$~\cite{Kruger:1999xa,Altmannshofer:2008dz} following the same
``cleanness'' principle applied to set the $P$-basis for the
CP-averaged observables~\cite{MatiasVirtoCP}. $A_{\rm CP}$ has been
measured at the $B$-factories~\cite{Wei:2009zv,:2012vwa} and,
recently, binned results and the most precise measurement to date of
the total have been provided by the LHCb
collaboration~\cite{LHCb:2012kz}. An interesting outcome of the latter
measurement is that the total (integrated) $A_{\rm CP}$ is negative and of order $\sim5\%$, with a sizable and negative contribution stemming from the lowest-$q^2$ bin. This seems to be in contradiction, at 1$\sigma$ level, with the SM prediction as it is obtained in QCD factorization. On the other hand, this type of substantial CP-asymmetries could be the smoking gun for some NPs scenarios (see e.g.~\cite{Alok:2011gv}). However, and as shown in the left-hand side panel of Fig.~\ref{fig:PlotACP} long-distance contributions of the light quarks can have an important effect in $A_{\rm CP}$. In fact, we obtain that the uncertainty produced by these can be rather large, cf. $A_{\rm CP}=(0^{+3}_{-4})\%$ in the [0.05, 2] GeV$^2$ bin, with an error band one order of magnitude larger than in QCD factorization.

As for $P_3^{{\rm CP}}$ we observe that it is practically zero in the
SM. This observable (in the form of $A_9$) has been tagged as a
benchmark for the detection of a non-standard weak phase that might
surface through the Wilson coefficient
$C_7^\prime$~\cite{Bobeth:2008ij,Altmannshofer:2008dz}, and our
findings show that it remains a clean null test, safe from
long-distance charm contributions and
contamination by light resonances as a consequence of being 
$\propto H_{V,A}^+$.
We investigate the sensitivity of this
observable, together with other combinations of the angular
coefficients $I_3$ and $\bar{I}_3$, to BSM values of $C_7^{\prime}$,
in Sec.~\ref{sec:sensitivity}.
      
\subsubsection{Non-vanishing lepton masses: The observables $M_1$ and $M_2$}
\label{sec:M1andM2}

Two more independent ``clean'' observables can be defined for the muonic mode at low $q^2$~\cite{Matias:2012xw}, \footnote{For the sake of making our discussion in this section more transparent, we slightly redefine these observables as introduced in that reference, factorizing out a $4m_\mu^2/q^2$ piece.}
\begin{equation}
M_1^\prime=\frac{1}{1-\beta^2}\frac{\beta^2I_{1s}-(2+\beta^2)I_{2s}}{4 I_{2s}},
\hspace{1cm}
M_2^\prime=-\frac{1}{1-\beta^2}\frac{\beta^2I_{1c}+I_{2c}}{I_{2c}},\label{Eq:Pis}
\end{equation}
which involve the pieces proportional to $4m_\mu^2/q^2$ present in $I_{1s}$ and $I_{1c}$,
\begin{eqnarray}
M^\prime_1 &=& \frac{1}{2}\frac{\left(|H_V^+|^2+|H_V^-|^2-(V\rightarrow
    A)\right)}{\left(|H_V^+|^2+|H_V^-|^2+(V\rightarrow A)\right)}, \nonumber
\\[2mm]
M^\prime_2 &=& \frac{q^2/(2m_\mu^2)(|H_P|^2+\beta^2 |H_S|^2)+\left(|H_V^0|^2-|H_A^0|^2\right)}{\left(|H_V^0|^2+|H_A^0|^2\right)}.\label{Eq:mlpieces}
\end{eqnarray}
  
The two questions that naturally arise are, first, what can we learn from a measurement of these observables; second, how these pieces enter in the formulas of the differential decay rate expressed in terms of a given basis of observables, which are usually defined in the $m_\ell\rightarrow0$ limit. To start off, notice that the contribution of these pieces to the differential decay rate vanishes except in the vicinity of the low-$q^2$ end-point, e.g. $4m_\mu^2/(0.5 {\rm GeV^2})\sim 0.09$. In this region it then follows that
\begin{equation}
M_1^\prime=\frac{1}{2}+\mathcal{O}(q^2/m_B^2)^2,\hspace{1cm} M_2^\prime=1+\mathcal{O}(q^2/m_B^2).\label{Eq:M12apps}
\end{equation}
To obtain the first relation it is sufficient to note that the photon pole in  $H^{\pm}_V$ makes $H^{\pm}_A$ to be of $\mathcal{O}(q^2/m_B^2)$ in comparison. Thus, this relation is completely model-independent, in the sense that it is insensitive to the long-range hadronic uncertainties as well as to the particular short-range structure of the decay. 
In order to derive the second relation, we first omit the effect of (pseudo)scalar operators. Then, using the corresponding expressions for the amplitudes it is straightforward to obtain    
\begin{eqnarray}
|H_V^0|^2+q^2/(2m_\mu^2)|H_P|^2-|H_A^0|^2=|H_V^0|^2+|H_A^0|^2+{\rm Rem.},
\end{eqnarray}
where 
\begin{eqnarray}
{\rm Rem.}=\frac{N^2\lambda}{2m_B^2q^2}|C_{10}-C_{10}^\prime|^2(S(q^2)-V_0(q^2))\simeq\mathcal{O}(q^2/m_B^2) 
\end{eqnarray}
due to the exact form-factor relation~(\ref{eq:Sq20}). The relation for $M_2^\prime$ is not model-independent as it can be broken by contributions from BSM scalar and pseudoscalar operators, although the size of these effects is very constrained by recent measurements of the $B_s\rightarrow \mu^+\mu^-$ decay rate. Indeed, if one uses the experimental upper bound given by the LHCb collaboration at 1-$\sigma$ level~\cite{:2012ct} and the SM prediction in Ref.~\cite{DeBruyn:2012wk} to constrain the size of these pieces (assuming $C_{10}=C_{10}^{\rm SM}$), one obtains that the relation for $M_2^\prime$ is broken only at the $\sim1\%$ level. 

\begin{figure}
\begin{center}
\includegraphics[scale=0.47]{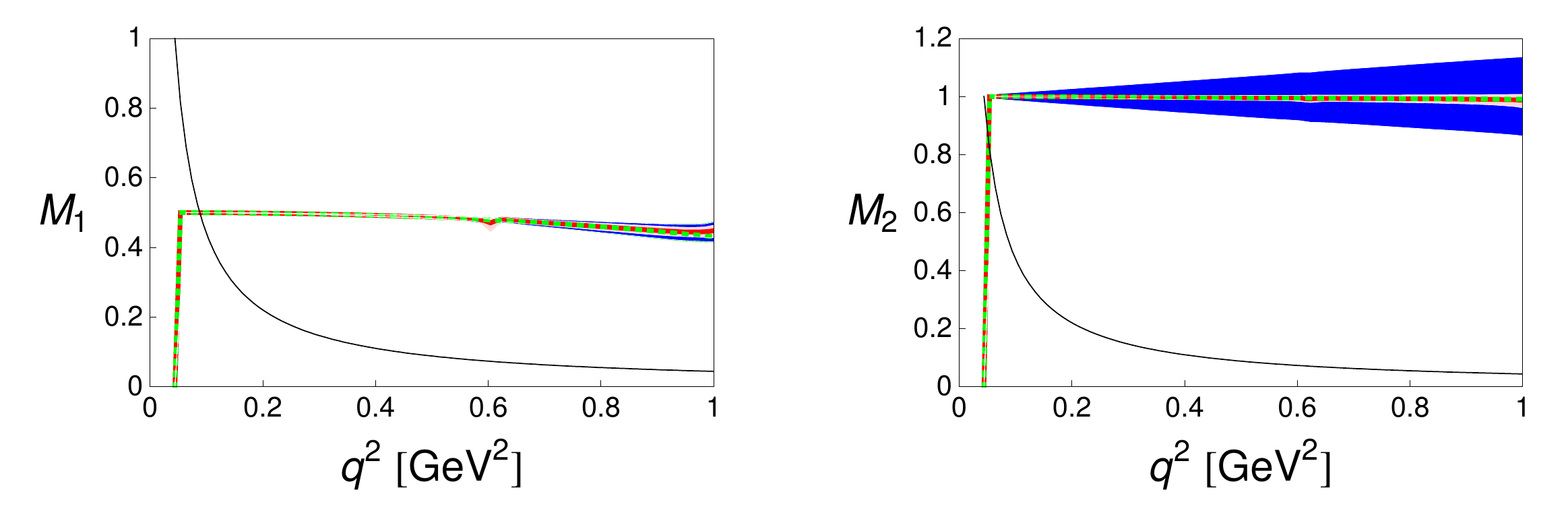}
\includegraphics[scale=0.47]{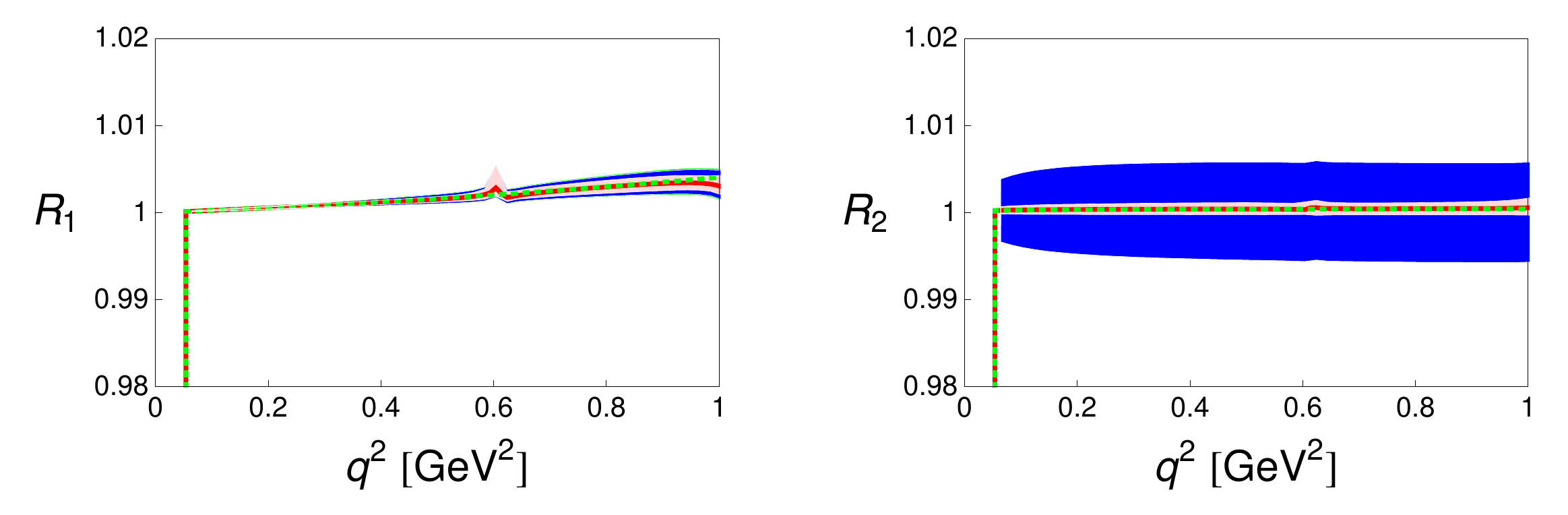}
\caption{Results for $M_{1,2}^{\prime}$ and $R_{1,2}$. The color code is as in Fig.~\ref{Fig:PlotIs}, whereas the black line is the suppression factor $4m_\mu^2/q^2$. \label{fig:PlotMs}}
\end{center}
\end{figure}

The Eqs.~(\ref{Eq:M12apps}) mean that, in practice, $M_{1,2}^\prime$ are not independent observables in the $q^2$ region in which the decay distribution has sensitivity to their effects. Taking this into account, one can generalize Eqs.~(\ref{Eq:RelsI121}) with expressions that are valid at finite $m_\ell$ and then, also, close to the low-$q^2$ end-point. Namely, we define, 
\begin{equation}
\tilde{I}_{1s}=\frac{1}{\beta^2}(4-\beta^2) I_{2s}, \hspace{1cm}\tilde{I}_{1c}=-\frac{1}{\beta^2}(2-\beta^2) I_{2c},\label{Eq:NewRels}
\end{equation}
that convert into the conventional formulas~(\ref{Eq:RelsI121}) in the massless limit. In the upper panel of Fig.~\ref{fig:PlotMs} we show the $q^2$ dependence of $M_1^\prime$ and $M_2^\prime$ at low $q^2$ using the same color code as in Figs.~\ref{Fig:PlotIs} and~\ref{fig:PlotsClean}. We superimpose in black the value of the factor $4m_\mu/q^2$ suppressing the contribution of these pieces to the decay rate. In the lower panel we explore the validity Eqs.~(\ref{Eq:NewRels}) by showing the ratios $R_{1,2}=\tilde{I}_{1s,2s}/I_{1s,2s}$. As one can check in the plots, these relations work at better than 1$\%$ of accuracy in the region of interest.

\subsection{The $\bar{B}^0\rightarrow \bar{K}^{*0}e^+e^-$ decay}

\begin{figure}
\begin{center}
\includegraphics[scale=0.6]{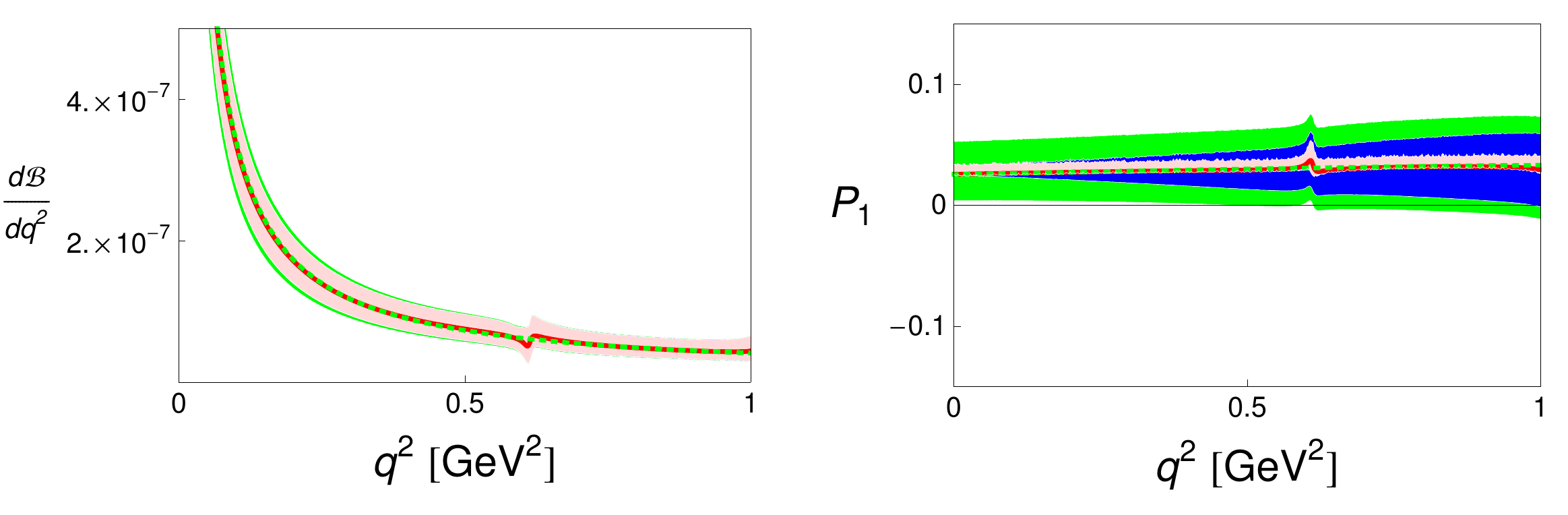}
\caption{Differential branching fraction and $P_1$ for the electronic mode at low $q^2$. The color code is as in Fig.~\ref{Fig:PlotIs}. \label{fig:PlotElectronic}}
\end{center}
\end{figure}

In comparison with the muonic mode, the electronic decay is
interesting because its low-$q^2$ end-point is at
$q^2=4m_e^2\simeq10^{-6}$ GeV$^2$. Hence, it presents an enhanced
sensitivity to the physics associated to the photon pole  becoming a
golden mode for probing possible BSM effects related to the magnetic
penguins operator $Q_{7}^{(\prime)}$~\cite{Grossman:2000rk}. On the
experimental side, $B-$factories have observed the electronic
mode~\cite{:2012vwa}, and prospects at LHCb are
discussed in~\cite{LHCbElectron}. This will also be an interesting
target for super flavour factories.
Our detailed error assessment of the
low-$q^2$ region in the muonic mode revealed that the low-$q^2$
end-point is
actually the least prone to form factors uncertainties due to the
constraints~(\ref{eq:Tplusq20}), at the same time as the hadronic
contributions, for CP-averaged observables, come out to be of
negligible size. Consequently,
the $\bar{B}^0\rightarrow \bar{K}^{*0}e^+e^-$ decay emerges as
highly relevant.

\begin{table}
\centering
\caption{Results and error budget of the integrated branching fraction and $P_1$ observable in the [0.0009, 1] GeV$^2$ bin for the electronic mode. \label{Table:binnedElectr}}
\begin{tabular}{|c|c|cccc|}
\hline
Obs.&Result&Hadronic& Fact.& $c$-quark&Light-quark\\
\hline
\raisebox{0.4ex}[15.pt]{$10^7\times\langle \frac{d\mathcal{B}}{dq^2}\rangle$}&$2.43^{+0.66}_{-0.47}$&$^{+0.50}_{-0.39}$&$^{+0.10}_{-0.05}$&$^{+0.42}_{-0.25}$&$\pm0.03$\\
\hline
\raisebox{0.4ex}[15.pt]{$10^2\times\langle P_1\rangle$}&$2.7^{+3.0}_{-2.7}$&$^{+0.8}_{-0.1}$&$^{+1.0}_{-1.2}$&$^{+2.7}_{-2.3}$&$\pm0.0$\\
\hline
\end{tabular}
\end{table}

In Fig.~\ref{fig:PlotElectronic} we show the SM predictions for the
differential branching fraction and the $P_1$ observable, which is
especially sensitive to NPs effects in the magnetic penguin
operators~\cite{Kruger:2005ep,Matias:2012xw}. In
Table~\ref{Table:binnedElectr} we show the integrated results for the
bin [0.0009, 1] GeV$^2$,  that is the one proposed
in~\cite{LHCbElectron}. As one can see, the integrated branching
fraction is larger than the one for the muonic case in its lowest
bin. Interestingly enough, our result agrees with the estimate
$\mathcal{B}\sim2.2\times 10^{-7}$ that is obtained assuming that the decay is entirely driven by the photon pole~\cite{Grossman:2000rk,LHCbElectron}. This dominance of the transverse amplitudes in this integrated decay rate is also reflected by a very small integrated longitudinal polarization in the same bin, namely $\langle F_L\rangle=0.077^{+0.047}_{-0.040}$.  
            
\subsection{Sensitivity to $C_7'$}
\label{sec:sensitivity}

The analysis of the low $q^2$ region of the $\bar{B}\rightarrow \bar{K}^*\ell^+\ell^-$ can provide tight constraints on NPs scenarios with right-handed flavour-changing neutral currents, specially those giving contributions to the chirally-flipped magnetic penguin operator $\mathcal{O}_7^\prime$. This is due to the fact that the angular coefficients $I_3$ and $I_9$, at low-$q^2$, are
\begin{equation}
I_3 \propto {\rm Re} \left(H_V^- H_V^{+*} \right),
\qquad
I_9 \propto {\rm Im} \left(H_V^-H_V^{+*} \right),
\end{equation}
where $H_V^+ \propto C_7'/q^2$,
so, approximately, they vanish unless $C_7^{\prime}\neq0$.
(Corrections involving $H_A^+$ are also suppressed by the smallness
of $C_9'$ and $C_{10}'$ in the Standard Model, but any BSM effects
generating $H_A^+$ are suppressed at $q^2 \approx 0$ due to the
absence of a photon pole.)
In the SM, small contributions to these observables are generated by the strange-quark mass and other effects quantified in this work as contributions to the $H_{V}^+$ helicity amplitude. Other decays and observables provide valuable and independent constraints on the $C_7$ and the $C_7^\prime$ planes, in particular the inclusive $B\rightarrow X_s\gamma$ decay and the isospin and the time-dependent CP-asymmetries in the exclusive $B\rightarrow K^*\gamma$ decay (see e.g.~\cite{DescotesGenon:2011yn}). The interest of the radiative decays onto higher-mass $K^*$ resonances has been also recently pointed out~\cite{Kou:2010kn}. In this work we focus on studying the sensitivity of the vicinity of low-$q^2$ end-point to the chirally-flipped Wilson coefficient $C_7^\prime$. A more comprehensive analysis should also consider studying new-physics effects in $C_7$~\cite{DescotesGenon:2011yn,Beaujean:2012uj,Altmannshofer:2011gn}, although $I_3$ and $I_9$ can be used to efficiently constrain also this Wilson coefficient only if $C_7^\prime$ is far from zero.

\begin{figure}
\begin{center}
\includegraphics[scale=0.6]{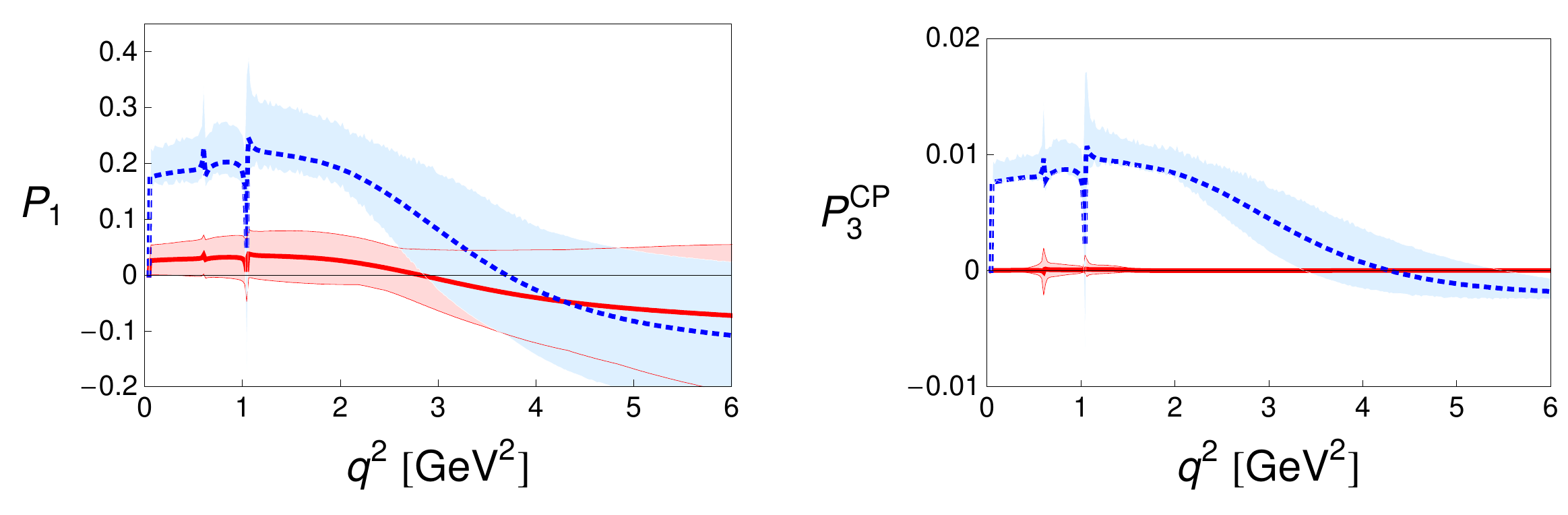}
\caption{Study of the sensitivity of the observables $P_1$ and $P_3^{\rm CP}$ to the a purely real or purely imaginary NPs contribution through $C_7^{\prime}$ (dashed line, blue bands). These are confronted with the SM expectation (solid line, red band). \label{fig:PlotSens}}
\end{center}
\end{figure}

In the context of CP-combinations, one can construct 4 independent
observables with these angular coefficients and their
CP-conjugates. However, $I_3$ and $I_9$ are a CP-odd and a CP-even
observable, respectively, and the combinations $\Delta_3$ and
$\Sigma_9$ become not very sensitive to the chirality of dilepton
pair. Therefore, only 1 CP-average and 1 CP-asymmetry, constructed
from $\Sigma_3$ and $\Delta_9$, in order, are sensitive to either the
real or imaginary parts of $C_7^{(\prime)}$. The corresponding
observables in the $P$-basis, $P_1$ and $P_3^{\rm CP}$, are a
convenient choice for their intrinsically reduced theoretical
uncertainty. In order to investigate the sensitivity of these
observables to $C_7^{\prime}$, we investigate two a priori fictitious scenarios: the first one in which $C_7^\prime$ is real and with a magnitude of 10$\%$ the value of $C_7(m_b)$ in the SM, and second where $C_7^\prime$ is a pure imaginary number with and absolute value of 1$\%$ the value of $C_7(m_b)$ in the SM. The result for $P_1$ in the first scenario is plotted as a dashed line confronted against the SM prediction represented as a solid line on the left panel of~\ref{fig:PlotSens}. The corresponding bands are the overall uncertainties in these scenarios calculated as those presented for the SM in the previous sections. On the right-hand side, we plot $P_3^{\rm CP}$ for the second scenario compared, again, with the SM prediction and following the same line and color code as on the left panel. 

The first remarkable outcome of this analysis is that, with the current theoretical uncertainties, $P_1$ is sensitive to a $C_7^\prime$ with a real part of about 10$\%\times C_7^{\rm SM}(m_b)$ only in the region between $q^2=4m_\ell^2$ and $q^2\simeq3$ GeV$^2$. The same conclusions apply to $P_3^{\rm CP}$ although, in this case, the attainable accuracy is much higher as it can be sensitive to complex phases with an absolute value below 1$\%\times C_7^{\rm SM}(m_b)$. This, in principle, is an important result as it singles out $P_3^{\rm CP}$ as a theoretically extremely clean observable to constrain (or favor) BSMs generating new CP-violating phases through $C_7^{\prime}$. This occurs despite the cautionary remarks made in Sec.~\ref{sec:CPasymmetries} about the CP-asymmetries. In fact important interference with strong phases, with a drastic reduction of the sensitivity, seems to appear only in the immediate surroundings of the resonance poles. Notice, however, that $P_3^{\rm CP}$ is an special observable as it vanishes in the SM so the cleanness of other CP-observables can not be concluded without a devoted study on their corresponding hadronic uncertainties.

\section{Conclusions and outlook}
\label{sect:conclusions}
We have performed a comprehensive analysis of angular observables
in the decays $\bar B \to \bar K^* \ell^+ \ell^-$, $\ell = \mu, e$,
paying particular attention to non-factorizable hadronic
uncertainties.
We exploit a suppression of the positive-helicity amplitudes in the Standard
Model that, as we have shown, still holds when taking into account
long-distance charm and light-quark non-factorizable effects in a
conservative way. As a result, we established that of those observables
that are ``clean'' in the factorizable approximation in the
large-energy/heavy-quark limit, the two observables $P_1 = A_T^{(2)}$
and $P_3^{\rm CP}$ are found to nearly vanish in the SM with
very small errors in the presence of the long-distance
non-factorizable corrections. At the same time they are highly sensitive to
right-handed currents. Importantly, we find that the lower end of
the low-$q^2$ range is not only theoretically clean, but provides the
best sensitivity to the Wilson coefficient $C_7'$, down to a
(theoretical) limit of below 10\% for the real part, and
as low as 1\% for the imaginary part.
This also raises the profile of the decay into electrons, $\ell = e$, which
is dominated by the photon pole, due to the small electron mass. We
feel that this process deserves further attention.
For other observables, which do not vanish for $H_V^+ =0$, the theoretical
control is more compromised by long-distance effects. Improving it
would seem to require further theoretical advances.

As a further result, we have found a way to generalise the
two well-known relations $3 I_{1s} = I_{2s}$ and $I_{1c} = - I_{2c}$,
which hold for vanishing lepton mass, to the massive case. This
may be of use in the experimental determination of the angular
distribution.

The formulation in terms of helicity amplitudes allowed us to show,
by adapting the LCSR formalism of \cite{Khodjamirian:2010vf},
a strong $\Lambda^3/(4 m_c^2 m_b)$ power suppression of
the long-distance contribution $h_+|_{c \bar c, \rm LD}$
from the charm part of the hadronic
weak Hamiltonian entering in the positive-helicity
amplitude $H_V^+$, which vanishes at leading power in QCDF.
Analogous contributions from the chromomagnetic operator are
suppressed by $\Lambda^2/m_b^2$.
Likewise, we exclude the possibility of non-negligible, light-resonance
long-distance contributions to $H_V^+$ once the $(\Lambda/m_b)^2$
suppression of the corresponding $\bar{B}\to \bar K^* V $ helicity
amplitude is taken into
account. More generally, we observe that, once data are binned, these
light-resonance contributions are small in all observables, even
though this is not true of the unbinned quantities.
On the other hand, the amplitudes $H_V^-$ and $H_V^0$ can
receive important long-distance charm-loop contributions,
at a level of $\mathcal O(10\%)$ of the leading-power QCDF result,
as is clear from the numerical values in
\cite{Khodjamirian:2010vf}. We have also argued that the much
smaller estimate of these contributions of \cite{Ball:2006cva}
is likely to suffer very large, incalculable corrections.
This limits the theoretical control over
the other ``clean'' observables, beyond $P_1$, $P_3$, and their CP
asymmetry counterparts, that, although independent of $H_V^0$,
involve $H_V^-$.

Our phenomenological analysis makes certain assumptions about
the values of helicity form factors and nonfactorizable corrections
to the helicity amplitudes. To get a better quantitative
understanding of the theoretical errors, it would be desirable
to have direct calculations of form factors in the helicity basis.
This should be straightforward within LCSR, or other methods.
It is also important to stress that  we have not derived a sum rule
for  $h_+|_{c \bar c, \rm LD}$, but only shown a parametric suppression
of it (and of the $Q_{8g}$ long-distance contributions.)
An attractive aspect of the LCSR method is that it avoids endpoint
divergences altogether, and a sum rule allowing for a quantitative
estimate could in principle be obtained. We feel that that is a path
worthwile pursuing.
On the other hand, for the light-quark non-factorizable effects
it would be interesting to go beyond the resonance model in
assessing the size (and helicity structure) of possible corrections to 
the QCDF results, perhaps by combining QCDF, LCSR, and
dispersion relations along the lines of \cite{Khodjamirian:2012rm}.
While this is unlikely to have a strong impact on $P_1$ or $P_3^{\rm CP}$,
it may be more important in other observables, and in particular for the
direct CP asymmetry $A_{\rm CP}$, which, as our resonance model
suggests, could be relatively sensitive to any large, ``soft'' strong
phases associated with light-hadron intermediate states.

\paragraph{Acknowledgment.} We would like to thank M.~Beneke,
C.~Bobeth, U.~Egede, G.~Hiller, J.~Matias, M.~Patel, M.-H.~Schune,
N.~Serra, R.~Zwicky,
and the participants of the \textit{Workshop on the physics reach of
rare exclusive $B$ decays} held at Sussex in September 2012, for useful
discussions. This work was funded by the UK Science and Technology
Facilities Council under grant numbers
ST/H004661/1 and  ST/J000477/1.
Support from SEPnet and NExT is acknowledged.
JMC acknowledges partial support from the
Spanish Ministerio de Econom\'ia y Competitividad and
European FEDER funds under the contract
FIS2011-28853-C02-01 and the Fundaci\'on Seneca project 11871/PI/09.

\paragraph{Note added.} After submission of this work to the arXiv,
a paper including a new LCSR approach to the matrix elements
of $Q_{8g}$ appeared \cite{Dimou:2012un}. It would be interesting to
adapt this method to the approach of the present paper.
We also note work on the effect of tensor operators in $\bar B \to K^*
\ell^+ \ell^-$ in the high-$q^2$ region \cite{Bobeth:2012vn}.

\appendix

\section{Notations and conventions}
\label{app:conventions}
\subsection{Polarisation vectors}
\label{app:polavectors}
For a particle moving in $+z$ direction with four-momentum
$k^\mu = (E, 0, 0, |{\bf k}|)$  we use
\begin{eqnarray}
 \epsilon_t^\mu &=& \frac{k^\mu}{\sqrt{k^2}} , \\
 \epsilon^\mu(\pm) &=& \frac{1}{\sqrt{2}} (0, \mp 1, - i, 0) , \\
 \epsilon^\mu(0) &=& \frac{1}{\sqrt{k^2}} (|{\bf k}|, 0, 0, E) .
\end{eqnarray}
In parentheses is the helicity, we sometimes alternatively
put this in a subscript. The subscript $t$ denotes ``timelike''
polarisation, which is a spin singlet.
The sign conventions ensure that the
helicities $\pm, 0$ form a triplet obeying conventional relative signs
and normalisations,
e.g. $[J^+]^{\mu}_{\;\;\nu} \epsilon^\nu(0) = \sqrt{2} \epsilon^\mu(1)$ .

The polarisation vectors satisfy the normalisation and completeness
relations
\begin{equation}
  \epsilon^\mu_a \epsilon^*_{b,\mu} = g_{ab},
    \qquad g_{ab} = \left\{ \begin{array}{ll}
                              -1, & a=b=\pm,0, \\
                              1, & a=b=t , \\
                              0, & \mbox{otherwise}
                            \end{array}\right. 
\end{equation}
\begin{equation}
  \eta^{\mu\nu} = \epsilon_t^\mu \epsilon_t^{\nu\, *}
     - \sum_{\lambda =\pm,0} \epsilon^\mu(\lambda) \epsilon^{\nu\,
       *}(\lambda)
 = \sum_{a,b=t,\pm,0} g_{ab} \epsilon_a^{\mu} \epsilon_b^{\nu \*} .
\end{equation}
For any other direction $\hat n$, we define the polarization vectors by
applying a standard rotation after boosting to the ``rest frame''
where $k^\mu = (\sqrt{k^2},0,0,0)$. In that frame,
\begin{equation}
  \epsilon^\mu(\hat n; \lambda) = \Big[ e^{-i \phi J_z} e^{-i \theta
    J_y} e^{+i \phi J_z} \Big]^{\mu}_{\;\;\nu}\, \epsilon^\nu(\hat z;
  \lambda) = \sum_{\lambda'} d^1_{\lambda' \lambda}(\theta)
    e^{i (\lambda-\lambda') \phi} \epsilon^\mu(\hat z; \lambda') ,
\end{equation}
where $\theta, \phi$ are the spherical coordinates of $\hat n$ and
$d^j_{m'm}(\theta)$ is the matrix element of the rotation operator
$e^{-i \theta J_y}$ (conventions as in \cite{Jacob:1959at}).
For arbitrary direction, the final form is usually more convenient.

For $\hat n = - \hat z$ (ie $\theta=\pi$) we rotate at $\phi = 0$. This means
$$
\epsilon(-\hat z; \pm) = \epsilon(\hat z; \mp) , \qquad
\epsilon(-\hat z; 0) = - \epsilon(\hat z; 0).
$$
Boosting back to the laboratory frame, we have
\begin{eqnarray}
 \epsilon^\mu_t(-\hat z) &=& \frac{1}{\sqrt{k^2}} (E, 0, 0, -|{\bf k}|), \\
 \epsilon^\mu(-\hat z; \pm) &=& \frac{1}{\sqrt{2}} (0, \pm1, - i, 0) 
                                = \epsilon^\mu(\hat z; \mp), \\
 \epsilon^\mu(-\hat z; 0) &=& \frac{1}{\sqrt{k^2}} (|{\bf k}|, 0, 0, - E) . 
\end{eqnarray}

\subsection{Standard Model parameters}
\label{app:SMparam}
\begin{table}[h]
\centering
\caption{Some of the SM parameters used in this work (in GeV).    
\label{Table:SM}}
\begin{tabular}{|cccccc|}
\hline
$M_W$&$\hat{m}_t(\hat{m}_t)$&$m_{b,{\rm PS}}$&$m_c$&$\hat{m}_s(2$ GeV$)$&$\Lambda_{\rm QCD}^{(5)}$\\
\hline
80.4&172(2)& 4.8(2)&1.3(1) &0.094(3) &0.214(8) \\
\hline
\end{tabular}
\end{table}

\begin{table}[h]
\centering
\caption{CKM parameters in the Wolfenstein parameterization used in this work~\cite{Charles:2004jd}.    
\label{Table:CKM}}
\begin{tabular}{|cccc|}
\hline
$\lambda$&$A$&$\bar{\rho}$&$\bar{\eta}$\\
\hline
0.22543(8)&0.805(20)&0.144(25)&0.342(16)\\
\hline
\end{tabular}
\end{table}   

\begin{table}[h]
\centering
\caption{Wilson coefficients of the SM at $\mu=4.8$ GeV, in the basis of~\cite{Chetyrkin:1996vx} and to NNLL accuracy.    
\label{tab:WilsonNum}}
\begin{tabular}{|cccccccccc|}
\hline
$C_1$ &$C_2$ &$C_3$ &$C_4$&$C_5$&$C_6$ &$C^{\rm eff}_7$ &$C^{\rm eff}_8$ &$C_9$&$C_{10}$\\
\hline
-0.144&1.060&0.011&-0.034&0.010&-0.040&-0.305&-0.168&4.24&-4.312\\
\hline
\end{tabular}
\end{table}

We use the PS-subtracted definition for the $b$-quark
mass~\cite{Beneke:2001at}. This is obtained from the $\overline{\rm MS}$-definition ($\hat{m}_b$) through the pole mass definition ($m_b$) using the subsequent identities
\begin{eqnarray} 
&\hat{m}_b(\mu)=m_b \left(1+\frac{\alpha_s(\mu) C_F}{4\pi}\left[3\log\frac{m_b^2}{\mu^2}-4\right]+\mathcal{O}(\alpha_s^2)\right),\label{Eq:mbMSbar}\\
&m_b=m_{b,{\rm PS}}(\mu_f)+\frac{4\alpha_s(\mu)}{3\pi}\mu_f\label{Eq:mbPS}.
\end{eqnarray}
The PS-mass at $\mu_f=2$ GeV is then identified with the pole mass,
$m_b\approx m_{b,{\rm PS}}(2$ GeV$)=4.78^{+0.20}_{-0.07} \simeq 4.8(2)$ GeV~\cite{Nakamura:2010zzi}. We use $m_b$ as the potential-subtracted definition unless it is stated otherwise. For the strange quark mass appearing at tree level we use the result obtained from LQCD calculations, $\hat{m}_s(2$ GeV$)=94(3)$ MeV~\cite{Colangelo:2010et}. Other SM parameters used in the calculation are listed in Tables~\ref{Table:SM} and~\ref{Table:CKM}. For $\alpha_s(\mu$) we use the three-loop renormalization-scale evolution~\cite{Chetyrkin:2000yt}, for $\alpha_{\rm em}(Q^2=0)\simeq1/137$ and for $G_F=1.16637(1)\times10^{-5}$ GeV$^{-2}$. Finally, we show in Table~\ref{tab:WilsonNum} the values of the Wilson coefficients of the SM at $\mu=m_b$ GeV, in the basis of~\cite{Chetyrkin:1996vx} and to NNLL accuracy.     

\subsection{Leading-twist LCDAs} 

We use the parameterization of the light vector-meson LCDAs of Ref.~\cite{Ball:2004rg} in terms of the wave functions $\Phi_{K^*,a}(u)$ and the semileptonic decay constants. However, we use the conventions and values of Refs.~\cite{Beneke:2004dp}, which have a sign difference in the odd-coefficients as compared with those of the LCSR calculation. Namely, 
\begin{equation}
\Phi_{K^*,a}=6u\bar{u}\left(1+3 a_{1,a}\xi+a_{2,a}\frac{3}{2}(5\xi^2-1)\right), \label{EQ:LCDAs}
\end{equation}
with $\bar{u}=(1-u)$, $\xi=2u-1$ and $a_{i,a}$ Gegenbauer
coefficients. There are two decay constants for a light vector
meson. The longitudinal constant can be obtained from experiment using
the branching fractions for the electromagnetic decays of the neutral
vector resonances and applying $SU(3)_F$ symmetry. We obtain
$f_{K^*}=220(5)$ MeV using the $\rho$ decay and so this value is
consistent with the one obtained from sum-rule
calculations~\cite{Ball:2004rg}. Sum rules also predict the tensor (transversal) decay constant, $f_{K^*,\perp}(1$~GeV$)=170(20)$ MeV. The latter is consistent with the LCSR determination $f_{K^*,\perp}(2$~GeV$)\approx 160$ MeV~\cite{Altmannshofer:2008dz}, although we have considerably increased the uncertainty. We evolve the transversal constant in the renormalization scale as
\begin{equation}
f_{K^*,\perp}(\mu)=f_{K^*,\perp}(\mu_0)\left(\frac{\alpha_s(\mu)}{\alpha_s(\mu_0)}\right)^{4/23}. \label{Eq:DCperpEv}
\end{equation}
For the $B$-meson, only two ``moments'' are needed,
\begin{eqnarray}
&& \lambda^{-1}_{B,+}=\int^\infty_0 d\omega\frac{\Phi_{B,+}(\omega)}{\omega}, \label{Eq:BMomp}\\
&&\lambda^{-1}_{B,-}(q^2)=\int^\infty_0 d\omega\frac{m_B\,\Phi_{B,-}(\omega)}{m_B\,\omega-q^2-i\epsilon}. \label{Eq:BMomm}
\end{eqnarray}
The wave functions $\Phi_{B,\pm}(\omega)$ are modelled such that they
fulfill certain relations derived from the equations of
motion~\cite{Beneke:2000wa,Beneke:2001at}. An important property is
that $\lambda^{-1}_{B,-}(q^2)$ diverges logarithmically as $q^2 \to 0$,
\begin{equation}
\stackrel{\hspace{1cm}q^2\rightarrow0}{\lambda^{-1}_{B,-}(q^2)\longrightarrow\infty},
\end{equation}
However, this IR divergence only afflicts the longitudinal (helicity-zero)
part of the decay amplitude, for which it indicates an enhanced long-distance
sensitivity at low $q^2$.
The model employed in Ref.~\cite{Beneke:2000wa,Beneke:2001at} leads to the following expressions for the $B$-meson moments:
\begin{eqnarray}
&&\lambda^{-1}_{B,+}=1/\omega_0, \label{Eq:BMomp2}\\
&&\lambda^{-1}_{B,-}(q^2)=\frac{\exp{\left(-q^2/(m_B\omega_0)\right)}}{\omega_0}\left(-{\rm Ei}(q^2/(m_B\omega_0))+i\pi\right),\label{Eq:BMomm2}
\end{eqnarray}
where $\omega_0=0.5(1)$ GeV. For $f_B$, we use an average of the current LQCD values with a generous error bar covering all of them $f_B=190(20)$ MeV~\cite{Na:2012kp}. In Table~\ref{Table:HP} we collect the numerical values for the hadronic parameters used in this work. 

\begin{table}[h]
\centering
\caption{Summary of the $K^*$ and $B$ parameters used in this work.    
\label{Table:HP}}
\begin{tabular}{|cccccc|}
\hline
$f_{K^*}$&$f_{K^*,\perp}$ (1 GeV)&$a_{1,a}$&$a_{2,a}$&$f_B$&$\lambda^{-1}_{B,+}$\\
\hline
220(5) MeV&170(20) MeV&0.2(2)&0.1(3)&190(20) MeV&2.0(5) GeV$^{-1}$\\
\hline
\end{tabular}
\end{table}

\subsection{Form factors} \label{app:formfactors}

$V$ denotes a vector meson, $\lambda$ its helicity,
$\epsilon_\lambda$ the corresponding
polarization vector, etc. $B$ is one of $B^+, B^0, B_s$. $q = p-k$.
\begin{eqnarray}
\langle V(k, \lambda) | \bar q \gamma_\mu b | B(p) \rangle
&=& \epsilon_{\mu\nu\rho\sigma} \epsilon^{*\nu}_\lambda p^\rho k^\sigma
\frac{2}{m_B+m_V} V(q^2) , \\
\langle V(k, \lambda) | \bar q \gamma_\mu \gamma_5 b | B(p) \rangle
&=& i (\epsilon_\lambda^* \cdot q) \frac{q_\mu}{q^2} 2\, m_V A_0(q^2)
\nonumber \\
&& + i (m_B+m_V) \left(\epsilon^{*\mu}_\lambda - \frac{(\epsilon^*_\lambda
  \cdot q) q^\mu}{q^2} \right) A_1(q^2)
\nonumber \\
&& - i  (\epsilon^*_\lambda \cdot q)\! \left( \frac{(2p-q)^\mu}{m_B\! +\! m_V}
             - (m_B\!-\!m_V) \frac{q^\mu}{q^2} \right) A_2(q^2) ,\quad \\
  \label{eq:t1def}
q^\nu \langle  V(k, \lambda) | \bar q \sigma_{\mu\nu} b | B(p) \rangle
&=& 2 \,i \, \epsilon_{\mu\nu\rho\sigma} \epsilon^{*\nu}_{\lambda} p^\rho
k^\sigma T_1(q^2) , \\
q^\nu \langle  V(k, \lambda) | \bar q \sigma_{\mu\nu} \gamma_5 b | B(p) \rangle
&=& \left( \epsilon^*_{\lambda;\mu} (m_B^2-m_V^2) - (\epsilon^* \cdot
  q) (2p-q)_\mu \right) T_2(q^2)
\nonumber \\
&& + (\epsilon^* \cdot q) \left( q_\mu - \frac{q^2}{m_B^2-m_V^2}
(2p-q)_\mu \right) T_3(q^2) .
\end{eqnarray}
The matrix element of the scalar density $\bar q b$ vanishes by
parity.
Furthermore, by PCAC,
\begin{eqnarray}
  i \langle V(k, \lambda) | \bar q \gamma_5 b | B(p) \rangle
&=& \frac{2\,m_V}{m_b + m_q} (\epsilon^* \cdot q) A_0(q^2) ,
\end{eqnarray}
\begin{eqnarray}
  \langle V(k, \lambda) | \bar q P_{L,R} b | B(p) \rangle
&=& \mp i \frac{m_V}{m_b + m_q} (\epsilon^* \cdot q) A_0(q^2) .
\end{eqnarray}

Our sign convention for the Levi-Civita symbol is that of
Bjorken and Drell, $\epsilon_{0123}=+1$.
Our conventions agree with those of \cite{Altmannshofer:2008dz}.
Some authors, in particular \cite{Beneke:2001at}, agree on the
physical meaning of the form factor symbols (in particular their
signs), but differ in phase conventions for the hadron states, such
that the form factor decompositions look different.

\section{Effective Lagrangian with vector resonances}
\label{app:resonancesdetails}

We use the anti-symmetric representation of the vector-meson fields. We explain here the particular conventions used in the model and we show some of the results for its contribution to the amplitude. Further details on the anti-symmetric description of the spin-1 fields can be found in Ref.~\cite{Ecker:1988te} whereas a discussion on the equivalence of this representation with any other one consistent with the symmetries and asymptotic behavior of QCD, can be found in Ref.~\cite{Ecker:1989yg}. 
We describe the $SU(3)$-multiplets of resonances using flavor-matrices of anti-symmetric tensor fields (e.g. $V_{\mu\nu}$ for the nonet of vectors). The interaction Lagrangian for resonances~\cite{Ecker:1988te,Ecker:1989yg}, at leading order of the chiral expansion in the light sector, is
\begin{equation}
\mathcal{L}^{(2)}_V=\frac{f_V}{4}\langle V^{\mu\nu}F^+_{\mu\nu}\rangle+i\frac{g_V}{2}\langle V^{\mu\nu}u_\mu u_\nu\rangle. \label{Eq:LagEffReson}
\end{equation}
In this Lagrangian, $u_\nu$ is the $vielbein$ introducing the pseudoscalar mesons ($\pi$, $K$, $\eta_8$) chirally coupled to the resonances fields and $F^+_{\mu\nu}$ is a generalized external-vector (electromagnetic) field strength, that in our case is $F^+_{\mu\nu}\simeq 2eQF_{\mu\nu}$, with $Q$ the charge operator in $SU(3)_F$-space. The respective pieces come accompanied by the coupling constants $f_V$ and $g_V$, which are defined in the chiral limit (the same as the mass term in the kinetic Lagrangian). The fields are normalized such that  
\begin{equation}
\langle 0|V^{\mu\nu}|V(p)\rangle=i\,m_V^{-1}\left(p^\mu\epsilon^\nu(p)-p^\nu\epsilon^\mu(p)\right).\label{Eq:NormalizationRes} 
\end{equation}
so the vector-resonance fields have the same convention as the one employed in the definition of the form factors (see Sec.~\ref{app:formfactors}). Finally, $V_{\mu\nu}$ includes the $\omega$ and $\phi$ mesons in ideal mixing, 
\begin{equation}
\omega=\frac{1}{\sqrt{2}}(\bar{u}u+\bar{d}d),\hspace{1cm}\phi=\bar{s}s.
\end{equation}

In the present case, we need only to consider the LHS of the Lagrangian in Eq.~(\ref{Eq:LagEffReson}), with the effective coupling $f_V$ receiving small $SU(3)_F$ corrections. We fix $f_V$ for each resonance using the measured $V\rightarrow e^+e^-$ decay ratios. An explicit calculation shows that the decay amplitude of an on-shell resonance $V$ with polarization $\lambda$ into an electron (positron) with momentum $q_-$ ($q_+$) and polarization $s$ ($s'$) is  
\begin{equation}
\mathcal{M}_{V\rightarrow e^+e^-}=i\,\frac{e^2 f_V Q_V}{m_V}\left(g^{\mu\nu}-\frac{q^\mu q^\nu}{m_V^2}\right)\bar{u}(q_-,s)\gamma_\mu v(q_+,s')\epsilon_\nu(\lambda),  \label{Eq:VepemAmp}
\end{equation}
where $Q_V$ depends on the charge of the constituent quarks in $V$, $Q_{\rho_0}=1/\sqrt{2}$, $Q_\omega=1/3\sqrt{2}$ and $Q_\phi=-1/3$, giving
\begin{equation}
\bar{\Gamma}_{V\rightarrow e^+e^-}=\frac{4\pi\alpha^2f_V^2Q_V^2}{3m_V}. \label{Eq:VepemDecayRete}
\end{equation}
In Table~\ref{Table:HPLightMR} we list the effective parameters necessary for the description of the light-resonance propagation and decay, namely the masses, the decay width, the branching fraction for the $V\rightarrow e^+e^-$ decay and the values of $f_V$. For the latter, we obtain the errors summing up in quadrature those of the parameters propagated linearly through Eq.~(\ref{Eq:VepemDecayRete}). We have also implemented the $SU(3)_F$-breaking explicitly in the meson masses by using the measured Breit-Wigner values~\cite{Nakamura:2010zzi}.  

\begin{table}[h]
\centering
\caption{Hadronic parameters of the light and neutral meson resonances.    
\label{Table:HPLightMR}}
\begin{tabular}{|c|cccc|}
\hline
$V$&$m_V$ [MeV]&$\bar{\Gamma}_V$ [MeV]&$\mathcal{B}_{V\rightarrow e^+e^-}\times 10^{-5}$&$f_V$ [MeV]\\
\hline
$\rho_0$&775.49(34)&149.1(8)&4.72(5)&221(1)\\
$\omega$&782.65(12)&8.49(8)&7.28(14)&198(2)\\
$\phi$&1019.455(20)&4.26(4)&29.54(30)&228(2)\\
\hline
\end{tabular}
\end{table}

On the other hand, the $\bar{B}\rightarrow V \bar{K}^*$ decay amplitude can be written as
\begin{equation}
\mathcal{M}_{\bar{B}\rightarrow V \bar{K}^*}=\epsilon^{*\mu}_V\epsilon_{K^*}^{*\nu}\mathcal{M}^V_{\mu\nu}, \label{Eq:BVVVertexStr}
\end{equation}
and the contribution of the resonances to the decay is then 
\begin{equation}
\mathcal{M}_{\bar{B}\rightarrow V(\rightarrow\ell^+\ell^-) \bar{K}^*}=\frac{4\pi\alpha_{\rm em}f_VQ_V}{m_V\left(q^2-m_V^2+im_V\Gamma_V\right)}\mathcal{M}^V_{\mu\nu}(\bar{\ell}\gamma^\mu\ell)\epsilon^{*\nu}_{K^*},\label{Eq:BVVllAmp}
\end{equation}
where we ignore the off-shell dependence of the real and imaginary parts of the pole position and of the $\bar{B}\rightarrow V \bar{K}^*$  amplitude, all of which we expect to have a negligible effect. The decay amplitude in Eq.~(\ref{Eq:BVVVertexStr}) is usually parameterized in terms of the three helicity amplitudes $H_V^{0,\pm}\propto\epsilon^{*\mu}_V(0,\pm)\epsilon_{K^*}^{*\nu}(0,\pm)\mathcal{M}^V_{\mu\nu}$~\cite{Beneke:2006hg}. In principle, 5 different independent observables of the non-leptonic decay are required to extract, up to a global phase, the real and complex parts of these amplitudes. One takes the decay rate plus two out of three polarization fractions, $f_L$, $f_\perp$, $f_\parallel$, and two phases, $\phi_\perp$, $\phi_\parallel$, 
\begin{eqnarray}
&&\Gamma=\frac{1}{16 m_B \pi} \left(|A_0|^2 +|A_\perp|^2+|A_\parallel|^2\right),\\
&&F_{L,\perp,\parallel}=\frac{|A_{0,\perp,\parallel}|^2}{|A_{0}|^2+|A_{\perp}|^2+|A_{\parallel}|^2},\\
&&\phi_{\perp,\parallel}={\rm arg}\frac{A_{\perp,\parallel}}{A_0}, \label{Eq:BVVObservables} 
\end{eqnarray}
where we have used the transversity basis for the amplitudes\footnote{We use a different sign convention for the definition of the transversal amplitude $A_\perp$ as compared with Ref.~\cite{Beneke:2006hg}.}. Not all the amplitudes are equally important and in fact, in naive factorization, there exists a hierarchy among them,
\begin{equation}
H_V^0:H_V^-:H_V^+=1:\frac{\Lambda}{m_b}:\left(\frac{\Lambda}{m_b}\right)^2,\label{Eq:HelVVHierarchy}
\end{equation}
which results from the $V-A$ nature of the weak interactions and the
approximate chiral-symmetry of the strong interactions at high
energies. However, $H_V^-$ does not factorize in QCDF, although
nonfactorizable corrections have been estimated (including modelling
of end-point-divergent convolutions) and found to be large, such that
$H_V^-$ can be comparable to $H_V^0$. No factorization formula for
$H_V^+$ is known, but for $\bar B^0 \to \phi \bar K^{*0}$ it has
been extracted from experiment and found to be small, consistent with
zero. This reduces to 3 the number of independent observables needed for the description of these non-leptonic decays. We summarize in the following the results for the two relevant amplitudes $H_V^0$ and $H_V^-$ in QCD factorization~\cite{Beneke:2006hg} and we choose a normalization of the different pieces involved that makes the comparison with the other contributions to the semileptonic decay more transparent. We also discuss the corresponding predictions for the decay rate, $F_L$, $\phi_\parallel$ and $A_{\rm CP}$ compared to experimental data.

\begin{table}
\centering
\caption{Results for the $\bar{B}\rightarrow V \bar{K}^*$ helicity amplitudes and on the relevant observables, which are compared to experimental data~\cite{Amhis:2012bh}.  
\label{Table:BVVResults}}
\begin{tabular}{|c|cc|cc|cc|}
\hline
&\multicolumn{2}{|c|}{\raisebox{0.3ex}[15.pt]{$\rho_0$}}&\multicolumn{2}{|c|}{\raisebox{0.3ex}[15.pt]{$\omega$}}&\multicolumn{2}{|c|}{\raisebox{0.3ex}[15.pt]{$\phi$}}\\
\hline
&Theo.&Expt.&Theo.&Expt.&Theo.&Expt.\\
\hline
$\mathcal{B}\times10^{6}$&2.7(1.2)&3.9(0.8)&3.3(1.4)&2.0(5)&9.4(3.8)&9.8(0.7)\\
\hline
$F_L$ [$\%$]&31(20)&40(14)&53(20)&70(13)&59(19)&48(3)\\
\hline 
$\phi_{\parallel}$ [$^{\oldstylenums{0}}$]&160(40)&$-$&120(31)&$-$&139(20)&136(8)\\
\hline
$A_{\rm CP}$&$-0.11(11)$&$-0.06(9)$&14(15)&0.45(25)&0&0.01(5)\\
\hline
\end{tabular}
\end{table}

The different amplitudes can be written as a sum of products of CKM matrix elements, factorizable coefficients containing form factors and decay constants and flavor amplitudes containing Wilson coefficients and perturbative corrections in the heavy quark limit. In case of the decays under study, we have~\cite{Beneke:2006hg}
\begin{equation}
H_{\bar{B}\rightarrow\rho_0\bar{K}^{*0}}=-\frac{m_B}{\sqrt{2}}\sum_{p=c,u}\frac{\lambda_p}{\lambda_t}\left(A_{\rho \bar{K}^*}\hat{\alpha}_4^{p}-A_{\bar{K}^*\rho}\left(\frac{3}{2}\alpha_{3,{\rm EW}}^{p}+\delta_{pu}\alpha_2\right)\right),\nonumber
\end{equation}
\begin{equation}
H_{\bar{B}\rightarrow\omega\bar{K}^{*0}}=\frac{m_B}{\sqrt{2}}\sum_{p=c,u}\frac{\lambda_p}{\lambda_t}\left(A_{\bar{K}^*\omega}(2\alpha_3^p+\delta_{pu}\alpha_2)+A_{\omega \bar{K}^*}\hat{\alpha}_4^{p}\right),\nonumber
\end{equation}
\begin{equation}
H_{\bar{B}\rightarrow\phi\bar{K}^{*0}}=m_BA_{\bar{K}^*\phi}\frac{\lambda_c}{\lambda_t}(\hat{\alpha}_4^c+\alpha_3^c),
\end{equation}
where $\lambda_i=V_{ib}V_{is}^*$ and we have omitted the polarization index. The values for the amplitudes $\alpha^{(p)}_i$, $\beta_i^{(p)}$ and $\hat{\alpha}_4$ that we use are listed in Table 2 of Ref.~\cite{Beneke:2006hg}. An exception is the value for $\hat{\alpha}_4^{c-}$ that is fixed with the polarization data on the $\bar{B}\rightarrow\phi\bar{K}^*$ decay (Eq. (39) in Ref.~\cite{Beneke:2006hg}). 

The factorizable coefficients are
\begin{equation}
A^0_{V_1V_2}=\;\frac{f_{V_{2}}}{f_{K^*}}A_0^{B\rightarrow V_1}(0),\hspace{1.5cm} A^\pm_{V_1V_2}=\;\frac{m_{V_{2}}}{m_B}\frac{f_{V_{2}}}{f_{K^*}}F_\pm^{B\rightarrow V_1}(0),\label{Eq:BVVFactCoeff}
\end{equation}
with $A_0^{B\rightarrow V_1}(0)$ and $F_\pm^{B\rightarrow V_1}(0)$ helicity vector form factors in the definition employed in  Ref.~\cite{Beneke:2006hg}. For the $B\rightarrow K^*$ form factors we use the values obtained in Sec.~\ref{sect:FFs}, whereas for the other transitions we take the values given in that reference. In Table~\ref{Table:BVVResults}, we show the numerical values for the helicity amplitudes for the decays of interest and the respective values for different observables compared with experimental data. The errors are obtained adding linearly those from the input quantities using Monte Carlo techniques. 

The contribution of the light resonances to the semileptonic decay amplitudes, in terms of those of the $\bar{B}\rightarrow V \bar{K}^*$ decay, can be expressed as
\begin{equation}
H_{{\rm sl},\;V}^{0,\pm}=\frac{\alpha_{\rm em}\;G_F\;\lambda_t}{\sqrt{2}}\frac{8\pi\,Q_V\,f_{K^*}\, f_V}{\left(q^2-m_V^2+im_V\Gamma_V\right)}\left(\frac{m_B}{m_V}\right)\,H_V^{0,\pm}.\label{Eq:BVVllHelAmp} 
\end{equation}
Notice that the amplitude $H_V^+$ is completely suppressed in the SM so that the amplitudes $H^+_{V}(\bar{B}\rightarrow \bar{K}^*\gamma^*)$ in the semileptonic decays are free from this hadronic pollution. On the other hand, and despite of the expected hierarchy in Eq.~(\ref{Eq:HelVVHierarchy}), the results in Table~\ref{Table:BVVResults} show that numerically $|H_V^-|\sim|H_V^0|$, leading to polarization fractions that are consistent with the experimental values. 

Finally, and as explained in Sec.~\ref{sec:lightquarks}, we make the light-quark contribution calculated in the VMD model to converge to the one obtained in QCDF around $q^2\simeq2$ GeV$^2$,
\begin{eqnarray}
\tilde a^{\rm had,\, lq}_\mu= \tilde a^{\rm had,\,
  lq,\,VMD}_\mu\,(1-f(q^2))+\tilde a^{\rm had,\, lq,\,QCDF}_\mu\,f(q^2),
\end{eqnarray}    
by using the function
\begin{equation}
f(q^2)=\frac{1}{1 + \exp{[-a (q^2 - 2)}]}.
\end{equation}
Our numerical results are insensitive to the specific choice of $a>0$, that we take to be $a=12$.


\begin{thebibliography}{100}

\bibitem{:2012ct}
R~Aaij et~al.
\newblock {First evidence for the decay $B_s \to \mu^+ \mu^-$}.
\newblock 2012, 1211.2674.

\bibitem{DeBruyn:2012wk}
Kristof De~Bruyn, Robert Fleischer, Robert Knegjens, Patrick Koppenburg, Marcel
  Merk, et~al.
\newblock {Probing New Physics via the $B^0_s\to \mu^+\mu^-$ Effective
  Lifetime}.
\newblock {\em Phys.Rev.Lett.}, 109:041801, 2012, 1204.1737.

\bibitem{Buras:2012ru}
Andrzej~J. Buras, Jennifer Girrbach, Diego Guadagnoli, and Gino Isidori.
\newblock {On the Standard Model prediction for $BR(B_{s,d} \to \mu^+ \mu^-$)}.
\newblock 2012, 1208.0934.

\bibitem{Wei:2009zv}
  J.~-T.~Wei {\it et al.}  [BELLE Collaboration],
  Phys.\ Rev.\ Lett.\  {\bf 103} (2009) 171801
  [arXiv:0904.0770 [hep-ex]].

\bibitem{Aaltonen:2011cn}
  T.~Aaltonen {\it et al.}  [CDF Collaboration],
  Phys.\ Rev.\ Lett.\  {\bf 106} (2011) 161801
  [arXiv:1101.1028 [hep-ex]].

\bibitem{LHCbNote}
{\bf LHCb}~Collaboration.
\newblock {Differential branching fraction and angular analysis of the
  $B^0\rightarrow K^{*0}\ell^+\ell^-$ decay}.
\newblock LHCb-CONF-2012-008.

\bibitem{Ritchie:2013mx}
  J.~L.~Ritchie [BABAR Collaboration],
  arXiv:1301.1700 [hep-ex].

\bibitem{Aaltonen:2011ja}
  T.~Aaltonen {\it et al.}  [CDF Collaboration],
  Phys.\ Rev.\ Lett.\  {\bf 108} (2012) 081807
  [arXiv:1108.0695 [hep-ex]].

\bibitem{Aubert:2008ps}
  B.~Aubert {\it et al.}  [BABAR Collaboration],
  Phys.\ Rev.\ Lett.\  {\bf 102} (2009) 091803
  [arXiv:0807.4119 [hep-ex]].

\bibitem{:2012vwa}
  J.~P.~Lees {\it et al.}  [BABAR Collaboration],
  Phys.\ Rev.\ D {\bf 86} (2012) 032012
  [arXiv:1204.3933 [hep-ex]].

\bibitem{Aaij:2012cq}
  RAaij {\it et al.}  [LHCb Collaboration],
  JHEP {\bf 1207} (2012) 133
  [arXiv:1205.3422 [hep-ex]].

\bibitem{Aaltonen:2011qs}
  T.~Aaltonen {\it et al.}  [CDF Collaboration],
  Phys.\ Rev.\ Lett.\  {\bf 107} (2011) 201802
  [arXiv:1107.3753 [hep-ex]].

\bibitem{Schaack:2012fsa}
  {\bf LHCb}~Collaboration, Talk at Moriond 2012, LHCb-CONF-2012-003;
  P.~N.~Schaack,
  CERN-THESIS-2012-064.

\bibitem{Ali:1991is}
Ahmed Ali, T.~Mannel, and T.~Morozumi.
\newblock {Forward backward asymmetry of dilepton angular distribution in the
  decay $b \to s l^+ l^-$}.
\newblock {\em Phys.Lett.}, B273:505--512, 1991.

\bibitem{Burdman:1998mk}
Gustavo Burdman.
\newblock {Short distance coefficients and the vanishing of the lepton
  asymmetry in $B \to V \ell^+ \ell^-$}.
\newblock {\em Phys.Rev.}, D57:4254--4257, 1998, hep-ph/9710550.

\bibitem{Melikhov:1998cd}
D.~Melikhov, N.~Nikitin, and S.~Simula.
\newblock {Probing right-handed currents in $B \to K^* l^+ l^-$ transitions}.
\newblock {\em Phys.Lett.}, B442:381--389, 1998, hep-ph/9807464.

\bibitem{Ali:1999mm}
Ahmed Ali, Patricia Ball, L.T. Handoko, and G.~Hiller.
\newblock {A Comparative study of the decays $B \to (K, K^{*}) \ell^+ \ell^-$
  in standard model and supersymmetric theories}.
\newblock {\em Phys.Rev.}, D61:074024, 2000, hep-ph/9910221.

\bibitem{Kruger:1999xa}
Frank Kruger, Lalit~M. Sehgal, Nita Sinha, and Rahul Sinha.
\newblock {Angular distribution and CP asymmetries in the decays $\bar B \to
  K^- \pi^+ e^- e^+$ and $\bar B \to \pi^- \pi^+ e^-e^+$}.
\newblock {\em Phys.Rev.}, D61:114028, 2000, hep-ph/9907386.

\bibitem{Kim:2000dq}
C.S. Kim, Yeong~Gyun Kim, Cai-Dian Lu, and Takuya Morozumi.
\newblock {Azimuthal angle distribution in $B \to K^* (\to K \pi) l^+ l^-$ at
  low invariant $m(l^+ l^-)$ region}.
\newblock {\em Phys.Rev.}, D62:034013, 2000, hep-ph/0001151.

\bibitem{Burdman:2000ku}
Gustavo Burdman and Gudrun Hiller.
\newblock {Semileptonic form-factors from $B \to K^* \gamma$ decays in the
  large energy limit}.
\newblock {\em Phys.Rev.}, D63:113008, 2001, hep-ph/0011266.

\bibitem{Grossman:2000rk}
Yuval Grossman and Dan Pirjol.
\newblock {Extracting and using photon polarization information in radiative B
  decays}.
\newblock {\em JHEP}, 0006:029, 2000, hep-ph/0005069.

\bibitem{Kim:2001xua}
C.S. Kim, Yeong~Gyun Kim, and Cai-Dian Lu.
\newblock {Possible supersymmetric effects on angular distributions in $B \to
  K^* (\to K \pi) l^+ l^-$ decays}.
\newblock {\em Phys.Rev.}, D64:094014, 2001, hep-ph/0102168.

\bibitem{Beneke:2001at}
M.~Beneke, T.~Feldmann, and D.~Seidel.
\newblock {Systematic approach to exclusive $B \to V l^+ l^-, V \gamma$
  decays}.
\newblock {\em Nucl.Phys.}, B612:25--58, 2001, hep-ph/0106067.

\bibitem{Bosch:2001gv}
Stefan~W. Bosch and Gerhard Buchalla.
\newblock {The Radiative decays $B \to V \gamma$ at next-to-leading order in
  QCD}.
\newblock {\em Nucl.Phys.}, B621:459--478, 2002, hep-ph/0106081.

\bibitem{Ali:2002qc}
Ahmed Ali and A.~Salim Safir.
\newblock {Helicity analysis of the decays $B \to K^{*} \ell^{+} \ell^{-}$ and
  $B \to \rho \ell \nu\_{\ell}$ in the large energy effective theory}.
\newblock {\em Eur.Phys.J.}, C25:583--601, 2002, hep-ph/0205254.

\bibitem{Faessler:2002ut}
Amand Faessler, T.~Gutsche, M.A. Ivanov, J.G. Korner, and Valery~E.
  Lyubovitskij.
\newblock {The Exclusive rare decays $B \to K(K^*) \bar{\ell} \ell$ and $B_c
  \to$ D(D*) $\bar{\ell} \ell$ in a relativistic quark model}.
\newblock {\em Eur.Phys.J.direct}, C4:18, 2002, hep-ph/0205287.

\bibitem{Kagan:2001zk}
Alexander~L. Kagan and Matthias Neubert.
\newblock {Isospin breaking in $B \to K* \gamma$ decays}.
\newblock {\em Phys.Lett.}, B539:227--234, 2002, hep-ph/0110078.

\bibitem{Feldmann:2002iw}
Thorsten Feldmann and Joaquim Matias.
\newblock {Forward-backward and isospin asymmetry for $B \to K^* l^+ l^-$ decay
  in the standard model and in supersymmetry}.
\newblock {\em JHEP}, 0301:074, 2003, hep-ph/0212158.

\bibitem{Beneke:2004dp}
M.~Beneke, Th. Feldmann, and D.~Seidel.
\newblock {Exclusive radiative and electroweak $b \to d$ and $b \to s$ penguin
  decays at NLO}.
\newblock {\em Eur.Phys.J.}, C41:173--188, 2005, hep-ph/0412400.

\bibitem{Grinstein:2004vb}
Benjamin Grinstein and Dan Pirjol.
\newblock {Exclusive rare $B \to K^* \ell^+ \ell^-$ decays at low recoil:
  Controlling the long-distance effects}.
\newblock {\em Phys.Rev.}, D70:114005, 2004, hep-ph/0404250.

\bibitem{Grinstein:2005ud}
Benjamin Grinstein and Dan Pirjol.
\newblock {Factorization in $B \to K \pi l^+ l^-$ decays}.
\newblock {\em Phys.Rev.}, D73:094027, 2006, hep-ph/0505155.

\bibitem{Kruger:2005ep}
Frank Kruger and Joaquim Matias.
\newblock {Probing new physics via the transverse amplitudes of $B^0 \to K^{*0}
  (\to K^- \pi^+) \ell^+ \ell^-$ at large recoil}.
\newblock {\em Phys.Rev.}, D71:094009, 2005, hep-ph/0502060.

\bibitem{Matias:2005rs}
Joaquim Matias.
\newblock {The Angular distribution of $B^0 \to K^{*0} (\to K^- \pi^+) \ell^+
  \ell^-$ at large recoil at large recoil in and beyond the SM}.
\newblock {\em PoS}, HEP2005:281, 2006, hep-ph/0511274.

\bibitem{Ball:2006cva}
Patricia Ball and Roman Zwicky.
\newblock {Time-dependent CP Asymmetry in $B \to K^* \gamma$ as a (Quasi) Null
  Test of the Standard Model}.
\newblock {\em Phys.Lett.}, B642:478--486, 2006, hep-ph/0609037.

\bibitem{Ball:2006eu}
Patricia Ball, Gareth~W. Jones, and Roman Zwicky.
\newblock {$B \to V \gamma$ beyond QCD factorisation}.
\newblock {\em Phys.Rev.}, D75:054004, 2007, hep-ph/0612081.

\bibitem{Lunghi:2006hc}
E.~Lunghi and J.~Matias.
\newblock {Huge right-handed current effects in $B\to K^*(K \pi)\ell^+\ell^-$
  in supersymmetry}.
\newblock {\em JHEP}, 0704:058, 2007, hep-ph/0612166.

\bibitem{Bobeth:2008ij}
Christoph Bobeth, Gudrun Hiller, and Giorgi Piranishvili.
\newblock {CP Asymmetries in bar $B \to \bar{K}^* (\to \bar{K} \pi) \bar{\ell}
  \ell$ and Untagged $\bar{B}_s$, $B_s \to \phi (\to K^{+} K^-) \bar{\ell}
  \ell$ Decays at NLO}.
\newblock {\em JHEP}, 0807:106, 2008, 0805.2525.

\bibitem{Egede:2008uy}
U.~Egede, T.~Hurth, J.~Matias, M.~Ramon, and W.~Reece.
\newblock {New observables in the decay mode $\bar B \to \bar K^{*0} l^+ l^-$}.
\newblock {\em JHEP}, 0811:032, 2008, 0807.2589.

\bibitem{Altmannshofer:2008dz}
Wolfgang Altmannshofer, Patricia Ball, Aoife Bharucha, Andrzej~J. Buras,
  David~M. Straub, et~al.
\newblock {Symmetries and Asymmetries of $B \to K^{*} \mu^{+} \mu^{-}$ Decays
  in the Standard Model and Beyond}.
\newblock {\em JHEP}, 0901:019, 2009, 0811.1214.

\bibitem{Egede:2009te}
U.~Egede, T.~Hurth, J.~Matias, M.~Ramon, and W.~Reece.
\newblock {The exclusive $B \to K^*(\to K \pi) l^+ l^-$ decay: CP conserving
  observables}.
\newblock {\em Acta Phys.Polon.}, B3:151--157, 2010, 0912.1339.

\bibitem{Egede:2009tp}
Ulrik Egede, Tobias Hurth, Joaquim Matias, Marc Ramon, and Will Reece.
\newblock {New physics reach of CP violating observables in the decay $B \to
  K^* l^+ l^-$}.
\newblock {\em PoS}, EPS-HEP2009:184, 2009, 0912.1349.

\bibitem{Bobeth:2010wg}
Christoph Bobeth, Gudrun Hiller, and Danny van Dyk.
\newblock {The Benefits of $\bar{B} \to \bar{K}^* l^+ l^-$ Decays at Low
  Recoil}.
\newblock {\em JHEP}, 1007:098, 2010, 1006.5013.

\bibitem{Bharucha:2010bb}
Aoife Bharucha and William Reece.
\newblock {Constraining new physics with $B \to K^* \mu^+ \mu^-$ in the early
  LHC era}.
\newblock {\em Eur.Phys.J.}, C69:623--640, 2010, 1002.4310.

\bibitem{Alok:2009tz}
Ashutosh~Kumar Alok, Amol Dighe, Diptimoy Ghosh, David London, Joaquim Matias,
  et~al.
\newblock {New-physics contributions to the forward-backward asymmetry in $B
  \to K^* \mu^+ \mu^-$}.
\newblock {\em JHEP}, 1002:053, 2010, 0912.1382.

\bibitem{Egede:2010zc}
Ulrik Egede, Tobias Hurth, Joaquim Matias, Marc Ramon, and Will Reece.
\newblock {New physics reach of the decay mode $\bar{B} \to
  \bar{K}^{*0}\ell^+\ell^-$}.
\newblock {\em JHEP}, 1010:056, 2010, 1005.0571.

\bibitem{Khodjamirian:2010vf}
A.~Khodjamirian, Th. Mannel, A.A. Pivovarov, and Y.-M. Wang.
\newblock {Charm-loop effect in $B \to K^{(*)} \ell^{+} \ell^{-}$ and $B\to
  K^*\gamma$}.
\newblock {\em JHEP}, 1009:089, 2010, 1006.4945.

\bibitem{Alok:2010zd}
Ashutosh~Kumar Alok, Alakabha Datta, Amol Dighe, Murugeswaran Duraisamy,
  Diptimoy Ghosh, et~al.
\newblock {New Physics in $b \to s \mu^+ \mu^-$: CP-Conserving Observables}.
\newblock 2010, 1008.2367.

\bibitem{Kou:2010kn}
E.~Kou, A.~{Le Yaouanc}, and A.~Tayduganov.
\newblock {Determining the photon polarization of the $b \to s \gamma$ using
  the $B \to K_1(1270) \gamma \to (K \pi \pi) \gamma$ decay}.
\newblock {\em Phys.Rev.}, D83:094007, 2011, 1011.6593.

\bibitem{Chang:2010zy}
Qin Chang, Xin-Qiang Li, and Ya-Dong Yang.
\newblock {$B \to K^{\ast} l^+ l^-$, $K l^+ l^-$ decays in a family
  non-universal $Z^{\prime}$ model}.
\newblock {\em JHEP}, 1004:052, 2010, 1002.2758.

\bibitem{ReeceThesis}
W.~R. Reece.
\newblock {Exploiting angular correlations in the rare decay $B \to K^* \mu^+
  \mu^-$ at LHCb}.
\newblock {\em CERN-THESIS-2010-095}, 2010.

\bibitem{Beylich:2011aq}
M.~Beylich, G.~Buchalla, and Th. Feldmann.
\newblock {Theory of $B \to K^{(*)} l^+l^-$ decays at high $q^2$: OPE and
  quark-hadron duality}.
\newblock {\em Eur.Phys.J.}, C71:1635, 2011, 1101.5118.
\newblock * Temporary entry *.

\bibitem{Bobeth:2011gi}
Christoph Bobeth, Gudrun Hiller, and Danny van Dyk.
\newblock {More Benefits of Semileptonic Rare $B$ Decays at Low Recoil: CP
  Violation}.
\newblock 2011, 1105.0376.
\newblock * Temporary entry *.

\bibitem{Becirevic:2011bp}
Damir Becirevic and Elia Schneider.
\newblock {On transverse asymmetries in $B \to K^* l^+l^-$}.
\newblock {\em Nucl.Phys.}, B854:321--339, 2012, 1106.3283.

\bibitem{Alok:2011gv}
Ashutosh~Kumar Alok, Alakabha Datta, Amol Dighe, Murugeswaran Duraisamy,
  Diptimoy Ghosh, et~al.
\newblock {New Physics in $b \to s \mu^+ \mu^-$: CP-Violating Observables}.
\newblock {\em JHEP}, 1111:122, 2011, 1103.5344.

\bibitem{Lu:2011jm}
Cai-Dian Lu and Wei Wang.
\newblock {Analysis of $B\to K^*_J (\to K \pi) \mu^+\mu^-$ in the higher kaon
  resonance region}.
\newblock {\em Phys.Rev.}, D85:034014, 2012, 1111.1513.

\bibitem{Wang:2011aa}
Ru-Min Wang, Yuan-Guo Xu, Yi-Long Wang, and Ya-Dong Yang.
\newblock {Revisiting $B_s\to\mu^+\mu^-$ and $B\to K^{(*)}\mu^+\mu^-$ decays in
  the MSSM with and without R-parity}.
\newblock {\em Phys.Rev.}, D85:094004, 2012, 1112.3174.

\bibitem{Matias:2012xw}
Joaquim Matias, Federico Mescia, Marc Ramon, and Javier Virto.
\newblock {Complete Anatomy of $\bar{B}_d \to \bar{K}^{* 0} (\to K \pi)l^+l^-$
  and its angular distribution}.
\newblock {\em JHEP}, 1204:104, 2012, 1202.4266.

\bibitem{Becirevic:2012dp}
Damir Becirevic and Andrey Tayduganov.
\newblock {Impact of $B\to K^\ast_0 \ell^+\ell^-$ on the New Physics search in
  $B\to K^\ast \ell^+\ell^-$ decay}.
\newblock 2012, 1207.4004.

\bibitem{Korchin:2012kz}
Alexander~Yu. Korchin and Vladimir~A. Kovalchuk.
\newblock {Contribution of vector resonances to the ${\bar B}_d^0 \to {\bar
  K}^{*0} \mu^+ \mu^-$ decay}.
\newblock {\em Eur.Phys.J.}, C72:2155, 2012, 1205.3683.

\bibitem{Das:2012kz}
Diganta Das and Rahul Sinha.
\newblock {New Physics Effects and Hadronic Form Factor Uncertainties in $B\to
  K^* \ell^+ \ell^-$}.
\newblock {\em Phys.Rev.}, D86:056006, 2012, 1205.1438.

\bibitem{Matias:2012qz}
Joaquim Matias.
\newblock {On the S-wave pollution of $B \to K^* l^+l^-$ observables}.
\newblock 2012, 1209.1525.

\bibitem{Blake:2012mb}
Thomas Blake, Ulrik Egede, and Alex Shires.
\newblock {The effect of S-wave interference on the $B^0 \to K^{\ast
  0}\ell^+\ell^-$ angular observables}.
\newblock 2012, 1210.5279.

\bibitem{Altmannshofer:2011gn}
Wolfgang Altmannshofer, Paride Paradisi, and David~M. Straub.
\newblock {Model-Independent Constraints on New Physics in $b \to s$
  Transitions}.
\newblock {\em JHEP}, 1204:008, 2012, 1111.1257.

\bibitem{DescotesGenon:2011yn}
Sebastien Descotes-Genon, Diptimoy Ghosh, Joaquim Matias, and Marc Ramon.
\newblock {Exploring New Physics in the $C_7-C_7'$ plane}.
\newblock {\em JHEP}, 1106:099, 2011, 1104.3342.

\bibitem{Beaujean:2012uj}
Frederik Beaujean, Christoph Bobeth, Danny van Dyk, and Christian Wacker.
\newblock {Bayesian Fit of Exclusive $b \to s \bar\ell\ell$ Decays: The
  Standard Model Operator Basis}.
\newblock {\em JHEP}, 1208:030, 2012, 1205.1838.

\bibitem{DescotesGenon:2012zf}
Sebastien Descotes-Genon, Joaquim Matias, Marc Ramon, and Javier Virto.
\newblock {Implications from clean observables for the binned analysis of $B
  \to K^*ll$ at large recoil}.
\newblock 2012, 1207.2753.

\bibitem{Altmannshofer:2012az}
Wolfgang Altmannshofer and David~M. Straub.
\newblock {Cornering New Physics in $b\to s$ Transitions}.
\newblock {\em JHEP}, 1208:121, 2012, 1206.0273.

\bibitem{Behring:2012mv}
Arnd Behring, Christian Gross, Gudrun Hiller, and Stefan Schacht.
\newblock {Squark Flavor Implications from $B\to K^* \ell^+ \ell^-$}.
\newblock {\em JHEP}, 1208:152, 2012, 1205.1500.

\bibitem{Ball:1998kk}
Patricia Ball and Vladimir~M. Braun.
\newblock {Exclusive semileptonic and rare $B$ meson decays in QCD}.
\newblock {\em Phys.Rev.}, D58:094016, 1998, hep-ph/9805422.

\bibitem{Charles:1998dr}
J.~Charles, A.~{Le Yaouanc}, L.~Oliver, O.~Pene, and J.C. Raynal.
\newblock {Heavy to light form-factors in the heavy mass to large energy limit
  of QCD}.
\newblock {\em Phys.Rev.}, D60:014001, 1999, hep-ph/9812358.

\bibitem{Beneke:2000wa}
M.~Beneke and T.~Feldmann.
\newblock {Symmetry breaking corrections to heavy to light $B$ meson
  form-factors at large recoil}.
\newblock {\em Nucl.Phys.}, B592:3--34, 2001, hep-ph/0008255.

\bibitem{Becirevic:2012dx}
Damir Becirevic, Emi Kou, Alain {Le Yaouanc}, and Andrey Tayduganov.
\newblock {Future prospects for the determination of the Wilson coefficient
  $C_{7\gamma}^\prime$}.
\newblock {\em JHEP}, 1208:090, 2012, 1206.1502.

\bibitem{Chetyrkin:1996vx}
Konstantin~G. Chetyrkin, Mikolaj Misiak, and Manfred Munz.
\newblock {Weak radiative $B$ meson decay beyond leading logarithms}.
\newblock {\em Phys.Lett.}, B400:206--219, 1997, hep-ph/9612313.

\bibitem{Bharucha:2010im}
Aoife Bharucha, Thorsten Feldmann, and Michael Wick.
\newblock {Theoretical and Phenomenological Constraints on Form Factors for
  Radiative and Semi-Leptonic $B$-Meson Decays}.
\newblock {\em JHEP}, 1009:090, 2010, 1004.3249.

\bibitem{Becirevic:2006nm}
Damir Becirevic, Vittorio Lubicz, and Federico Mescia.
\newblock {An Estimate of the $B \to K^* \gamma$ form factor}.
\newblock {\em Nucl.Phys.}, B769:31--43, 2007, hep-ph/0611295.

\bibitem{Liu:2011raa}
Zhaofeng Liu, Stefan Meinel, Alistair Hart, Ron~R. Horgan, Eike~H. Muller,
  et~al.
\newblock {A Lattice calculation of $B \to K^{(*)}$ form factors}.
\newblock 2011, 1101.2726.

\bibitem{Ball:2004rg}
Patricia Ball and Roman Zwicky.
\newblock {$B_{d,s} \to \rho, \omega, K^*, \phi$ decay form-factors from
  light-cone sum rules revisited}.
\newblock {\em Phys.Rev.}, D71:014029, 2005, hep-ph/0412079.

\bibitem{Bauer:2000yr}
Christian~W. Bauer, Sean Fleming, Dan Pirjol, and Iain~W. Stewart.
\newblock {An Effective field theory for collinear and soft gluons: Heavy to
  light decays}.
\newblock {\em Phys.Rev.}, D63:114020, 2001, hep-ph/0011336.

\bibitem{Bauer:2001yt}
Christian~W. Bauer, Dan Pirjol, and Iain~W. Stewart.
\newblock {Soft collinear factorization in effective field theory}.
\newblock {\em Phys.Rev.}, D65:054022, 2002, hep-ph/0109045.

\bibitem{Beneke:2002ph}
M.~Beneke, A.P. Chapovsky, M.~Diehl, and T.~Feldmann.
\newblock {Soft collinear effective theory and heavy to light currents beyond
  leading power}.
\newblock {\em Nucl.Phys.}, B643:431--476, 2002, hep-ph/0206152.

\bibitem{Beneke:2002ni}
M.~Beneke and T.~Feldmann.
\newblock {Multipole expanded soft collinear effective theory with nonAbelian
  gauge symmetry}.
\newblock {\em Phys.Lett.}, B553:267--276, 2003, hep-ph/0211358.

\bibitem{Beneke:2003pa}
M.~Beneke and T.~Feldmann.
\newblock {Factorization of heavy to light form-factors in soft collinear
  effective theory}.
\newblock {\em Nucl.Phys.}, B685:249--296, 2004, hep-ph/0311335.

\bibitem{Beneke:2005gs}
M.~Beneke and D.~Yang.
\newblock {Heavy-to-light $B$ meson form-factors at large recoil energy:
  Spectator-scattering corrections}.
\newblock {\em Nucl.Phys.}, B736:34--81, 2006, hep-ph/0508250.

\bibitem{Beneke:2004rc}
M.~Beneke, Y.~Kiyo, and D.s. Yang.
\newblock {Loop corrections to subleading heavy quark currents in SCET}.
\newblock {\em Nucl.Phys.}, B692:232--248, 2004, hep-ph/0402241.

\bibitem{Becher:2004kk}
Thomas Becher and Richard~J. Hill.
\newblock {Loop corrections to heavy-to-light form-factors and evanescent
  operators in SCET}.
\newblock {\em JHEP}, 0410:055, 2004, hep-ph/0408344.

\bibitem{Kirilin:2005xz}
G.G. Kirilin.
\newblock {Loop corrections to the form-factors in $B \to \pi l \nu$ decay}.
\newblock 2005, hep-ph/0508235.

\bibitem{Colangelo:1995jv}
P.~Colangelo, F.~{De Fazio}, Pietro Santorelli, and E.~Scrimieri.
\newblock {QCD sum rule analysis of the decays $B \to K \ell^{+} \ell^{-}$ and
  $B \to K^{*} \ell^{+} \ell^{-}$}.
\newblock {\em Phys.Rev.}, D53:3672--3686, 1996, hep-ph/9510403.

\bibitem{Ivanov:2007cw}
Mikhail~A. Ivanov, Jurgen~G. Korner, Sergey~G. Kovalenko, and Craig~D. Roberts.
\newblock {$B$ to light-meson transition form-factors}.
\newblock {\em Phys.Rev.}, D76:034018, 2007, nucl-th/0703094.

\bibitem{Bagan:1997bp}
E.~Bagan, Patricia Ball, and Vladimir~M. Braun.
\newblock {Radiative corrections to the decay $B \to \pi e \nu$ and the heavy
  quark limit}.
\newblock {\em Phys.Lett.}, B417:154--162, 1998, hep-ph/9709243.

\bibitem{TFtalk}
T.~Feldmann.
\newblock {Talk at Workshop on the Physics Reach of Rare and Exclusive
  semileptonic $B$ decays, University of Sussex, September 10-11, 2012}.

\bibitem{Beneke:2001ev}
M.~Beneke, G.~Buchalla, M.~Neubert, and Christopher~T. Sachrajda.
\newblock {QCD factorization in $B \to \pi K, \pi \pi$ decays and extraction of
  Wolfenstein parameters}.
\newblock {\em Nucl.Phys.}, B606:245--321, 2001, hep-ph/0104110.

\bibitem{Voloshin:1996gw}
M.B. Voloshin.
\newblock {Large ${\cal O} (m_c^{-2})$ nonperturbative correction to the
  inclusive rate of the decay $B \to X_s \gamma$}.
\newblock {\em Phys.Lett.}, B397:275--278, 1997, hep-ph/9612483.

\bibitem{Ligeti:1997tc}
Zoltan Ligeti, Lisa Randall, and Mark~B. Wise.
\newblock {Comment on nonperturbative effects in $\bar B \to X_s \gamma$}.
\newblock {\em Phys.Lett.}, B402:178--182, 1997, hep-ph/9702322.

\bibitem{Khodjamirian:2012rm}
A.~Khodjamirian, Th. Mannel, and Y.-M. Wang.
\newblock {$B \to K \ell^{+}\ell^{-}$ decay at large hadronic recoil}.
\newblock 2012, 1211.0234.

\bibitem{Weinberg:1978kz}
Steven Weinberg.
\newblock {Phenomenological Lagrangians}.
\newblock {\em Physica}, A96:327, 1979.

\bibitem{Gasser:1983yg}
J.~Gasser and H.~Leutwyler.
\newblock {Chiral Perturbation Theory to One Loop}.
\newblock {\em Annals Phys.}, 158:142, 1984.

\bibitem{Ecker:1988te}
G.~Ecker, J.~Gasser, A.~Pich, and E.~de~Rafael.
\newblock {The Role of Resonances in Chiral Perturbation Theory}.
\newblock {\em Nucl.Phys.}, B321:311, 1989.

\bibitem{Ecker:1989yg}
G.~Ecker, J.~Gasser, H.~Leutwyler, A.~Pich, and E.~de~Rafael.
\newblock {Chiral Lagrangians for Massive Spin 1 Fields}.
\newblock {\em Phys.Lett.}, B223:425, 1989.

\bibitem{Oller:1998hw}
J.A. Oller, E.~Oset, and J.R. Pelaez.
\newblock {Meson meson interaction in a nonperturbative chiral approach}.
\newblock {\em Phys.Rev.}, D59:074001, 1999, hep-ph/9804209.

\bibitem{Oller:1998zr}
J.A. Oller and E.~Oset.
\newblock {N/D description of two meson amplitudes and chiral symmetry}.
\newblock {\em Phys.Rev.}, D60:074023, 1999, hep-ph/9809337.

\bibitem{Cirigliano:2011ny}
Vincenzo Cirigliano, Gerhard Ecker, Helmut Neufeld, Antonio Pich, and Jorge
  Portoles.
\newblock {Kaon Decays in the Standard Model}.
\newblock {\em Rev.Mod.Phys.}, 84:399, 2012, 1107.6001.

\bibitem{Bijnens:2010ws}
Johan Bijnens and Ilaria Jemos.
\newblock {Hard Pion Chiral Perturbation Theory for $B\to\pi$ and $D\to\pi$
  Formfactors}.
\newblock {\em Nucl.Phys.}, B840:54--66, 2010, 1006.1197.

\bibitem{Colangelo:2012ew}
Gilberto Colangelo, Massimiliano Procura, Lorena Rothen, Ramon Stucki, and
  Jaume Tarrus.
\newblock {On the factorization of chiral logarithms in the pion form factors}.
\newblock 2012, 1208.0498.

\bibitem{'tHooft:1973jz}
Gerard {'t Hooft}.
\newblock {A Planar Diagram Theory for Strong Interactions}.
\newblock {\em Nucl.Phys.}, B72:461, 1974.

\bibitem{Lichard:1997ya}
Peter Lichard.
\newblock {Some implications of meson dominance in weak interactions}.
\newblock {\em Phys.Rev.}, D55:5385--5407, 1997, hep-ph/9702345.

\bibitem{Korchin:2011ze}
Alexander~Yu. Korchin and Vladimir~A. Kovalchuk.
\newblock {Asymmetries in ${\bar B}_d^0 \to {\bar K}^{*0} e^+ e^-$ decay and
  contribution of vector resonances}.
\newblock 2011, 1111.4093.

\bibitem{Beneke:2006hg}
Martin Beneke, Johannes Rohrer, and Deshan Yang.
\newblock {Branching fractions, polarisation and asymmetries of $B \to VV$
  decays}.
\newblock {\em Nucl.Phys.}, B774:64--101, 2007, hep-ph/0612290.

\bibitem{Kagan:2004uw}
Alexander~L. Kagan.
\newblock {Polarization in $B \to VV$ decays}.
\newblock {\em Phys.Lett.}, B601:151--163, 2004, hep-ph/0405134.

\bibitem{Aubert:2004xc}
Bernard Aubert et~al.
\newblock {Measurement of the $B^0 \to \phi K^0$ decay amplitudes}.
\newblock {\em Phys.Rev.Lett.}, 93:231804, 2004, hep-ex/0408017.

\bibitem{Chen:2005zv}
K.-F. Chen et~al.
\newblock {Measurement of polarization and triple-product correlations in $B
  \to \phi K*$ decays}.
\newblock {\em Phys.Rev.Lett.}, 94:221804, 2005, hep-ex/0503013.

\bibitem{SerraPrivate}
N.~Serra.
\newblock {Private communication.}

\bibitem{MatiasVirtoCP}
Joaquim Matias, {S. Descotes-Genon}, and {Javier Virto}.
\newblock {To appear.}

\bibitem{LHCb:2012kz}
R~Aaij et~al.
\newblock {Measurement of the CP asymmetry in $B^0 \to K^{*0} \mu^+ \mu^-$
  decays}.
\newblock 2012, 1210.4492.

\bibitem{LHCbElectron}
J.~Lefran\c{c}ois and M.H. Schune.
\newblock {Measuring the photon polarization in $b\to s \gamma$ using the $B
  \to K^* e^+ e^-$ decay channel}.
\newblock 2009.

\bibitem{Jacob:1959at}
M.~Jacob and G.C. Wick.
\newblock {On the general theory of collisions for particles with spin}.
\newblock {\em Annals Phys.}, 7:404--428, 1959.

\bibitem{Charles:2004jd}
J.~Charles et~al.
\newblock {CP violation and the CKM matrix: Assessing the impact of the
  asymmetric $B$ factories}.
\newblock {\em Eur.Phys.J.}, C41:1--131, 2005, hep-ph/0406184.

\bibitem{Nakamura:2010zzi}
K.~Nakamura et~al.
\newblock {Review of particle physics}.
\newblock {\em J.Phys.G}, G37:075021, 2010.

\bibitem{Colangelo:2010et}
Gilberto Colangelo, Stephan Durr, Andreas Juttner, Laurent Lellouch, Heinrich
  Leutwyler, et~al.
\newblock {Review of lattice results concerning low energy particle physics}.
\newblock {\em Eur.Phys.J.}, C71:1695, 2011, 1011.4408.

\bibitem{Chetyrkin:2000yt}
K.G. Chetyrkin, Johann~H. Kuhn, and M.~Steinhauser.
\newblock {RunDec: A Mathematica package for running and decoupling of the
  strong coupling and quark masses}.
\newblock {\em Comput.Phys.Commun.}, 133:43--65, 2000, hep-ph/0004189.

\bibitem{Na:2012kp}
Heechang Na, Chris~J. Monahan, Christine~T.H. Davies, Ron Horgan, G.~Peter
  Lepage, et~al.
\newblock {The $B$ and $B_s$ Meson Decay Constants from Lattice QCD}.
\newblock {\em Phys.Rev.}, D86:034506, 2012, 1202.4914.

\bibitem{Amhis:2012bh}
Y.~Amhis et~al.
\newblock {Averages of $b$-hadron, $c$-hadron, and $\tau$-lepton properties as
  of early 2012}.
\newblock 2012, 1207.1158.

\bibitem{Dimou:2012un}
  M.~Dimou, J.~Lyon and R.~Zwicky,
  arXiv:1212.2242 [hep-ph].

\bibitem{Bobeth:2012vn}
  C.~Bobeth, G.~Hiller and D.~van Dyk,
  arXiv:1212.2321 [hep-ph].

\end{thebibliography}
\end{document}